\documentclass[12pt]{article}

\usepackage{color, colortbl}
\usepackage{captcont}
\usepackage{lineno}
\usepackage{epsfig}
\usepackage{epstopdf}
\usepackage{amsmath}
\usepackage{hhline}
\usepackage{amssymb}
\usepackage{mathptmx}
\usepackage[scaled=.92]{helvet}
\usepackage{courier}
\usepackage{cite}
\usepackage{rotating}
\usepackage{url}
\usepackage{multirow}
\usepackage{arydshln}
\usepackage{bm} 

\newcommand{\redcc}{$\sigma_{\rm red}^{c\overline{c}}\, $}
\newcommand{\redbb}{$\sigma_{\rm red}^{b\overline{b}}\, $}

\newcommand\reading[1]{{\color{black} #1}}
\newlength{\dinwidth}
\newlength{\dinmargin}
\def\Journal#1#2#3#4{{#1} {\bf #2}, (#3) #4}

\def\EPJC{{Eur. Phys. J.} {\bf C}}

\def\be{\begin{equation}}
\def\ee{\end{equation}}
\def\bea{\begin{eqnarray}}
\def\eea{\end{eqnarray}}
\def\etal{{\it et~al.}}

\begin{document}  
\newcommand{\pom}{{I\!\!P}}
\newcommand{\slowpi}{\pi_{\mathit{slow}}}
\newcommand{\fiidiii}{F_2^{D(3)}}
\newcommand{\fiidiiiarg}{\fiidiii\,(\beta,\,Q^2,\,x)}
\newcommand{\n}{1.19\pm 0.06 (stat.) \pm0.07 (syst.)}
\newcommand{\nz}{1.30\pm 0.08 (stat.)^{+0.08}_{-0.14} (syst.)}
\newcommand{\fiidiiiful}{F_2^{D(4)}\,(\beta,\,Q^2,\,x,\,t)}
\newcommand{\fiipom}{\tilde F_2^D}
\newcommand{\ALPHA}{1.10\pm0.03 (stat.) \pm0.04 (syst.)}
\newcommand{\ALPHAZ}{1.15\pm0.04 (stat.)^{+0.04}_{-0.07} (syst.)}
\newcommand{\fiipomarg}{\fiipom\,(\beta,\,Q^2)}
\newcommand{\pomflux}{f_{\pom / p}}
\newcommand{\nxpom}{1.19\pm 0.06 (stat.) \pm0.07 (syst.)}
\newcommand {\gapprox}
   {\raisebox{-0.7ex}{$\stackrel {\textstyle>}{\sim}$}}
\newcommand {\lapprox}
   {\raisebox{-0.7ex}{$\stackrel {\textstyle<}{\sim}$}}
\def\gsim{\,\lower.25ex\hbox{$\scriptstyle\sim$}\kern-1.30ex%
\raise 0.55ex\hbox{$\scriptstyle >$}\,}
\def\lsim{\,\lower.25ex\hbox{$\scriptstyle\sim$}\kern-1.30ex%
\raise 0.55ex\hbox{$\scriptstyle <$}\,}
\newcommand{\pomfluxarg}{f_{\pom / p}\,(x_\pom)}
\newcommand{\dsf}{\mbox{$F_2^{D(3)}$}}
\newcommand{\dsfva}{\mbox{$F_2^{D(3)}(\beta,Q^2,x_{I\!\!P})$}}
\newcommand{\dsfvb}{\mbox{$F_2^{D(3)}(\beta,Q^2,x)$}}
\newcommand{\dsfpom}{$F_2^{I\!\!P}$}
\newcommand{\gap}{\stackrel{>}{\sim}}
\newcommand{\lap}{\stackrel{<}{\sim}}
\newcommand{\fem}{$F_2^{em}$}
\newcommand{\tsnmp}{$\tilde{\sigma}_{NC}(e^{\mp})$}
\newcommand{\tsnm}{$\tilde{\sigma}_{NC}(e^-)$}
\newcommand{\tsnp}{$\tilde{\sigma}_{NC}(e^+)$}
\newcommand{\st}{$\star$}
\newcommand{\sst}{$\star \star$}
\newcommand{\ssst}{$\star \star \star$}
\newcommand{\sssst}{$\star \star \star \star$}

\newcommand{\tw}{\theta_W}
\newcommand{\sw}{\sin{\theta_W}}
\newcommand{\cw}{\cos{\theta_W}}
\newcommand{\sww}{\sin^2{\theta_W}}
\newcommand{\cww}{\cos^2{\theta_W}}
\newcommand{\trm}{m_{\perp}}
\newcommand{\trp}{p_{\perp}}
\newcommand{\trmm}{m_{\perp}^2}
\newcommand{\trpp}{p_{\perp}^2}
\newcommand{\alp}{\alpha_s}

\newcommand{\alps}{\alpha_s}
\newcommand{\sqrts}{$\sqrt{s}$}
\newcommand{\LO}{$O(\alpha_s^0)$}
\newcommand{\Oa}{$O(\alpha_s)$}
\newcommand{\Oaa}{$O(\alpha_s^2)$}
\newcommand{\PT}{p_{\perp}}
\newcommand{\JPSI}{J/\psi}
\newcommand{\sh}{\hat{s}}
\newcommand{\uh}{\hat{u}}
\newcommand{\MP}{m_{J/\psi}}
\newcommand{\PO}{I\!\!P}
\newcommand{\xbj}{\ensuremath{x_{\rm Bj}} }
\newcommand{\xpom}{x_{\PO}}
\newcommand{\ttbs}{\char'134}
\newcommand{\xpomlo}{3\times10^{-4}}  
\newcommand{\xpomup}{0.05}  
\newcommand{\dgr}{^\circ}
\newcommand{\pbarnt}{\,\mbox{{\rm pb$^{-1}$}}}
\newcommand{\gev}{\,\mbox{GeV}}
\newcommand{\WBoson}{\mbox{$W$}}
\newcommand{\fbarn}{\,\mbox{{\rm fb}}}
\newcommand{\fbarnt}{\,\mbox{{\rm fb$^{-1}$}}}
%
%
\newcommand{\qsq}{\ensuremath{Q^2} }
\newcommand{\gevsq}{\ensuremath{\mathrm{GeV}^2} }
\newcommand{\et}{\ensuremath{E_t^*} }
\newcommand{\rap}{\ensuremath{\eta^*} }
\newcommand{\gp}{\ensuremath{\gamma^*}p }
\newcommand{\dsiget}{\ensuremath{{\rm d}\sigma_{ep}/{\rm d}E_t^*} }
\newcommand{\dsigrap}{\ensuremath{{\rm d}\sigma_{ep}/{\rm d}\eta^*} }
\newcommand{\pdfhq}{HERAPDF-HQMASS}

\begin{titlepage}

\noindent
\noindent
\begin{flushleft}
{\tt DESY 18-037    \hfill    ISSN 0418-9833} \\
{\tt \today\\}
\end{flushleft}

\vspace{1.5cm}
\begin{center}
\begin{Large}
{\bf Combination and QCD analysis of charm and beauty production cross-section measurements in deep inelastic {\boldmath $ep$\unboldmath} scattering at HERA}
\unboldmath
\vspace{2cm}

\end{Large}
\end{center}

\def\gev{\rm GeV}
\def\ie{\it i.e.}
\def\etal{\hbox{$\it et~al.$}}
\def\clb#1 {(#1 Coll.),}

\hyphenation{do-mi-nant
}
\begin{center}
{\Large The H1 and ZEUS Collaborations}
\end{center}
\vspace{3.5cm}

\begin{abstract}
\noindent
Measurements of open charm and beauty production cross sections in deep inelastic $ep$ scattering at HERA from the H1 and ZEUS Collaborations are combined. 
Reduced cross sections
are obtained in the kinematic range of negative four-momentum transfer squared of the photon $2.5$~GeV$^2\le Q^2 \le 2000$ GeV$^2$ 
and Bjorken scaling variable $3 \cdot 10^{-5} \le x_{\rm Bj} \le 5 \cdot 10^{-2}$. 
The combination method accounts for the correlations of the statistical and systematic uncertainties among the different datasets. 
Perturbative QCD calculations are compared to the combined data.
A next-to-leading order QCD analysis is performed using these data together with the combined inclusive deep inelastic scattering cross sections from HERA.
The running charm- and beauty-quark masses are determined as
$m_c(m_c) = 1.290^{+0.046}_{-0.041} {\rm (exp/fit)}$ ${}^{+0.062}_{-0.014}  {\rm (model)}$ ${}^{+0.003}_{-0.031} {\rm (parameterisation)}$ GeV
and 
$m_b(m_b) = 4.049^{+0.104}_{-0.109} {\rm (exp/fit)}$ ${}^{+0.090}_{-0.032} {\rm (model)}$ ${}^{+0.001}_{-0.031} {\rm (parameterisation)}~{\rm GeV}$.

\end{abstract}
\vspace{1.5cm}
\begin{center}
Submitted to EPJC
\end{center}
\end{titlepage}
\newpage

{\small\raggedright
H.~Abramowicz$^{\mathrm{52},\mathrm{a1}}$,
I.~Abt$^{\mathrm{38}}$,
L.~Adamczyk$^{\mathrm{22}}$,
M.~Adamus$^{\mathrm{62}}$,
R.~Aggarwal$^{\mathrm{8},\mathrm{a2}}$,
V.~Andreev$^{\mathrm{34}}$,
S.~Antonelli$^{\mathrm{5}}$,
V.~Aushev$^{\mathrm{26}}$,
A.~Baghdasaryan$^{\mathrm{65}}$,
K.~Begzsuren$^{\mathrm{58}}$
% ...

\newpage
\footnotesize\begin{description}\setlength{\parsep}{0em}\setlength{\itemsep}{0em}
\item[$^{1}$] 
 I. Physikalisches Institut der RWTH, Aachen, Germany
\item[$^{2}$] 
 {Institute of Physics and Technology of Ministry of Education and Science of Kazakhstan, Almaty, Kazakhstan}
\item[$^{3}$] 
 Inter-University Institute for High Energies ULB-VUB, Brussels and Universiteit Antwerpen, Antwerp, Belgium$^{\mathrm{b1}}$
\item[$^{4}$] 
 School of Physics and Astronomy, University of Birmingham, Birmingham, UK$^{\mathrm{b2}}$
\item[$^{5}$] 
 {INFN Bologna, Bologna, Italy{}}$^{\mathrm{b3}}$
\item[$^{6}$] 
 {Physikalisches Institut der Universit\"at Bonn, Bonn, Germany{}}$^{\mathrm{b4}}$
\item[$^{7}$] 
 Horia Hulubei National Institute for R\&D in Physics and Nuclear Engineering (IFIN-HH) , Bucharest, Romania$^{\mathrm{b5}}$
\item[$^{8}$] 
 {Panjab University, Department of Physics, Chandigarh, India}
\item[$^{9}$] 
 {Calabria University, Physics Department and INFN, Cosenza, Italy{}}$^{\mathrm{b3}}$
\item[$^{10}$] 
 {GSI Helmholtzzentrum f\"{u}r Schwerionenforschung GmbH, Darmstadt, Germany}
\item[$^{11}$] 
 STFC, Rutherford Appleton Laboratory, Didcot, Oxfordshire, UK$^{\mathrm{b2}}$
\item[$^{12}$] 
 Institut f\"ur Physik, TU Dortmund, Dortmund, Germany$^{\mathrm{b6}}$
\item[$^{13}$] 
 Joint Institute for Nuclear Research, Dubna, Russia
\item[$^{14}$] 
 CERN, Geneva, Switzerland
\item[$^{15}$] 
 Irfu/SPP, CE Saclay, Gif-sur-Yvette, France
\item[$^{16}$] 
 {School of Physics and Astronomy, University of Glasgow, Glasgow, United Kingdom{}}$^{\mathrm{b2}}$
\item[$^{17}$] 
 II. Physikalisches Institut, Universit\"at G\"ottingen, G\"ottingen, Germany
\item[$^{18}$] 
 Institut f\"ur Experimentalphysik, Universit\"at Hamburg, Hamburg, Germany$^{\mathrm{b6},\mathrm{b7}}$
\item[$^{19}$] 
 {Deutsches Elektronen-Synchrotron DESY, Hamburg, Germany}
\item[$^{20}$] 
 Physikalisches Institut, Universit\"at Heidelberg, Heidelberg, Germany$^{\mathrm{b6}}$
\item[$^{21}$] 
 {The Henryk Niewodniczanski Institute of Nuclear Physics, Polish Academy of \ Sciences, Krakow, Poland}$^{\mathrm{b15}}$
\item[$^{22}$] 
 {AGH University of Science and Technology, Faculty of Physics and Applied Computer Science, Krakow, Poland}
\item[$^{23}$] 
 {Department of Physics, Jagellonian University, Krakow, Poland}
\item[$^{24}$] 
 {National Centre for Particle Physics, Universiti Malaya, 50603 Kuala Lumpur, Malaysia{}}$^{\mathrm{b8}}$
\item[$^{25}$] 
 {Institute for Nuclear Research, National Academy of Sciences, Kyiv, Ukraine}
\item[$^{26}$] 
 {Department of Nuclear Physics, National Taras Shevchenko University of Kyiv, Kyiv, Ukraine}
\item[$^{27}$] 
 Department of Physics, University of Lancaster, Lancaster, UK$^{\mathrm{b2}}$
\item[$^{28}$] 
 Department of Physics, University of Liverpool, Liverpool, UK$^{\mathrm{b2}}$
\item[$^{29}$] 
 School of Physics and Astronomy, Queen Mary, University of London, London, UK$^{\mathrm{b2}}$
\item[$^{30}$] 
 {Physics and Astronomy Department, University College London, London, United Kingdom{}}$^{\mathrm{b2}}$
\item[$^{31}$] 
 Aix Marseille Universit\'{e}, CNRS/IN2P3, CPPM UMR 7346, 13288 Marseille, France
\item[$^{32}$] 
 Departamento de Fisica Aplicada, CINVESTAV, M\'erida, Yucat\'an, M\'exico$^{\mathrm{b9}}$
\item[$^{33}$] 
 Institute for Theoretical and Experimental Physics, Moscow, Russia$^{\mathrm{b10}}$
\item[$^{34}$] 
 Lebedev Physical Institute, Moscow, Russia
\item[$^{35}$] 
 Institute for Information Transmission Problems RAS, Moscow, Russia$^{\mathrm{b11}}$
\item[$^{36}$] 
 Moscow Institute of Physics and Technology, Dolgoprudny, Moscow Region, Russian Federation$^{\mathrm{b12}}$
\item[$^{37}$] 
 {Lomonosov Moscow State University, Skobeltsyn Institute of Nuclear Physics, Moscow, Russia{}}$^{\mathrm{b13}}$
\item[$^{38}$] 
 {Max-Planck-Institut f\"ur Physik, M\"unchen, Germany}
\item[$^{39}$] 
 {Department of Physics, York University, Ontario, Canada M3J 1P3{}}$^{\mathrm{b14}}$
\item[$^{40}$] 
 LAL, Universit\'e Paris-Sud, CNRS/IN2P3, Orsay, France
\item[$^{41}$] 
 {Department of Physics, University of Oxford, Oxford, United Kingdom{}}$^{\mathrm{b2}}$
\item[$^{42}$] 
 {INFN Padova, Padova, Italy{}}$^{\mathrm{b3}}$
\item[$^{43}$] 
 {Dipartimento di Fisica e Astronomia dell' Universit\`a and INFN, Padova, Italy{}}$^{\mathrm{b3}}$
\item[$^{44}$] 
 LLR, Ecole Polytechnique, CNRS/IN2P3, Palaiseau, France
\item[$^{45}$] 
 {Department of Physics, Temple University, Philadelphia, PA 19122, USA}
\item[$^{46}$] 
 Faculty of Science, University of Montenegro, Podgorica, Montenegro$^{\mathrm{b16}}$
\item[$^{47}$] 
 Institute of Physics, Academy of Sciences of the Czech Republic, Praha, Czech Republic$^{\mathrm{b17}}$
\item[$^{48}$] 
 Faculty of Mathematics and Physics, Charles University, Praha, Czech Republic$^{\mathrm{b17}}$
\item[$^{49}$] 
 {Department of Particle Physics and Astrophysics, Weizmann Institute, Rehovot, Israel}
\item[$^{50}$] 
 Dipartimento di Fisica Universit\`a di Roma Tre and INFN Roma~3, Roma, Italy
\item[$^{51}$] 
 Institute for Nuclear Research and Nuclear Energy, Sofia, Bulgaria
\item[$^{52}$] 
 {Raymond and Beverly Sackler Faculty of Exact Sciences, School of Physics, \ Tel Aviv University, Tel Aviv, Israel{}}$^{\mathrm{b18}}$
\item[$^{53}$] 
 {Polytechnic University, Tokyo, Japan{}}$^{\mathrm{b19}}$
\item[$^{54}$] 
 {Department of Physics, Tokyo Institute of Technology, Tokyo, Japan{}}$^{\mathrm{b19}}$
\item[$^{55}$] 
 {Universit\`a di Torino and INFN, Torino, Italy{}}$^{\mathrm{b3}}$
\item[$^{56}$] 
 {Universit\`a del Piemonte Orientale, Novara, and INFN, Torino, Italy{}}$^{\mathrm{b3}}$
\item[$^{57}$] 
 {Institute of Particle and Nuclear Studies, KEK, Tsukuba, Japan{}}$^{\mathrm{b19}}$
\item[$^{58}$] 
 Institute of Physics and Technology of the Mongolian Academy of Sciences, Ulaanbaatar, Mongolia
\item[$^{59}$] 
 Universit\'e Claude Bernard Lyon 1, CNRS/IN2P3, Villeurbanne, France
\item[$^{60}$] 
 Paul Scherrer Institut, Villigen, Switzerland
\item[$^{61}$] 
 {Faculty of Physics, University of Warsaw, Warsaw, Poland}
\item[$^{62}$] 
 {National Centre for Nuclear Research, Warsaw, Poland}
\item[$^{63}$] 
 Department of Physics and Astronomy, Purdue University 525 Northwestern Ave, West Lafayette, USA
\item[$^{64}$] 
 Fachbereich C, Universit\"at Wuppertal, Wuppertal, Germany
\item[$^{65}$] 
 Yerevan Physics Institute, Yerevan, Armenia
\item[$^{66}$] 
 {Meiji Gakuin University, Faculty of General Education, Yokohama, Japan{}}$^{\mathrm{b19}}$
\item[$^{67}$] 
 {Deutsches Elektronen-Synchrotron DESY, Zeuthen, Germany}
\item[$^{68}$] 
 Institut f\"ur Teilchenphysik, ETH, Z\"urich, Switzerland$^{\mathrm{b20}}$
\item[$^{69}$] 
 Physik-Institut der Universit\"at Z\"urich, Z\"urich, Switzerland$^{\mathrm{b20}}$
\item[$^{\dagger}$] 
 Deceased
\end{description}
\medskip
\begin{description}\setlength{\parsep}{0em}\setlength{\itemsep}{0em}\item[$^{\mathrm{a1}}$] 
 Also at Max Planck Institute for Physics, Munich, Germany, External Scientific Member
\item[$^{\mathrm{a2}}$] 
 Now at DST-Inspire Faculty, Department of Technology, SPPU, India
\item[$^{\mathrm{a3}}$] 
 Now at University of Bath, United Kingdom
\item[$^{\mathrm{a4}}$] 
 Also at Lodz University, Poland
\item[$^{\mathrm{a5}}$] 
 Now at Rockefeller University, New York, NY 10065, USA
\item[$^{\mathrm{a6}}$] 
 Now at INFN Roma, Italy
\item[$^{\mathrm{a7}}$] 
 Alexander von Humboldt Professor; also at DESY and University of Oxford
\item[$^{\mathrm{a8}}$] 
 Now at Sant Longowal Institute of Engineering and Technology, Longowal, Punjab, India
\item[$^{\mathrm{a9}}$] 
 Also at Agensi Nuklear Malaysia, 43000 Kajang, Bangi, Malaysia
\item[$^{\mathrm{a10}}$] 
 Member of Lodz University, Poland
\item[$^{\mathrm{a11}}$] 
 Also at LAPP, Universit\'e de Savoie, CNRS/IN2P3, Annecy-le-Vieux, France
\item[$^{\mathrm{a12}}$] 
 Also at University of Bristol, United Kingdom
\item[$^{\mathrm{a13}}$] 
 Now at European X-ray Free-Electron Laser facility GmbH, Hamburg, Germany
\item[$^{\mathrm{a14}}$] 
 Now at Polish Air Force Academy in Deblin
\item[$^{\mathrm{a15}}$] 
 Also at Ulaanbaatar University, Ulaanbaatar, Mongolia
\item[$^{\mathrm{a16}}$] 
 Also supported by DESY
\item[$^{\mathrm{a17}}$] 
 Now at Kobe University, Japan
\end{description}
\medskip
\begin{description}\setlength{\parsep}{0em}\setlength{\itemsep}{0em}\item[$^{\mathrm{b1}}$] 
 Supported by FNRS-FWO-Vlaanderen, IISN-IIKW and IWT and by Interuniversity Attraction Poles Programme, Belgian Science Policy
\item[$^{\mathrm{b2}}$] 
 Supported by the UK Science and Technology Facilities Council, and formerly by the UK Particle Physics and Astronomy Research Council
\item[$^{\mathrm{b3}}$] 
 Supported by the Italian National Institute for Nuclear Physics (INFN)
\item[$^{\mathrm{b4}}$] 
 Supported by the German Federal Ministry for Education and Research (BMBF), under contract No.\ 05 H09PDF
\item[$^{\mathrm{b5}}$] 
 Supported by the Romanian National Authority for Scientific Research under the contract PN 09370101
\item[$^{\mathrm{b6}}$] 
 Supported by the Bundesministerium f\"ur Bildung und Forschung, FRG, under contract numbers 05H09GUF, 05H09VHC, 05H09VHF, 05H16PEA
\item[$^{\mathrm{b7}}$] 
 Supported by the SFB 676 of the Deutsche Forschungsgemeinschaft (DFG)
\item[$^{\mathrm{b8}}$] 
 Supported by HIR grant UM.C/625/1/HIR/149 and UMRG grants RU006-2013, RP012A-13AFR and RP012B-13AFR from Universiti Malaya, and ERGS grant ER004-2012A from the Ministry of Education, Malaysia
\item[$^{\mathrm{b9}}$] 
 Supported by CONACYT, M\'exico, grant 48778-F
\item[$^{\mathrm{b10}}$] 
 Russian Foundation for Basic Research (RFBR), grant no 1329.2008.2 and Rosatom
\item[$^{\mathrm{b11}}$] 
 Russian Foundation for Sciences, project no 14-50-00150
\item[$^{\mathrm{b12}}$] 
 Ministery of Education and Science of Russian Federation contract no 02.A03.21.0003
\item[$^{\mathrm{b13}}$] 
 Partially supported by RF Presidential grant NSh-7989.2016.2
\item[$^{\mathrm{b14}}$] 
 Supported by the Natural Sciences and Engineering Research Council of Canada (NSERC)
\item[$^{\mathrm{b15}}$] 
 Partially Supported by Polish Ministry of Science and Higher Education, grant DPN/N168/DESY/2009
\item[$^{\mathrm{b16}}$] 
 Partially Supported by Ministry of Science of Montenegro, no. 05-1/3-3352
\item[$^{\mathrm{b17}}$] 
 Supported by the Ministry of Education of the Czech Republic under the project INGO-LG14033
\item[$^{\mathrm{b18}}$] 
 Supported by the Israel Science Foundation
\item[$^{\mathrm{b19}}$] 
 Supported by the Japanese Ministry of Education, Culture, Sports, Science and Technology (MEXT) and its grants for Scientific Research
\item[$^{\mathrm{b20}}$] 
 Supported by the Swiss National Science Foundation
\item[$^{\mathrm{b21}}$] 
 Supported by the Polish National Science Centre (NCN) grant no. DEC-2014/13/B/ST2/02486
\end{description}}

\newpage
 \section{Introduction}
\label{sec:intro}

Measurements of open charm and beauty production in neutral current (NC) deep inelastic electron\footnote{In 
this paper the term `electron' 
denotes both electron and positron
.}--proton scattering (DIS) at HERA provide important input 
for tests of the theory of strong interactions, quantum 
chromodynamics ({QCD}). 
Measurements at HERA \cite{
h196,
zeusdstar97,
h1gluon,
zd97,
h1f2c,
zd00,
h1dmesons,
h1vertex05,
h1ltt_hera1,
h1dstar_hera1,
zeusdmesons,
zd0dp,
zmu,
h1ltt_hera2,
h1dstarhighQ2,
zeusdpluslambda,
h1cbjets,
h1dstar_hera2,
zeusdplus_hera2,
zeusdstar_hera2,
zeusvtx,
zel,
zmu_hera1,
bgl
} 
have shown that heavy-flavour production
in DIS proceeds predominantly via the boson-gluon-fusion process, $\gamma g\rightarrow {\rm Q\overline{Q}}$, where Q is the heavy quark. The cross section therefore depends strongly on the gluon distribution in the proton and the heavy-quark mass. 
This mass provides a 
hard 
scale for the applicability of perturbative QCD (pQCD). 
However, other hard scales are also present in this process: the transverse momenta of the outgoing quarks and the negative four momentum squared, $Q^2$, of the exchanged photon. The presence of several hard scales 
complicates the calculation of heavy-flavour production in pQCD. Different approaches have been developed to cope with the multiple scale problem inherent in this process.
In this paper, the massive fixed-flavour-number scheme (FFNS) \cite{riemersma,gjr,ct10f3,mstw08f3,abkm09,abkm09b,abm11,abmp16,am11}
and different implementations of the variable-flavour-number scheme (VFNS) \cite{rt_opt,FONLL,fonllb_and_c,nnlonllx} are considered.

At HERA, different flavour tagging methods are applied for charm and beauty cross-section measurements:
the full reconstruction of $D$ or $D^{*\pm}$ mesons
~\cite{
h196,zeusdstar97,zd97,h1f2c,zd00,h1dstar_hera1,zeusdmesons,zd0dp,h1dstarhighQ2,h1dstar_hera2,
zeusdpluslambda,
zeusdplus_hera2,
zeusdstar_hera2
}, which is almost exclusively sensitive to charm production;
the lifetime of heavy-flavoured hadrons
\cite{h1dmesons,h1vertex05,h1ltt_hera1,h1ltt_hera2,zeusvtx}
and their semi-leptonic decays
\cite{zmu,zel,zmu_hera1}, both enabling the measurement of the charm and beauty cross section simultaneously.
In general, the different methods explore different regions of the heavy-quark phase space and show different dependencies on sources of systematic uncertainties.
Therefore, by using different tagging techniques a more complete picture of heavy-flavour production is obtained. 

In this paper, a simultaneous combination of charm and beauty production cross-section measurements
is presented. This analysis is an extension of the previous H1 and ZEUS combination 
of charm measurements in DIS~\cite{HERAcharmcomb}, including new charm and beauty data~\cite{zeusdplus_hera2,zeusdstar_hera2,zeusvtx,zel,zmu_hera1,zmu,h1ltt_hera2} and extracting combined beauty cross sections for the first time. 
As a result, a single consistent dataset from HERA of reduced charm and beauty cross sections in DIS is obtained, including all correlations. This dataset covers the kinematic range of photon virtuality  $2.5$ GeV$^2 \le Q^2 \le 2000$ GeV$^2$ 
and Bjorken scaling variable $3 \cdot 10^{-5} \le \xbj \le 5 \cdot 10^{-2}$.

The procedure follows the method used previously \cite{glazov,H1comb,DIScomb,HERAcharmcomb,sasha}.  
The correlated systematic uncertainties and the normalisation of the different measurements are accounted for 
such that one consistent dataset is obtained. Since different experimental techniques of charm and beauty tagging have been employed using different detectors and methods of kinematic reconstruction, this combination leads to a significant reduction of statistical and systematic uncertainties with respect to the individual measurements.
The simultaneous combination of charm and beauty cross-section measurements reduces the correlations between them and hence also the uncertainties.
The combined reduced charm cross sections of the previous analysis~\cite{HERAcharmcomb} are superseded by the new results presented in this paper. 

The combined data are compared to theoretical predictions obtained in the FFNS
at next-to-leading order (NLO, $O(\alpha_s^2)$) QCD using HERAPDF2.0~\cite{HERAPDF20}, ABKM09~\cite{abkm09,abkm09b} and ABMP16~\cite{abmp16} parton distribution functions (PDFs), 
and to approximate next-to-next-to-leading order (NNLO, $O(\alpha_s^3)$) using ABMP16~\cite{abmp16} PDFs. 
In addition, QCD calculations in the RTOPT~\cite{rt_opt} VFNS at NLO and approximate NNLO are compared with the data. 
The NLO calculations are at $O(\alpha_s^2)$ except for the massless parts of the coefficient functions, which are at $O(\alpha_s)$;  the NNLO calculations are one order of $\alpha_s$ higher.
A comparison is also made to predictions of two variants of the FONLL-C scheme~\cite{FONLL,fonllb_and_c} ($O(\alpha_s^3)$ (NNLO) in the PDF evolution, $O(\alpha_s^2)$ in all coefficient functions):
the default scheme, which includes next-to-leading-log (NLL) resummation of quasi-collinear final state gluon radiation, and a variant which includes NLL low-$x$ resummation in the PDFs and the matrix elements (NLLsx) ~\cite{nnlonllx} in addition.

The new data are subjected to a QCD analysis together with the final inclusive DIS cross-section data from HERA~\cite{HERAPDF20}
allowing for the determination at NLO of  the running charm- and beauty-quark masses, as defined from the QCD Lagrangian in the modified minimum-subtraction ($\overline{\rm MS}$) scheme.

The paper is organised as follows. In section~\ref{theory}, the reduced heavy-flavour cross section is defined and the theoretical framework 
is briefly introduced. The heavy-flavour tagging methods, the data samples 
and the combination method are presented in section~\ref{sec:data}. The resulting reduced charm and beauty cross sections are presented in section~\ref{sec:results} and in section~\ref{sect:comp-thoery} they are compared with theoretical calculations based on existing PDF sets and with existing predictions at NLO and at NNLO in the FFNS and VFNS. 
In section~\ref{sec:qcd}, the NLO QCD analysis is described and the measurement of the running masses of the charm and beauty quarks in the $\overline{\rm MS}$ scheme at NLO is presented. 
The conclusions are given in section~\ref{sec:summary}.

\section{Heavy-flavour production in DIS}
\label{theory}
In this paper, charm and beauty production via NC DIS are considered. In the kinematic range explored by the analysis of the data presented here, $Q^2$ is much smaller than $M_Z^2$, such that the virtual photon exchange dominates. 
Contributions from $Z$ exchange and $\gamma Z$ interference are small and therefore neglected.
The cross section for the production of a heavy flavour of type Q, with Q being either beauty, $b$, or  charm, $c$, may then be written in terms of the heavy-flavour contributions to the structure functions $F_2$ and $F_{\rm L}$, $F^{\rm Q\overline{Q}}_2(x_{\rm Bj},Q^2)$ and 
$F^{\rm Q\overline{Q}}_{\rm L}(x_{\rm Bj},Q^2)$, as
 \begin{equation}
\frac{{\rm d}^2 \sigma^{\rm Q\overline{Q}}}{{\rm d}x_{\rm Bj} {\rm d}Q^2} = \frac{2\pi \alpha^2(Q^2)}{x_{\rm Bj}Q^4} ( [1+(1-y)^2]F^{\rm Q\overline{Q}}_2 (x_{\rm Bj},Q^2) -y^2F_{\rm L}^{\rm Q\overline{Q}} (x_{\rm Bj},Q^2) )\ ,
\label{eqn:xsect}
\end{equation}
where  $y$ denotes the lepton inelasticity.
The superscripts $\rm Q\overline{Q}$ indicate the presence of a heavy quark pair in the final state.
The cross section ${{\rm d}^2 \sigma^{\rm Q\overline{Q}}}/{{\rm d}x_{\rm Bj} {\rm d}Q^2}$ is given at the
Born level without QED and electroweak radiative corrections, except for the running electromagnetic coupling, $\alpha(Q^2)$.

In this paper, the results are presented in terms of reduced cross sections, defined as follows:
\begin{eqnarray}
\sigma_{\rm red}^{\rm Q\overline{Q}}&=&\frac{{\rm d}^2\sigma^{\rm Q\overline{Q}}}{{\rm d}x_{\rm Bj} {\rm d}Q^2} \cdot \frac{x_{\rm Bj}Q^4}{2\pi\alpha^2(Q^2)\,(1+(1-y)^2)}\cr
&=&F_2^{\rm Q\overline{Q}}-\frac{y^2}{1+(1-y)^2}F^{\rm Q\overline{Q}}_{\rm L}.
\end{eqnarray}
In the kinematic range addressed, 
the expected contribution from the exchange of longitudinally polarised photons, $F_{\rm L}^{\rm Q\overline{Q}}$, is small. In charm production it is expected to reach a few per cent at high $y$~\cite{daum_fl}. The structure functions $F_2^{\rm Q\overline{Q}}$ and $F_{\rm L}^{\rm Q\overline{Q}}$
are calculated to the same order (in most cases $O(\alpha_s^2)$) in all calculations explicitly performed in
this paper.
 
 Various theoretical approaches can be used to describe heavy-flavour production in DIS.
At values of $Q^2$ not very much larger than the heavy-quark mass, $m_{\rm Q}$, heavy flavours are predominantly produced dynamically by the photon-gluon-fusion process. The creation of a $\rm Q\overline{Q}$ pair  sets a lower limit of $2m_{\rm Q}$ to the mass of the hadronic final state. This low mass cutoff affects the kinematics and the higher order corrections in the phase space accessible at HERA. Therefore, a careful theoretical treatment of the heavy-flavour masses is mandatory for the pQCD analysis of heavy-flavour production as well as for the determination of the PDFs of the proton from data including heavy flavours.

In this paper, the FFNS is used for pQCD calculations for the corrections of measurements to the full phase space and in the QCD fits. In this scheme, heavy quarks are always treated as massive and therefore are not considered as partons in the proton. The number of (light) active flavours in the PDFs, $n_f$, is set to three and heavy quarks are produced only in the hard-scattering process. The leading-order (LO) contribution to heavy-flavour production ($O(\alpha_s)$ in the coefficient functions) is the photon-gluon-fusion process. The NLO massive coefficient functions using on-shell mass renormalisation (pole masses)~\cite{riemersma} were adopted by many global QCD analysis 
groups~\cite{abm11,gjr,ct10f3,mstw08f3}, providing PDFs derived from this scheme. 
They were extended to the $\overline{\rm MS}$ 
scheme~\cite{abkm09b}, using scale dependent (running) heavy-quark masses. 
The advantages of performing heavy-flavour calculations in the $\overline{\rm MS}$ scheme are reduced scale uncertainties and improved theoretical precision of the mass definition~\cite{am11,bgl}.
In all FFNS heavy-flavour calculations presented in this paper, the default renormalisation scale $\mu_r$ and factorisation scale $\mu_f$ are set to $\mu_r=\mu_f=\sqrt{Q^2+4m_{\rm Q}^2}$, where $m_{\rm Q}$ is the appropriate pole or running mass.

For the extraction of the combined reduced cross sections of charm and beauty production, it is necessary to predict
inclusive cross sections as well as exclusive cross sections with certain phase-space restrictions applied. For this purpose, 
 the FFNS at NLO is used to calculate inclusive~\cite{riemersma} and exclusive~\cite{hvqdis} quantities in the pole-mass scheme. This is currently the only scheme for which exclusive NLO calculations are available. 

The QCD analysis at next-to-leading order\footnote{The analysis is restricted to NLO because the NNLO calculations~\cite{Alekhin:2013nda} are not yet complete.}
 including the extraction of the heavy-quark running masses is performed in the FFNS with the OPENQCDRAD programme~\cite{openqcdrad} in the \textsc{xFitter} (former \textsc{HERAFitter}) framework~\cite{xfitter}. In OPENQCDRAD, heavy-quark production is calculated either using the $\rm{\overline{MS}}$ or the pole-mass scheme of heavy-quark masses. In this paper, the  $\rm{\overline{MS}}$ scheme is adopted. 
 
Predictions from different variants of the VFNS are also compared to the data.
The expectations from the NLO and approximate NNLO RTOPT~\cite{rt_opt} implementation as used for HERAPDF2.0 \cite{HERAPDF20} are confronted with both the charm and beauty cross sections while the  FONLL-C calculations~\cite{fonllb_and_c,nnlonllx} are compared to the charm data only.
In the VFNS, heavy quarks are treated as massive at small $Q^2$ up to $Q^2\approx O(m_{\rm Q}^2)$ and as massless at  $Q^2\gg m_{\rm Q}^2$, with interpolation prescriptions between the two regimes which avoid double counting of common terms. In the FONLL-C calculations, the massive part of the charm coefficient functions is treated at NLO ($O(\alpha_s^2)$) while the massless part and the PDFs are treated at NNLO ($O(\alpha_s^2)$ and $O(\alpha_s^3$), respectively).  
In addition to the default FONLL-C scheme the NLLsx variant~\cite{nnlonllx} is considered.

\section{Combination of H1 and ZEUS measurements}
\label{sec:data}
The different charm- and beauty-tagging methods exploited at HERA enable a comprehensive study of heavy-flavour production in NC DIS.

Using fully reconstructed $D$ or $D^{*\pm}$ mesons gives the best signal-to-background ratio for measurements of the charm production process. Although the branching ratios of beauty hadrons to $D$ and $D^{*\pm}$ mesons are large, the contributions from beauty production to the observed $D$ or $D^{*\pm}$ meson samples are small for several reasons. 
Firstly, beauty production in $ep$ collisions is suppressed relative to charm production by a factor $1/4$ due to the quark's electric charge coupling to the photon. 
Secondly, the photon-gluon-fusion cross section depends on the invariant mass of the outgoing partons, $\hat{s}$, which has a threshold value of $4m_{\rm Q}^2$. Because the beauty-quark mass, $m_b$, is about three times the charm-quark mass, $m_c$, beauty production is significantly suppressed.
Thirdly, in beauty production $D$ and $D^{*\pm}$ mesons originate from the fragmentation of charm quarks that are produced by the weak decay of $B$ mesons.  
Therefore the momentum fraction of the beauty quark carried by the $D$ or $D^{*\pm}$ meson is small, so that the mesons often remain undetected.

Fully inclusive analyses based on the lifetime of the heavy-flavoured mesons are sensitive to both charm and beauty production. 
Although the first two reasons  given above for the suppression of beauty production relative to charm production also hold in this case, sensitivity to beauty production can be enhanced by several means.  
The proper lifetime of $B$ mesons is on average a factor of $2$ to $3$ that of $D$ mesons~\cite{pdg2016}.
Therefore, the charm and beauty contributions can be disentangled by using observables directly sensitive to the lifetime of the decaying heavy-flavoured hadrons. 
The separation can be further improved by the simultaneous use of observables sensitive to the mass of the heavy-flavoured hadron: the relative transverse momentum, $p_{\rm T}^{\rm rel}$, of the particle with respect to the flight direction of the decaying heavy-flavoured hadron; the number of tracks with lifetime information; the invariant mass calculated from the charged particles attached to a secondary-vertex candidate. 

The analysis of lepton production is sensitive to semi-leptonic decays of both charm and beauty hadrons. When taking into account the fragmentation fractions of the heavy quarks as well as the fact that in beauty production leptons may originate both from the $b\rightarrow c$ and the $c\rightarrow s$ transitions, the semi-leptonic branching fraction of $B$ mesons is about twice that of $D$ mesons~\cite{pdg2016}. 
Because of the large masses of $B$ mesons and the harder fragmentation of beauty quarks compared to charm quarks, leptons originating directly from the $B$ decays have on average higher momenta than those produced in $D$ meson decays. Therefore, the experimentally observed fraction of beauty-induced leptons is enhanced relative to the observed charm-induced fraction. 
Similar methods as outlined in the previous paragraph are then used to further facilitate the separation of the charm and beauty contributions on a statistical basis.  

While the measurement of fully reconstructed  $D$ or $D^{*\pm}$ mesons yields the cleanest charm production sample, it suffers from small branching fractions and significant losses, because all particles from the  $D$ or $D^{*\pm}$ meson decay have to be measured. Fully inclusive and semi-inclusive-lepton analyses, which are sensitive to both charm and beauty production, profit from larger branching fractions and better coverage in polar angle. However, they are affected by a worse signal to background ratio and the large statistical correlations between charm and beauty measurements inherent to these methods.

\subsection{Data samples}
\label{subsec:data}

The H1\cite{h1} and ZEUS~\cite{zeus} detectors were general purpose
instruments which consisted of tracking systems surrounded by
electromagnetic and hadronic calorimeters and muon detectors, ensuring
close to $4\pi$ coverage of the $ep$ interaction region. Both detectors
were equipped with high-resolution silicon vertex detectors~\cite{H1cst,ZEUSmvd}. 

The datasets included in the combination are listed in table~\ref{tab:data}.
The data have been obtained from both the HERA I (in the years 1992--2000) and 
HERA II (in the years 2003--2007) data-taking periods. 
The combination includes measurements using
different tagging techniques: 
the reconstruction of particular 
decays of $D$ mesons~\cite{h1dstar_hera1,h1dstar_hera2,h1dstarhighQ2,zd97,zd00,zd0dp,zeusdplus_hera2,zeusdstar_hera2} (datasets $2-7, 9, 10$), 
the inclusive analysis of 
tracks exploiting lifetime information~\cite{h1ltt_hera2,zeusvtx} (datasets $1, 11$) and 
the reconstruction of electrons and muons from 
heavy-flavour semileptonic decays~\cite{zmu,zel,zmu_hera1} (datasets $8, 12, 13$).

The datasets $1$ to $8$ have already been used in the previous combination~\cite{HERAcharmcomb} of charm cross-section measurements, while the datasets $9$ to $13$ are included for the first time in this analysis. 
Dataset $9$ of the current analysis supersedes one dataset of  the previous charm combination (dataset $8$ in table~1 of~\cite{HERAcharmcomb}), because the earlier analysis was based on a subset of only about $30\,\%$ of the final statistics collected during the HERA II running period.

For the inclusive lifetime analysis~\cite{h1ltt_hera2}~(dataset $1$) the reduced cross sections $\sigma_{\rm red}^{ c\overline{c}}$ and $\sigma_{\rm red}^{ b\overline{b}}$ are taken directly from the publication.
For all other measurements, the combination starts from the measured double-differential visible cross sections $\sigma_{\rm vis,bin}$ in bins of $Q^2$ and either \xbj or $y$, where the visibility is defined by the particular range of transverse momentum $p_T$ and pseudorapidity\footnote{The pseudorapidity is defined as $\eta=-\ln\tan\frac{\Theta}{2}$, where the polar angle $\Theta$ is defined with respect to the proton direction in the laboratory frame.
}
 $\eta$ of the $D$ meson, lepton or jet as given in the corresponding publications.
In case of inclusive $D$ meson cross sections, small beauty contributions as estimated in the corresponding papers are subtracted. 
Consistent with equation~(\ref{eqn:xsect}),
all published visible cross-section measurements are corrected to Born level apart from the running of $\alpha$, i.e. they include corrections for 
radiation of real photons from the incoming and outgoing lepton
using the HERACLES programme~\cite{heracles}. 
QED corrections to the incoming and outgoing quarks are judged to be negligible and are therefore not considered.
All cross sections are updated using the most recent hadron decay branching ratios~\cite{pdg2016}.


\subsection{Extrapolation of  visible cross sections to $ \boldmath\boldsymbol \sigma_{\rm red}^{\rm Q\overline{Q}} \unboldmath$}
\label{sect:extract}
Except for dataset $1$ of table~\ref{tab:data}, for which only measurements expressed in the full phase space are available, the visible cross sections 
 $\sigma_{\rm vis, bin}$ measured in a limited phase space are converted to reduced cross sections $\sigma_{\rm red}^{\rm Q\overline{Q}}$ using a common theory.
 The reduced cross section of a heavy flavour Q at a reference ($x_{\rm Bj}, Q^2$) point is extracted  according to 
\begin{equation}
\sigma_{\rm red}^{\rm Q\overline{Q}}(x_{\rm Bj},Q^2)=\sigma_{\rm vis, bin}
\frac{\sigma_{\rm red}^{\rm Q\overline{Q}, th}(x_{\rm Bj},Q^2)}{\sigma^{\rm th}_{\rm vis, bin}}.\label{formel_ftc}
\end{equation}
The programme for heavy-quark production in DIS, HVQDIS~\cite{hvqdis}, is used with running $\alpha$ to calculate the theory predictions for
 $\sigma_{\rm red}^{\rm Q\overline{Q}, th}(x_{\rm Bj},Q^2)$ 
 and 
$\sigma^{\rm th}_{\rm vis, bin}$ in the NLO FFNS. 
Since the ratio in equation~(\ref{formel_ftc}) describes the extrapolation from the visible phase space in  $p_T$ and $\eta$ of the heavy-flavour tag to the full phase space, only the shape of the cross-section predictions in $p_T$ and $\eta$ 
is relevant for the corrections, while theory uncertainties related to normalisation cancel.  

In pQCD, $\sigma_{\rm red}^{\rm th}$ can be written as a convolution integral of proton PDFs with hard matrix elements. For the identification of heavy-flavour production, however, specific particles used for tagging have to be measured in the hadronic final state.
This requires that in the calculation of $\sigma_{\rm vis}^{\rm th}$, the convolution includes 
the proton PDFs, the hard matrix elements and the fragmentation functions. In the case of the HVQDIS programme, non-perturbative fragmentation functions are used.  
The different forms of the convolution integrals for $\sigma_{\rm red}^{\rm th}$ and $\sigma_{\rm vis}^{\rm th}$ necessitate the consideration of  different sets of theory parameters.

The following parameters are used in these NLO  calculations 
and are varied within the quoted limits to estimate the uncertainties in the predictions introduced by these parameters:
\begin{itemize}
 \item The {\bf renormalisation and factorisation scales}
are taken as $\mu_{\rm r}=\mu_{\rm f}=\sqrt{Q^2+4m_{\rm Q}^2}$.  
The scales are varied simultaneously up or down by a factor of two.
 \item The {\bf pole masses of the charm and  beauty quarks} are set to $m_c=1.50 \pm 0.15$~GeV, $m_b=4.50 \pm 0.25$~GeV, respectively. 
 These variations also affect the values of the renormalisation and factorisation scales. 
 \item For the {\bf strong coupling constant}, the value $\alpha_s^{n_f=3}(M_Z) = 0.105 \pm 0.002$ is chosen, which corresponds to
 $\alpha_s^{n_f=5}(M_Z) = 0.116 \pm 0.002$.
\item The {\bf proton PDFs} are described by a 
series of {FFNS} variants of the HERAPDF1.0 set~\cite{HERAcharmcomb,DIScomb} 
at NLO determined within the \textsc{xFitter} framework. 
No heavy-flavour measurements were included in the determination of these PDF sets. These PDF sets are 
those used in the previous combination~\cite{HERAcharmcomb} which were calculated for  $m_c=1.5\pm0.15$~GeV,~ $\alpha_s^{n_f=3}(M_Z) = 0.105 \pm 0.002$ and simultaneous variations of the renormalisation and factorisation scales up or down by a factor two. 
For the determination of the PDFs, the beauty-quark mass was fixed at $m_b=4.50$~GeV. 
The renormalisation and factorisation scales were set to $\mu_r=\mu_f=Q$ for the light flavours and to 
$\mu_r=\mu_f=\sqrt{Q^2+4m_{\rm Q}^2}$ for the heavy flavours. 
For all parameter settings considered, the respective HERAPDF1.0 set is used. 
As a cross check of the extrapolation procedure, the cross sections are also evaluated with the $3$-flavour NLO versions of the HERAPDF2.0 set (FF3A)~\cite{HERAPDF20}; the differences are found to be smaller than the PDF-related cross-section uncertainties. 
\end{itemize}

For the calculation of $\sigma_{\rm vis}^{\rm th}$, assumptions have been made on the fragmentation of the heavy quarks into particular hadrons and, when necessary, on the subsequent decays of the heavy flavoured hadrons into the particles used for tagging. 
In the calculation of $\sigma_{\rm vis}^{\rm th}$ the following settings and parameters are used in addition to those needed for $\sigma_{\rm red}^{\rm th}$ and are varied within the quoted limits:
\begin{itemize}
\item The {\bf charm fragmentation function} is described by the Kartvelishvili function~\cite{kart} controlled by a single parameter $\alpha_K$ to describe the longitudinal fraction of the charm-quark momentum transferred to the $D$ or $D^{*\pm}$ meson. 
Depending on the invariant mass $\hat{s}$ of the outgoing parton system, different values of $\alpha_K$ and their uncertainties  are used as measured at HERA~\cite{h1frag,zeusfrag} for $D^{*\pm}$ mesons. 
The variation of $\alpha_K$ as a function of $\hat{s}$ observed in $D^{*\pm}$ measurements has been adapted 
to the longitudinal-fragmentation function of ground state $D$ mesons not originating from $D^{*\pm}$ decays~\cite{HERAcharmcomb}. 
Transverse fragmentation is modelled by assigning to the charmed hadron a transverse momentum $k_T$ with respect to the direction of the charmed quark with an average value of $\langle k_T\rangle=0.35\pm0.15$~GeV~\cite{HERAcharmcomb}.
\item The {\bf charm fragmentation fractions} of a charm quark into a specific charmed hadron and their uncertainties are taken from~\cite{ffc2016}.
\item The{ \bf beauty fragmentation function} is parameterised according to Peterson et al.~\cite{Peterson:1982ak} with $\epsilon_{b}=0.0035 \pm 0.0020$~\cite{fragb0035}.
\item The {\bf branching ratios of {\boldsymbol\boldmath $D$} and {\boldmath$ D^{*\pm}$} mesons} into the specific decay channels analysed and their uncertainties are taken from~\cite{pdg2016}.
\item The {\bf branching fractions of semi-leptonic decays} of heavy quarks to a muon or electron and their uncertainties are taken from~\cite{pdg2016}.
\item The {\bf decay spectra of leptons originating from charmed hadrons} are modelled according to~\cite{SLspectrum}. 
\item The {\bf  decay spectra for beauty hadrons into leptons} are taken from the PYTHIA\cite{pythia} Monte Carlo (MC) programme, mixing direct semi-leptonic decays and cascade decays through charm according to the measured branching ratios~\cite{pdg2016}. It is checked that the MC describes BELLE and BABAR data~\cite{SLbeauty} well.
\item When necessary for the extrapolation procedure, {\bf parton-level jets} are reconstructed using the same clustering algorithms as used on detector level, 
and the cross sections are corrected for jet-hadronisation effects using corrections derived in the original papers~\cite{zeusvtx,zmu_hera1}.%
\footnote{Since no such corrections are provided, an uncertainty of $5\%$ is assigned to cover the untreated hadronisation effects~\cite{zmu_hera1}.}
\end{itemize}
While the central values for the extrapolation factors
${\sigma_{\rm red}^{\rm Q\overline{Q}, th}(x_{\rm Bj},Q^2)}/{\sigma^{\rm th}_{\rm vis, bin}}$ (see equation~\ref{formel_ftc})
are obtained in the FFNS
pole-mass scheme at NLO, their uncertainties are calculated such that they 
should cover potential deviations from the unknown `true' QCD result. 
The resulting reduced cross sections, with these uncertainties included, thus can be compared
to calculations in any QCD scheme to any order.

\subsection{Combination method}
\label{sec:comb}

The quantities to be combined are the reduced charm and beauty cross sections, \redcc~ and \redbb, respectively.
The combined cross sections are determined at common ($\xbj, Q^2$)  grid points. 
For \redcc{}, the grid is chosen to be the same as in~\cite{HERAcharmcomb}. 
The results are given for a  centre-of-mass energy of $\sqrt{s}=318$~GeV.
When needed, the measurements are transformed to the common grid $(\xbj,Q^2)$ points using inclusive NLO FFNS calculations \cite{riemersma}. The uncertainties on the resulting scaling factors are found to be negligible.

The combination is based on the $\chi^2$-minimisation procedure~\cite{glazov} used previously~\cite{H1comb,DIScomb,HERAcharmcomb,HERAPDF20}. The total $\chi^2$ is defined as
\begin{equation}
\chi^2_{\rm exp}\left(\boldsymbol{m},\boldsymbol{b}\right)=
\sum_{e}\ \left[\sum_{i}
\frac{\left(m^i-\sum\nolimits_j\gamma_j^{\ i,e}m^ib_j-\mu^{i,e}\right)^2}
{\left(\mu^{i,e}\cdot\delta_{i,e,{\rm stat}}\right)^2+\left(m^i\cdot\delta_{i,e,{\rm uncorr}}\right)^2} \right]
+ \sum_j{b_j}^2.
\end{equation}
The three sums run over the different input datasets $e$, listed in table~\ref{tab:data}, the 
($\xbj, Q^2$)  grid points $i$, for which the measured cross sections $\mu^{i,e}$ are combined to the cross sections $m^i$, and the sources $j$ of the shifts $b_j$ in units of standard deviations of the correlated uncertainties.
The correlated uncertainties comprise the correlated systematic uncertainties and the statistical correlation between the charm and beauty cross-section measurements. 
The quantities $\gamma_j^{\ i,e}$, $\delta_{i,e,{\rm stat}}$ and $\delta_{i,e,{\rm uncorr}}$ denote the relative correlated systematic, relative statistical and relative uncorrelated systematic uncertainties, respectively.
The components of the vector $\boldsymbol{m}$ are the combined cross sections $m^i$ while those of the vector  $\boldsymbol{b}$  are the shifts $b_j$.

In the present analysis, the correlated and uncorrelated systematic uncertainties are 
predominantly of multiplicative nature, i.e.~they are proportional to the expected cross sections $m^i$.
The statistical uncertainties are mainly background dominated and thus are treated as constant. 
All experimental systematic uncertainties are treated as independent between H1 and ZEUS. 
For the datasets 1, 8 and 11 of table~\ref{tab:data}, statistical correlations between charm and beauty cross sections are accounted for 
as reported in the original papers. 
Where necessary, the statistical correlation factors are corrected to take into account differences in the kinematic region 
of the charm and beauty measurements (dataset 11) or binning schemes (dataset 1), using theoretical predictions calculated with the HVQDIS programme.
The consistent treatment of the correlations of statistical and systematic uncertainties, 
including the correlations between the charm and beauty data sets where relevant, 
yields a significant reduction of the overall uncertainties of the combined data, as detailed in the following section.

\section{Combined cross sections}
\label{sec:results}

\begin{figure}[h]
\center
\epsfig{file=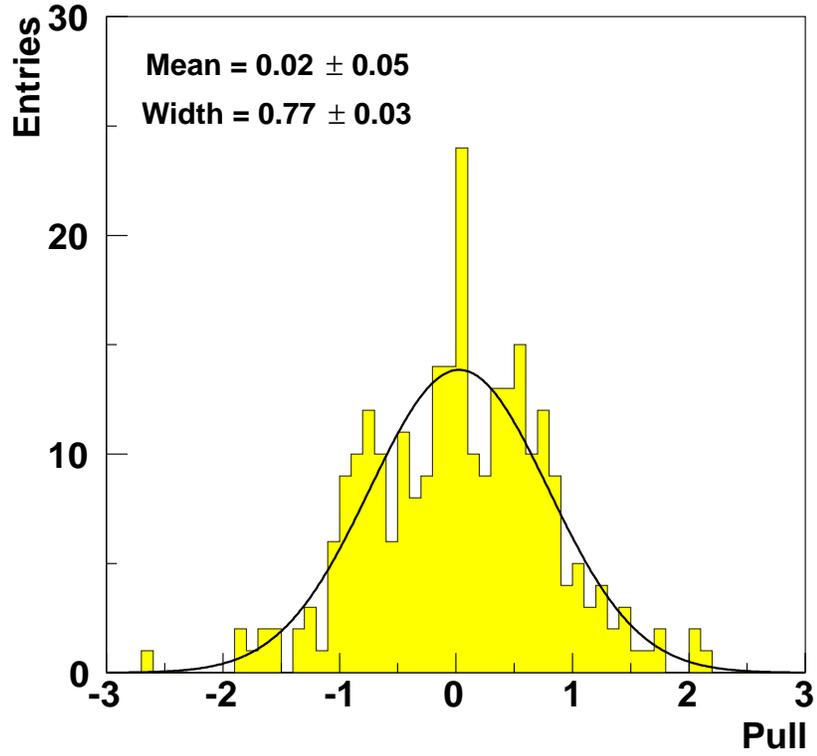 ,width=0.7\textwidth}
\caption{The pull distribution for the combination of the charm and beauty reduced cross sections. The solid line shows a fit of a Gaussian to the pull distribution. The mean and the width quoted are the results from the fit.}
\label{fig:pull} 
\end{figure}

The values of the combined cross sections \redcc and \redbb, together with the statistical, the uncorrelated and correlated systematic and the total uncertainties, are listed in tables~\ref{tab:sigcc} and \ref{tab:sigbb}.
A total of $209$ charm and $57$ beauty data points are combined simultaneously to obtain $52$ reduced charm and $27$ reduced beauty cross-section measurements.
A  $\chi^2$ value of $149$ for $187$ degrees of freedom (d.o.f.) is obtained in the combination, indicating good consistency of the input datasets. 
The distribution of pulls of the $266$ input data points with respect to the  combined cross sections is presented in figure~\ref{fig:pull}. It is consistent with a Gaussian around zero without any significant outliers. 
The observed width of the pull distribution is smaller than unity which indicates a conservative estimate of the  systematic uncertainties.

There are $167$ sources of correlated uncertainties in total. 
These are $71$ experimental systematic sources, $16$ sources due to the extrapolation procedure 
(including the uncertainties on the fragmentation fractions and branching ratios) 
and $80$ statistical charm and beauty correlations. 
The sources of correlated systematic and extrapolation uncertainties are listed in the appendix, together with the cross-section shifts induced by the sources and the reduction factors of the uncertainties, obtained as a result of the combination. Both quantities are given in units of $\sigma$ of the original uncertainties.
All shifts of the systematic sources with respect to their nominal values are smaller than $1.5\sigma$.
Several systematic uncertainties are reduced significantly -- by up to factors of two or more. The reductions are due to the different heavy-flavour tagging methods applied and to the fact that for a given process (charm or beauty production), an unique cross section is probed by the different measurements at a given $(x_{\rm Bj},Q^2)$ point.
Those uncertainties for which large reductions have been observed already in the previous analysis~\cite{HERAcharmcomb} are reduced to at least the same level in the current combination, some
are further significantly reduced due to the inclusion of new precise data~\cite{zeusdplus_hera2,zeusdstar_hera2,zeusvtx}. 
The shifts and reductions obtained for the $80$ statistical correlations between charm and beauty cross sections are not shown.  Only small reductions in the range of $10\%$ are observed and these reductions are  independent of $\xbj$ and $Q^2$.
The cross-section tables of the combined data together with the full information on the uncertainties can be found elsewhere~\cite{resulturl}.
 
\begin{figure}[h]
\center
\epsfig{file=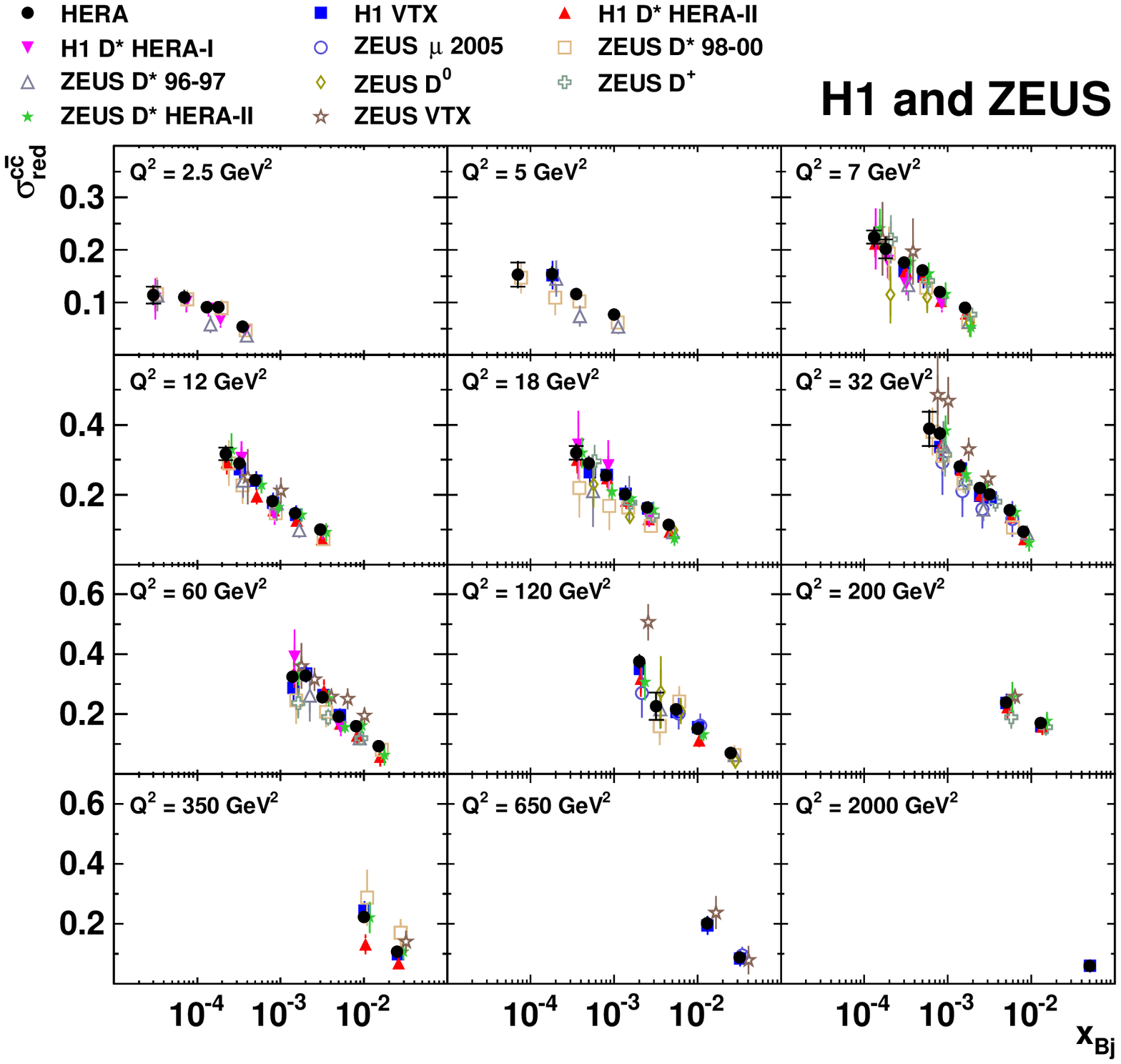 ,width=1.0\textwidth}
\caption{Combined measurements of the reduced charm production cross sections, \redcc, (full circles) as a function of \xbj for different values of $Q^2$. 
	The inner error bars indicate the uncorrelated part of the uncertainties and 
	the outer error bars represent the total uncertainties. 
  The input measurements with their total uncertainties are also shown by different markers. 
   For better visibility the individual input data are slightly displaced in \xbj towards larger values.}
\label{fig:data-charm} 
\end{figure}

\begin{figure}[h]
\center
\epsfig{file=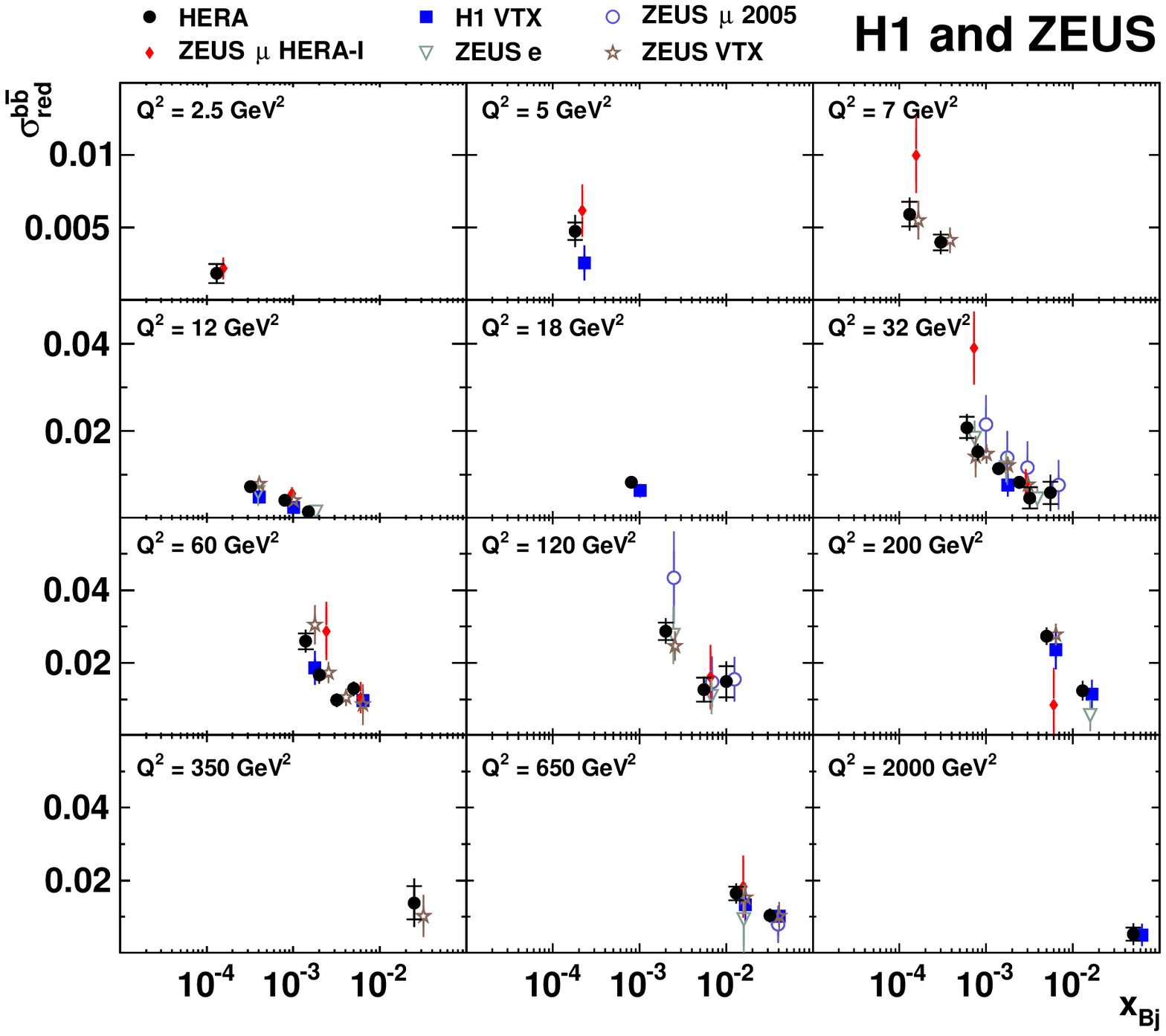 ,width=1.0\textwidth}
\caption{Combined measurements of the reduced beauty production cross sections, \redbb, (full circles) as a function of \xbj for different values of $Q^2$. 
	The inner error bars indicate the uncorrelated part of the uncertainties and 
	the outer error bars represent the total uncertainties. 
  The input measurements with their total uncertainties are also shown by different markers. 
   For better visibility the individual input data are slightly displaced in \xbj towards larger values.}
\label{fig:data-beauty} 
\end{figure}

\begin{figure}[h]
\center
\setlength{\unitlength}{1cm}
\begin{picture}(15.,20.)
\put(2.,-0.5){\epsfig{file=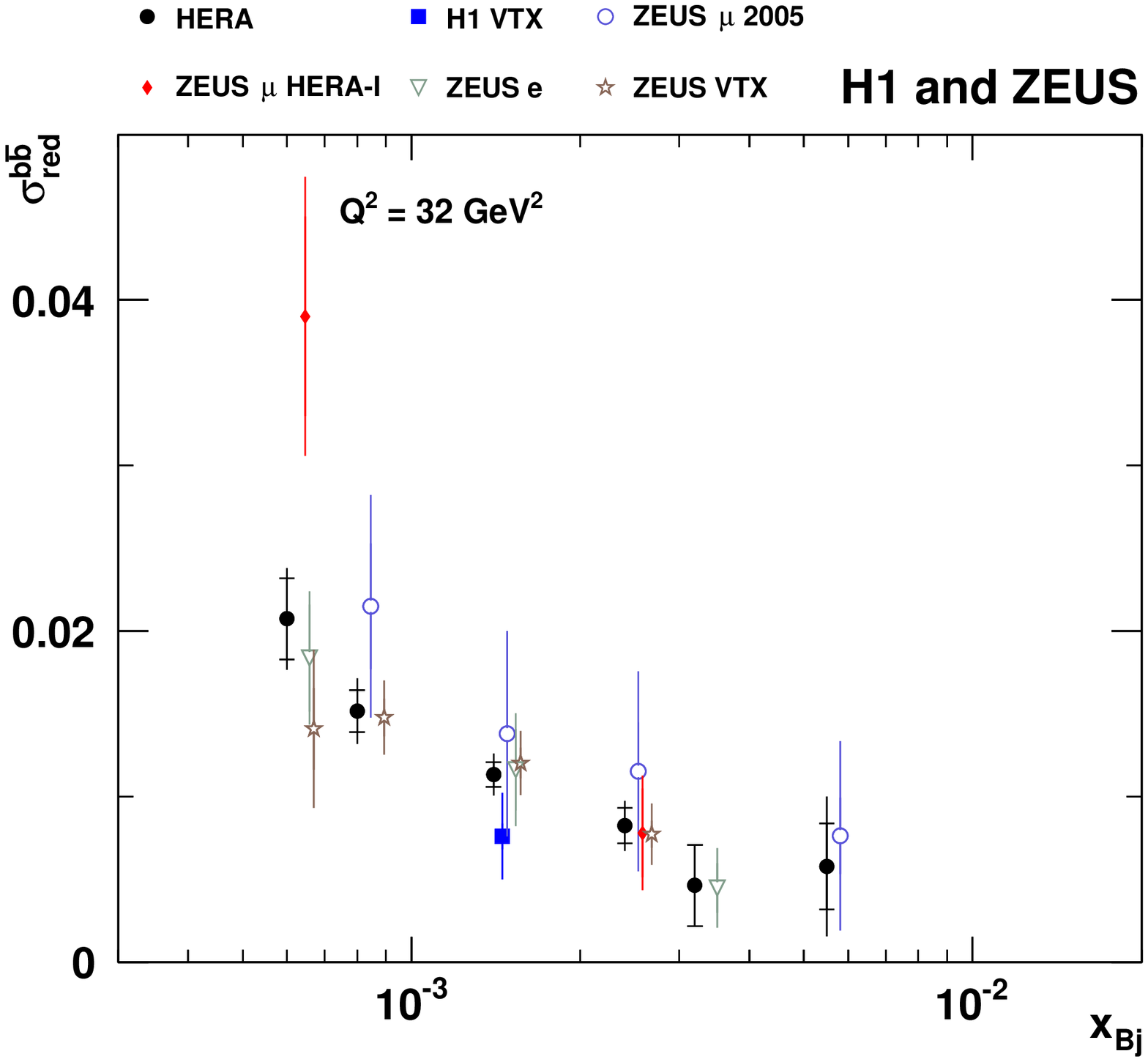,width=0.7\textwidth}}
\put(2,9.8){\epsfig{file=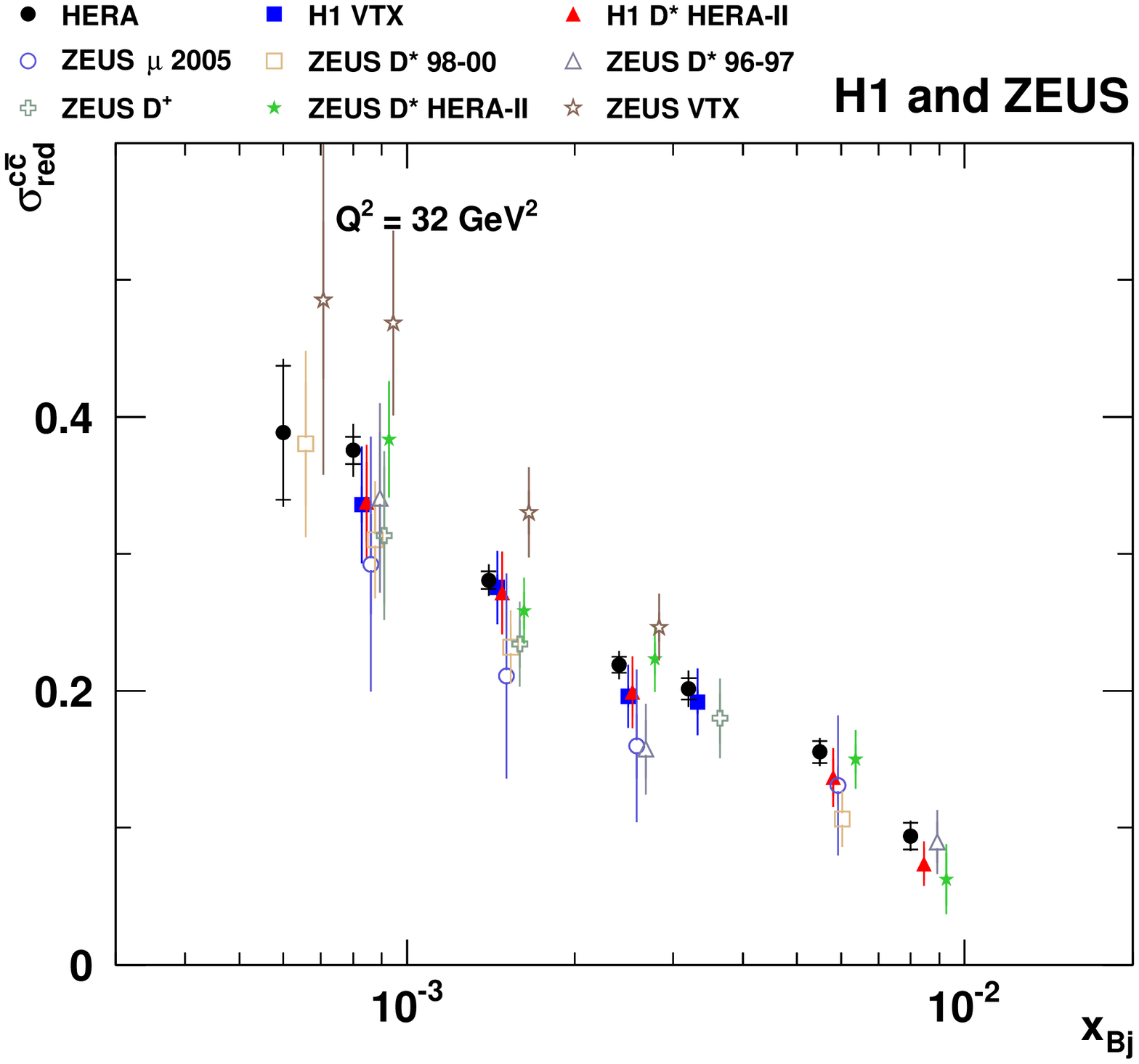,width=0.7\textwidth}}
\put(11.5,7.5){\Large(b)}
\put(11.5,17.8){\Large(a)}
\end{picture}
\caption{Reduced cross sections as a function of \xbj at $Q^2 = 32$ GeV$^2$ for (a) charm and (b) beauty production. The combined cross sections (full circles) are compared to
the input measurements shown by different markers. 
	For the combined measurements, the inner error bars indicate the uncorrelated part of the uncertainties and 
	the outer error bars represent the total uncertainties. 
  For better visibility the individual input data are slightly displaced in \xbj towards larger values.}
\label{fig:data-32GeV} 
\end{figure}

The combined reduced cross sections \redcc and \redbb are shown as a function of $\xbj$ in bins of $Q^2$
together with the input H1 and ZEUS data in figures~\ref{fig:data-charm} and \ref{fig:data-beauty}, respectively.
The combined cross sections are significantly more precise than any of the individual input datasets for charm as well as for beauty production. This is illustrated in figure~\ref{fig:data-32GeV}, where the charm and beauty measurements for $Q^2 = 32$~GeV$^2$ are shown. 
The uncertainty of the combined reduced charm cross section is $9\%$ on average and 
 reaches values of about $5\%$ or better in the region $12$~GeV$^2\le Q^2\le60$~GeV$^2$.  The uncertainty of the combined reduced beauty cross section is about $25\%$ on average and reaches about $15\%$ at small $\xbj$ and~$12$~GeV$^2\le Q^2\le200$~GeV$^2$.

\begin{figure}[h]
\center
\epsfig{file=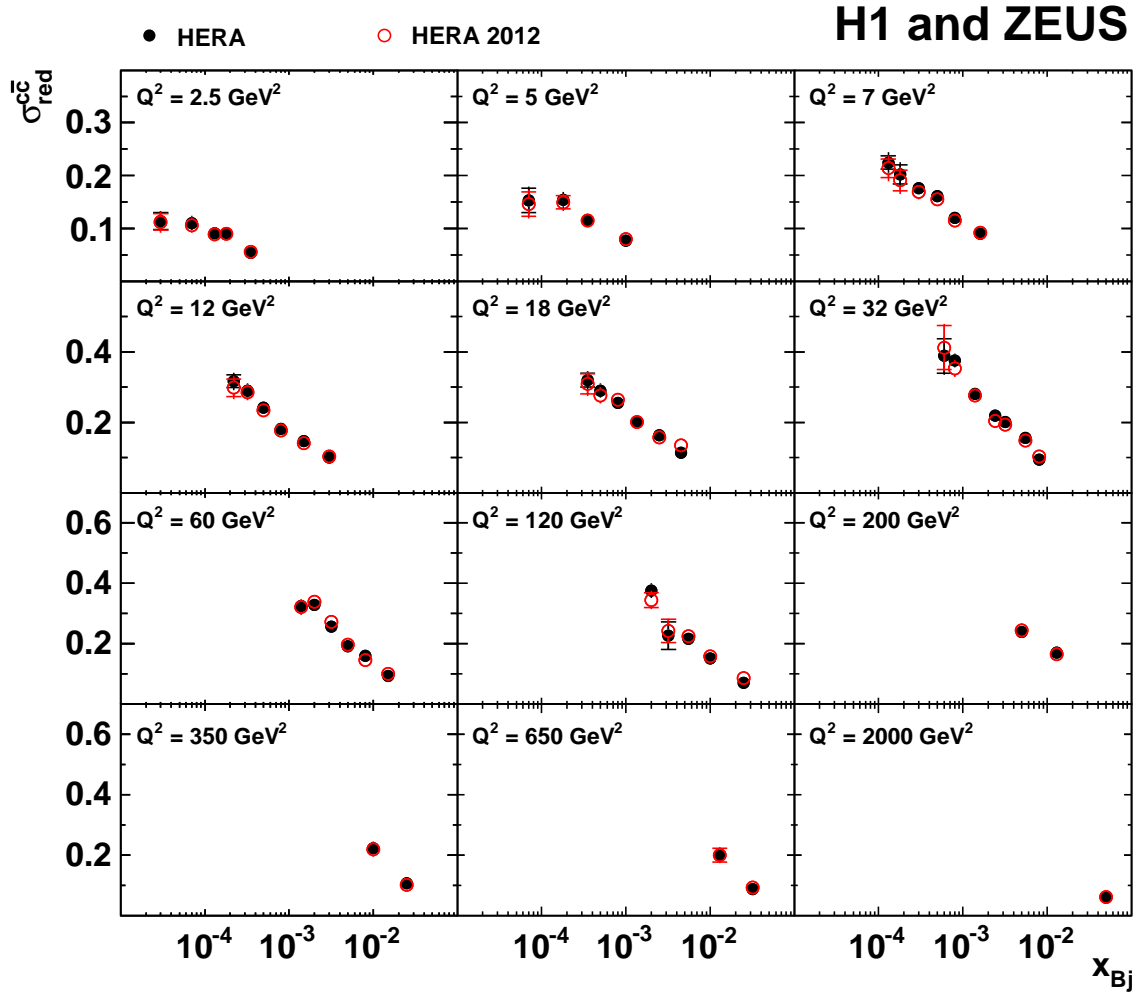 ,width=1.0\textwidth}
\caption{Combined reduced cross sections, \redcc, (full circles) as a function of \xbj for given values of $Q^2$, 
  compared to the results of the previous combination~\cite{HERAcharmcomb}, denoted as `HERA 2012' (open circles).}
\label{fig:data-charm-2012} 
\end{figure}

In figure~\ref{fig:data-charm-2012}, the new combined reduced charm cross sections are compared to the results of the previously published combination~\cite{HERAcharmcomb}.   
Good consistency between the different combinations can be observed. 
A detailed analysis of the cross-section measurements 
reveals a relative improvement in precision of about $20\,\%$ on average with respect to the previous measurements. The improvement reaches about $30\,\%$ in the range $7$~GeV$^2\le Q^2\le 60$~GeV$^2$, where the newly added datasets (datasets $9-11$ in table~\ref{tab:data}) contribute with high precision.

\section{Comparison with theory predictions}
\label{sect:comp-thoery} 
The combined heavy-flavour data are compared
with calculations using various schemes and PDF sets. Predictions using the FFNS~\cite{riemersma,gjr,ct10f3,mstw08f3,abkm09,abkm09b,abm11,abmp16} and the VFNS~\cite{rt_opt,FONLL,fonllb_and_c,nnlonllx} are considered, focussing on results using HERAPDF2.0 PDF sets. The data are also compared to FFNS predictions based on different variants of PDF sets at NLO and approximate NNLO provided by the ABM group~\cite{abmp16,abkm09}.
In the case of the VFNS, recent calculations of the NNPDF group based on the NNPDF3.1sx PDF set~\cite{nnlonllx} at NNLO, which specifically aim for 
 a better description of the DIS structure functions at small $\xbj$ and 
$Q^2$,
are also confronted with the measurements.
The calculations in the FFNS based on the HERAPDF2.0 FF3A PDF set will be considered as reference calculations in the subsequent parts of the paper.

\subsection{FFNS predictions}
\label{sect:ffns-predictions}
\begin{figure}[h]
\center
\epsfig{file=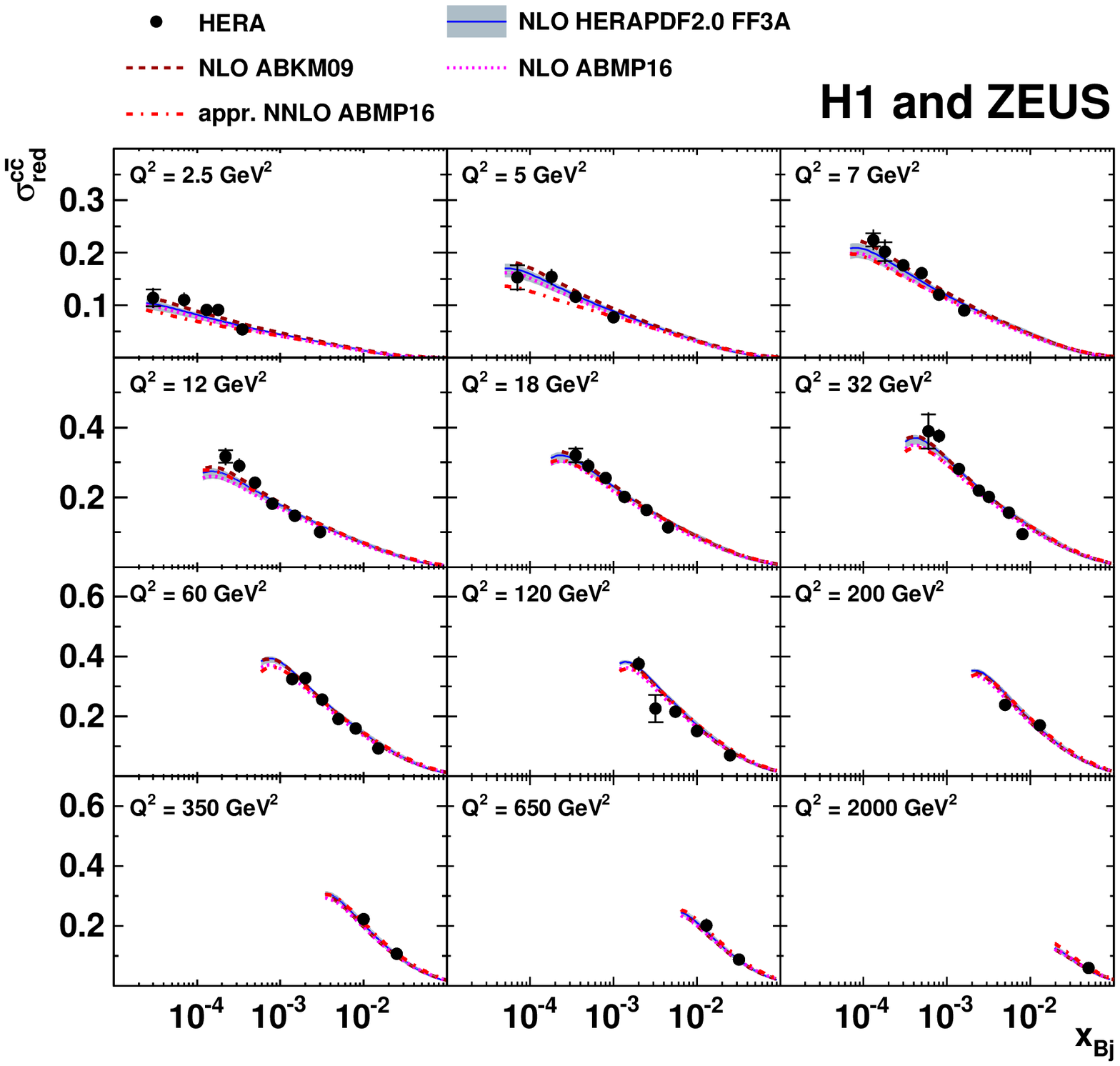 ,width=1.0\textwidth}
\caption{Combined reduced charm cross sections, \redcc, (full circles) as a function of \xbj for given values of $Q^2$, 
  compared to the NLO QCD FFNS predictions based on the HERAPDF2.0 FF3A (solid lines),  ABKM09 (dashed lines) and ABMP16 (dotted lines) PDF sets. Also shown is the approximate NNLO prediction using ABMP16 (dashed-dotted lines). The shaded bands on the HERAPDF2.0 FF3A predictions show the theory uncertainties obtained by adding PDF, scale and charm-quark mass uncertainties in quadrature.
 }
\label{fig:data-charm-theory} 
\end{figure}

\begin{figure}[h]
\center
\epsfig{file=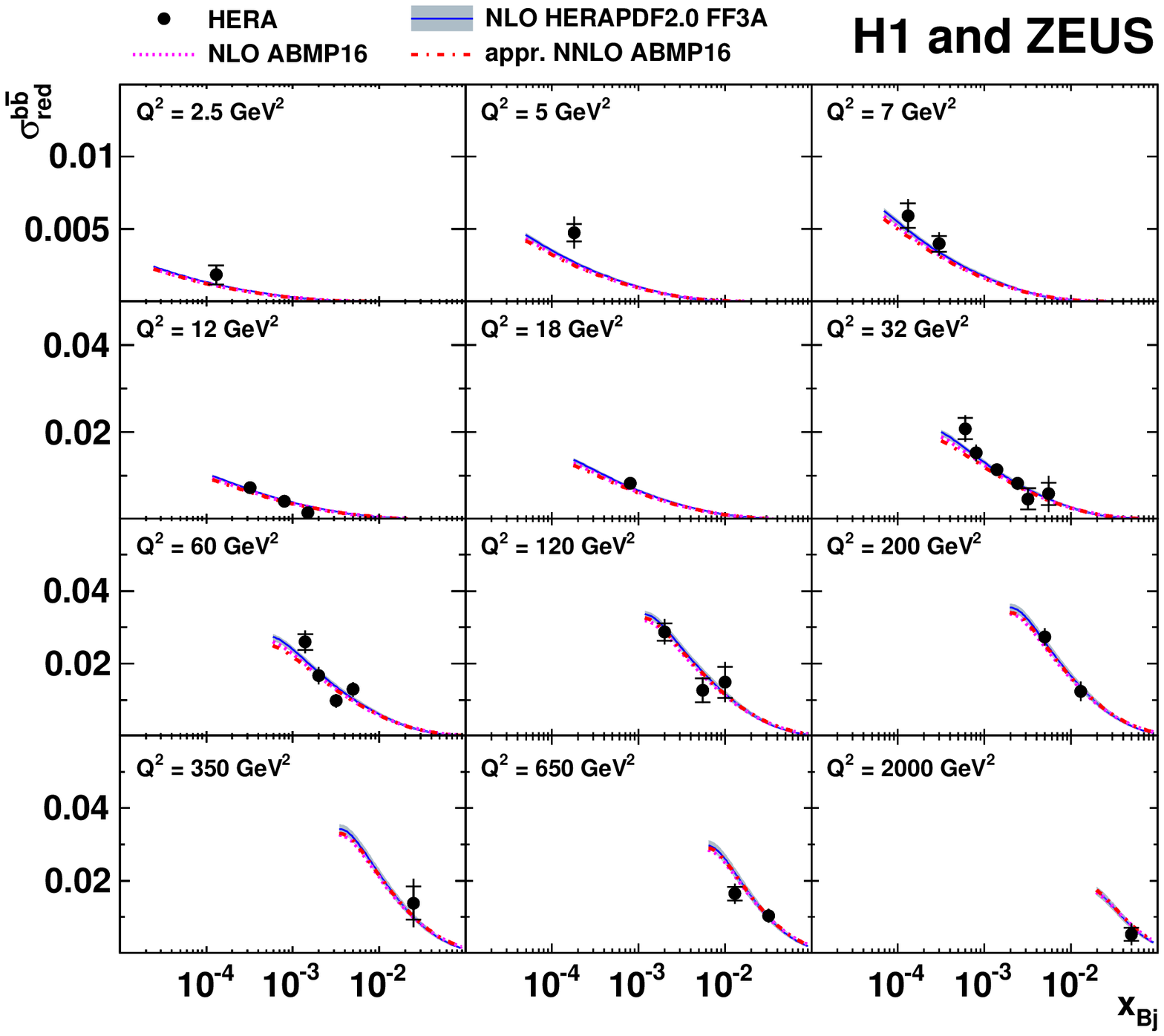 ,width=1.0\textwidth}
\caption{Combined reduced beauty cross sections, \redbb, (full circles) as a function of \xbj for given values of $Q^2$, 
 compared to the NLO QCD FFNS predictions based on the HERAPDF2.0 FF3A (solid lines) and ABMP16 (dotted lines) PDF sets. Also shown is the approximate NNLO prediction using ABMP16 (dashed-dotted lines). The shaded bands on the HERAPDF2.0 FF3A predictions show the theory uncertainties obtained by adding PDF, scale and beauty-quark mass uncertainties in quadrature.}
\label{fig:data-beauty-theory} 
\end{figure}

In figures~\ref{fig:data-charm-theory} and \ref{fig:data-beauty-theory}, theoretical predictions of the FFNS in the $\overline{\rm MS}$ running mass scheme
are compared to the combined reduced cross sections \redcc and \redbb, respectively. 
The theoretical predictions are obtained within the open-source QCD fit framework 
for PDF determination \textsc{xFitter}~\cite{xfitter}, which uses the \textsc{OPENQCDRAD} programme~\cite{openqcdrad} for the cross-section calculations.
The running heavy-flavour masses are set to the world average values~\cite{pdg2016} of~$m_c(m_c)=1.27\pm0.03$~GeV and $m_b(m_b)=4.18\pm0.03$~GeV. 
The predicted reduced cross sections are calculated
using the HERAPDF2.0 FF3A~\cite{HERAPDF20} and ABMP16~\cite{abmp16} NLO PDF sets 
 using NLO $(O(\alpha_s^2))$ coefficient functions and the 
ABMP16~\cite{abmp16} NNLO PDF set using approximate NNLO coefficient 
functions. 
The charm data are also compared to NLO predictions based on the ABKM09~\cite{abkm09} NLO PDF set used
in the previous analysis~\cite{HERAcharmcomb} of combined charm data. This PDF set was determined using a charm-quark mass of $m_c(m_c)=1.18$~GeV.
The PDF sets considered were extracted without explicitly using heavy-flavour data from HERA with the exception of the ABMP16 set, in which the HERA charm data from the previous combination~\cite{HERAcharmcomb} and some of the beauty data~\cite{h1ltt_hera2,zeusvtx} have been included.
For the predictions based on the HERAPDF2.0 FF3A set, theory uncertainties are given which are calculated by adding in quadrature the uncertainties from the PDF set, simultaneous variations of $\mu_r$ and $\mu_f$ by a factor of two up or down and the variation of the quark masses within the quoted uncertainties. 

The FFNS calculations 
reasonably describe the charm data (figure~\ref{fig:data-charm-theory}) although in the kinematic range where the data are very precise, the data show a  \xbj dependence somewhat steeper than 
predicted by
the calculations. For the different PDF sets and QCD orders considered,
the predictions are quite similar at larger $Q^2$ while some differences can be observed at smaller $Q^2$ or  $x_{\rm Bj}$.
For beauty production (figure~\ref{fig:data-beauty-theory}) the predictions are in good agreement with the data within the considerably larger experimental uncertainties. 

\begin{figure}[h]
\center
\epsfig{file=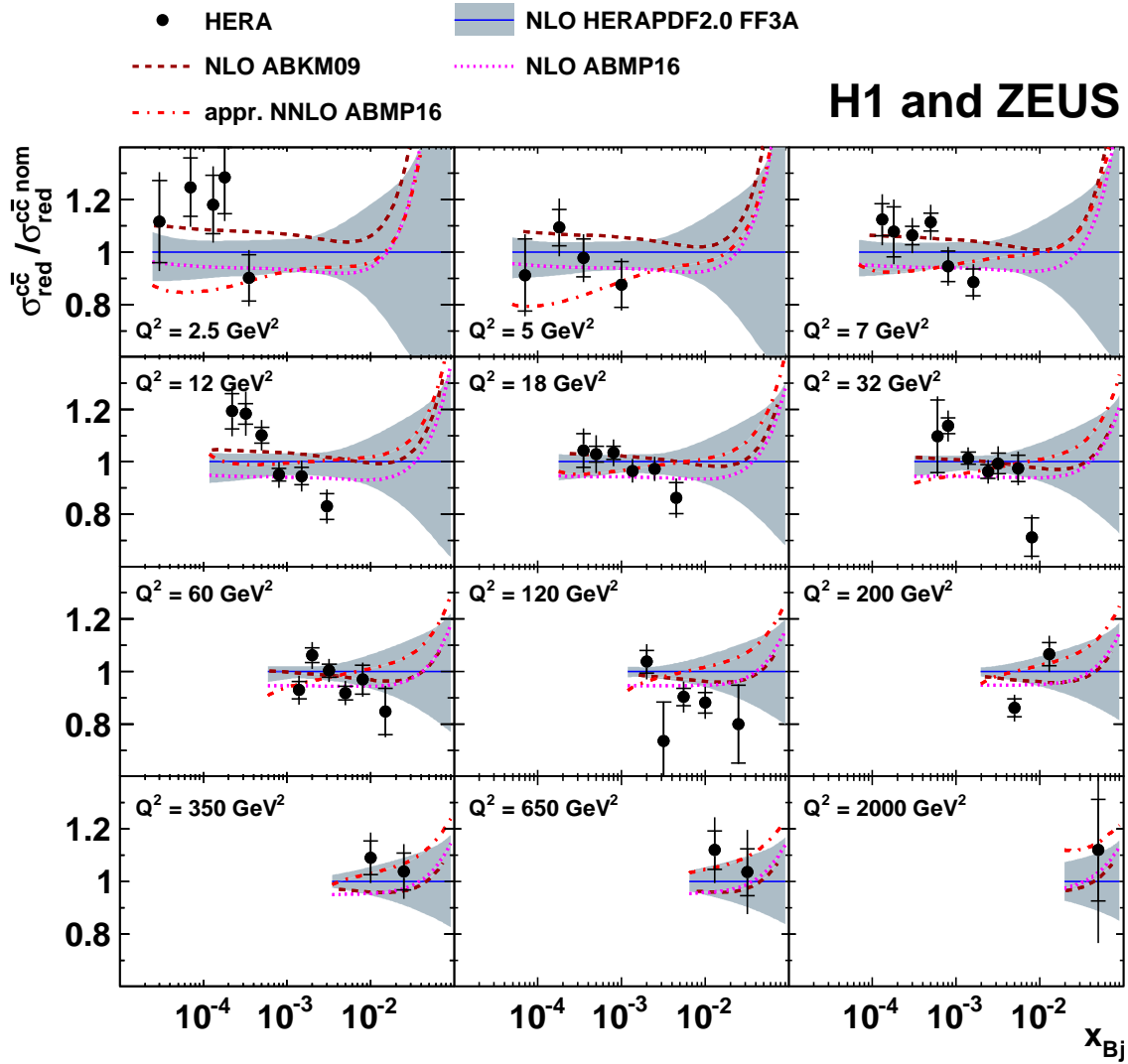 ,width=1.0\textwidth}
\caption{Ratio of reduced charm cross sections, \redcc, as a function of \xbj for given values of $Q^2$ for the combined data~(full circles) and the NLO (dashed and dotted lines) and approximate NNLO (dashed-dotted lines) QCD FFNS predictions, obtained using various PDFs, as in figure~\ref{fig:data-charm-theory}, with respect to the FFNS NLO predictions, $\sigma^{ c\overline{c}\ {\rm nom}}_{\rm red}$, obtained using HERAPDF2.0 FF3A (solid lines with shaded uncertainty bands).}
\label{fig:data-charm-theory-ff-ratio} 
\end{figure}

The description of the charm-production data is  illustrated
further in figure~\ref{fig:data-charm-theory-ff-ratio}, which shows the ratios of the reduced cross sections for data, ABKM09 and ABMP16 at NLO and approximate NNLO with respect to the NLO reduced cross sections predicted in the FFNS using the HERAPDF2.0 FF3A set. 
For $Q^2\ge 18$~GeV$^2$, the theoretical predictions are similar to each other in the kinematic region accessible at HERA. In this region, the predictions based on the different PDF sets and orders are well within the theoretical uncertainties obtained for the HERAPDF2.0 FF3A set. 
Towards smaller $Q^2$ and $x_{\rm Bj}$, some differences in the predictions become evident. 
In the region of $7$~GeV$^2\le Q^2\le120$~GeV$^2$, the theory tends to be below the data at small \xbj and above the data at large $x_{\rm Bj}$, independent of the PDF set and order used. 

\begin{figure}[h]
\center
\epsfig{file=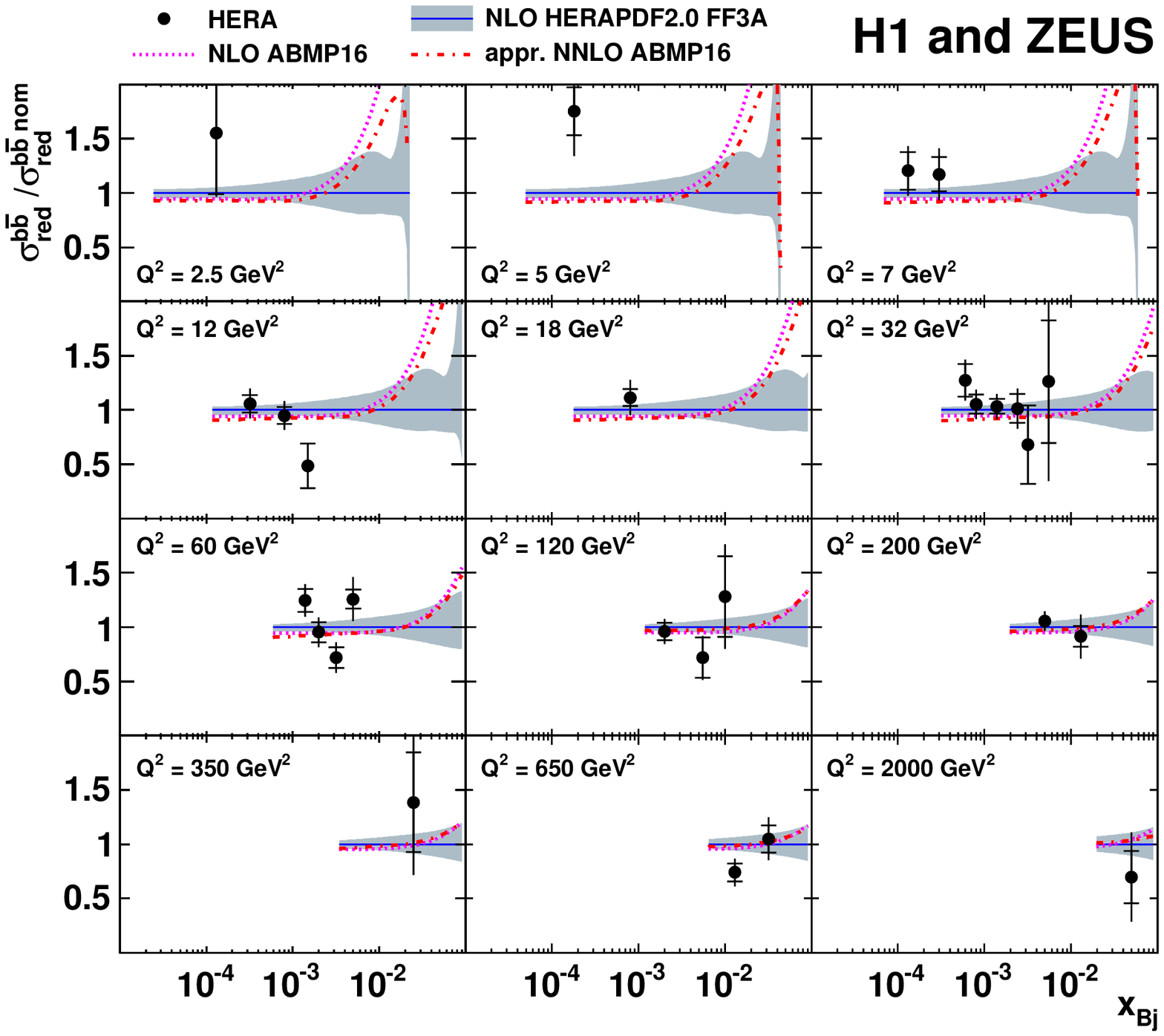 ,width=1.0\textwidth}
\caption{
Ratio of reduced beauty cross sections, \redbb, as a function of \xbj for given values of $Q^2$ for the combined data~(full circles) and the NLO (dashed and dotted lines) and approximate NNLO (dashed-dotted lines) QCD FFNS predictions, obtained using the same PDF sets as in figure~\ref{fig:data-beauty-theory}, with respect to the FFNS NLO predictions, $\sigma^{ b\overline{b}\ {\rm nom}}_{\rm red}$, obtained using HERAPDF2.0 FF3A (solid lines with shaded uncertainty bands).}
\label{fig:data-beauty-theory-ff-ratio} 
\end{figure}

In figure~\ref{fig:data-beauty-theory-ff-ratio}, the corresponding ratios are shown for the beauty data. In the kinematic region accessible at HERA, the predictions are very similar to each other. Within the experimental uncertainties, the data are well described by all calculations. 

 \subsection{VFNS predictions}
\label{sect:vfns-predictions}
\begin{figure}[h]
\center
\epsfig{file=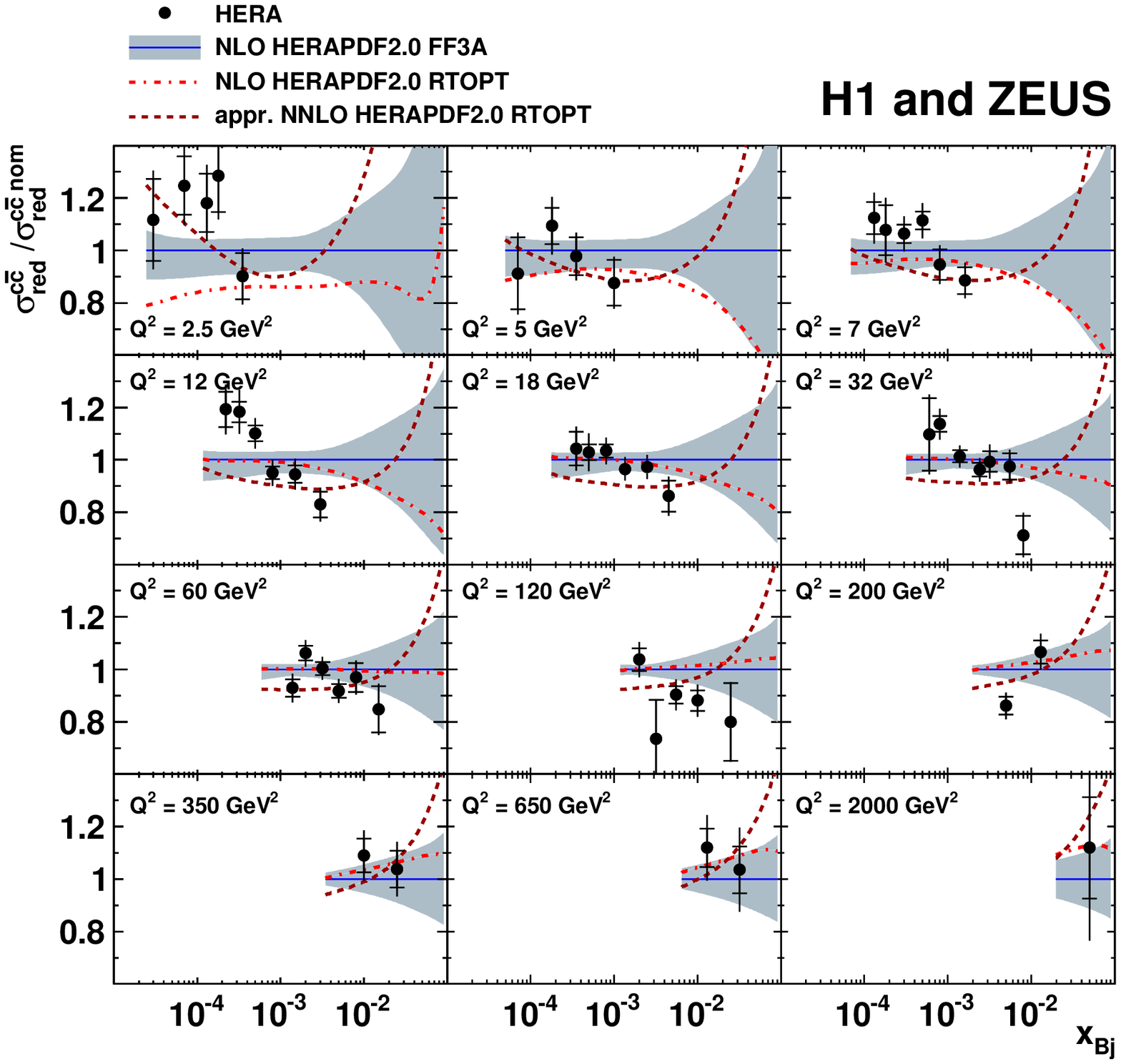 ,width=1.0\textwidth}
\caption{Ratio of reduced charm cross sections, \redcc, as a function of \xbj for given values of $Q^2$ for the combined data~(full circles) and the NLO (dashed-dotted lines) and approximate NNLO (dashed lines) VFNS predictions, obtained using the respective HERAPDF2.0 PDF sets, with respect to the FFNS NLO predictions, $\sigma^{ c\overline{c}\ {\rm nom}}_{\rm red}$, obtained using HERAPDF2.0 FF3A (solid lines with shaded uncertainty bands). The uncertainties for the VFNS predictions (not shown) are of similar size to those presented for the FFNS calculation.}
\label{fig:data-charm-theory-vf-ratio} 
\end{figure}

In figure~\ref{fig:data-charm-theory-vf-ratio}, predictions of the RTOPT~\cite{rt_opt} NLO and approximate NNLO VFNS using the corresponding NLO and NNLO HERAPDF2.0 PDF sets are compared to the charm measurements. As in figure~\ref{fig:data-charm-theory-ff-ratio}, the ratio of data and theory predictions to the reference calculations are shown. While the NLO VFNS predictions are in general consistent with 
 both the data cross sections and the reference calculations, the approximate NNLO cross sections show somewhat  larger differences,
about $10\%$ smaller than the reference cross sections in the region $12$~GeV$^2\le Q^2\le 120$~GeV$^2$. On the other hand,
at $Q^2\le 7$~GeV$^2$ the $x_{\rm Bj}$ slopes of the NNLO VFNS predictions tend 
to describe the data somewhat better than the reference calculations.
Overall, the NLO and approximate NNLO VFNS predictions describe the data 
about equally well, but not better than the reference FFNS calculations.

\begin{figure}[h]
\center
\epsfig{file=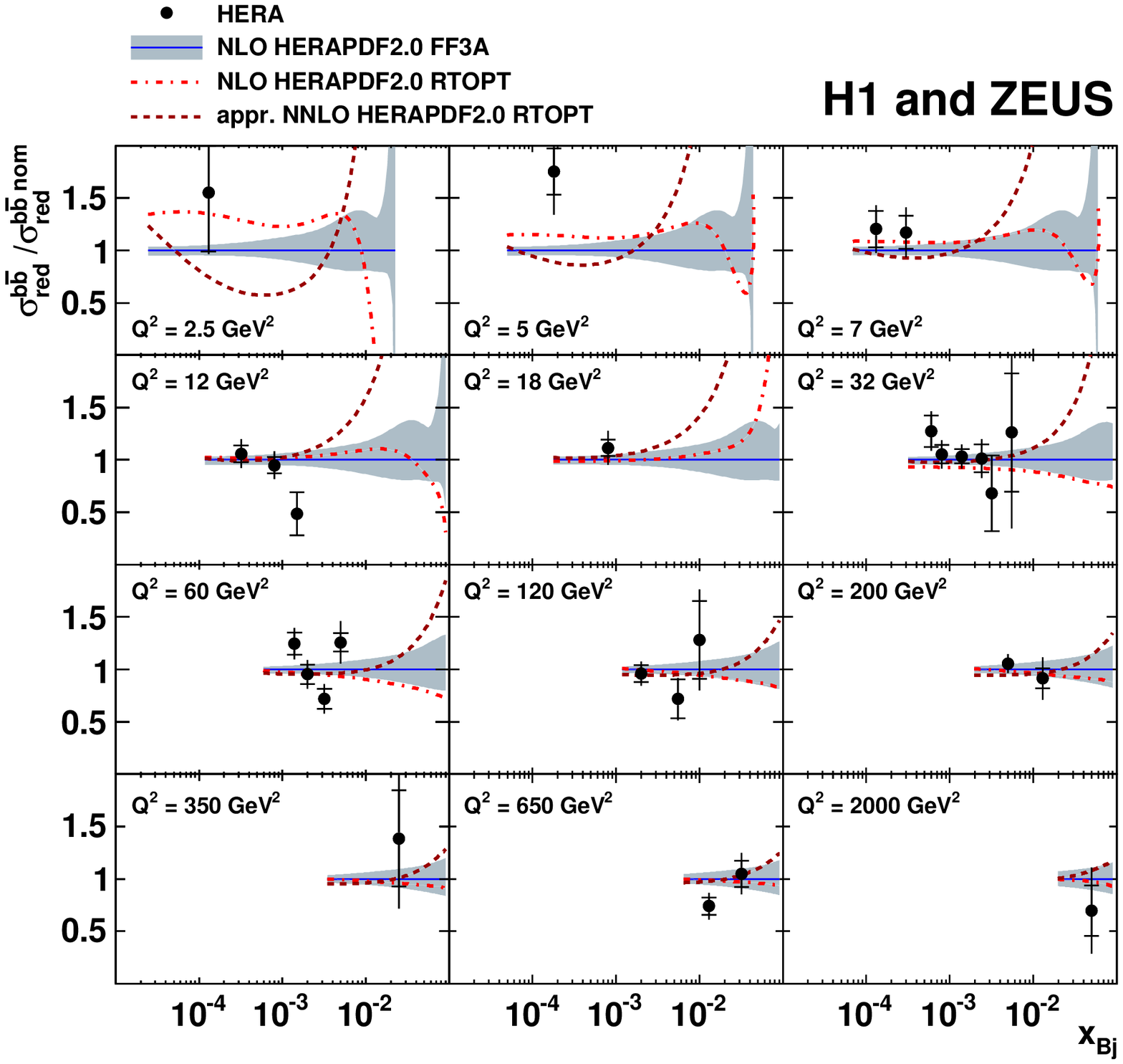 ,width=1.0\textwidth}
\caption{Ratio of reduced beauty cross sections, \redbb, as a function of \xbj for given values of $Q^2$ for the combined data~(full circles) and the NLO (dashed-dotted lines) and approximate NNLO (dashed lines) VFNS predictions, obtained using the respective HERAPDF2.0 PDF sets, with respect to the FFNS NLO predictions, $\sigma^{ b\overline{b}\ {\rm nom}}_{\rm red}$, obtained using HERAPDF2.0 FF3A (solid lines with shaded uncertainty bands). The uncertainties for the VFNS predictions (not shown) are of similar size to those presented for the FFNS calculation.}
\label{fig:data-beauty-theory-vf-ratio} 
\end{figure}

In figure~\ref{fig:data-beauty-theory-vf-ratio}, the same ratios as in the preceding paragraph are shown for beauty production.
In the kinematic region accessible in DIS beauty production at HERA, the differences between the different calculations are small in comparison to the experimental uncertainties of the measurements.

\begin{figure}[h]
\center
\epsfig{file=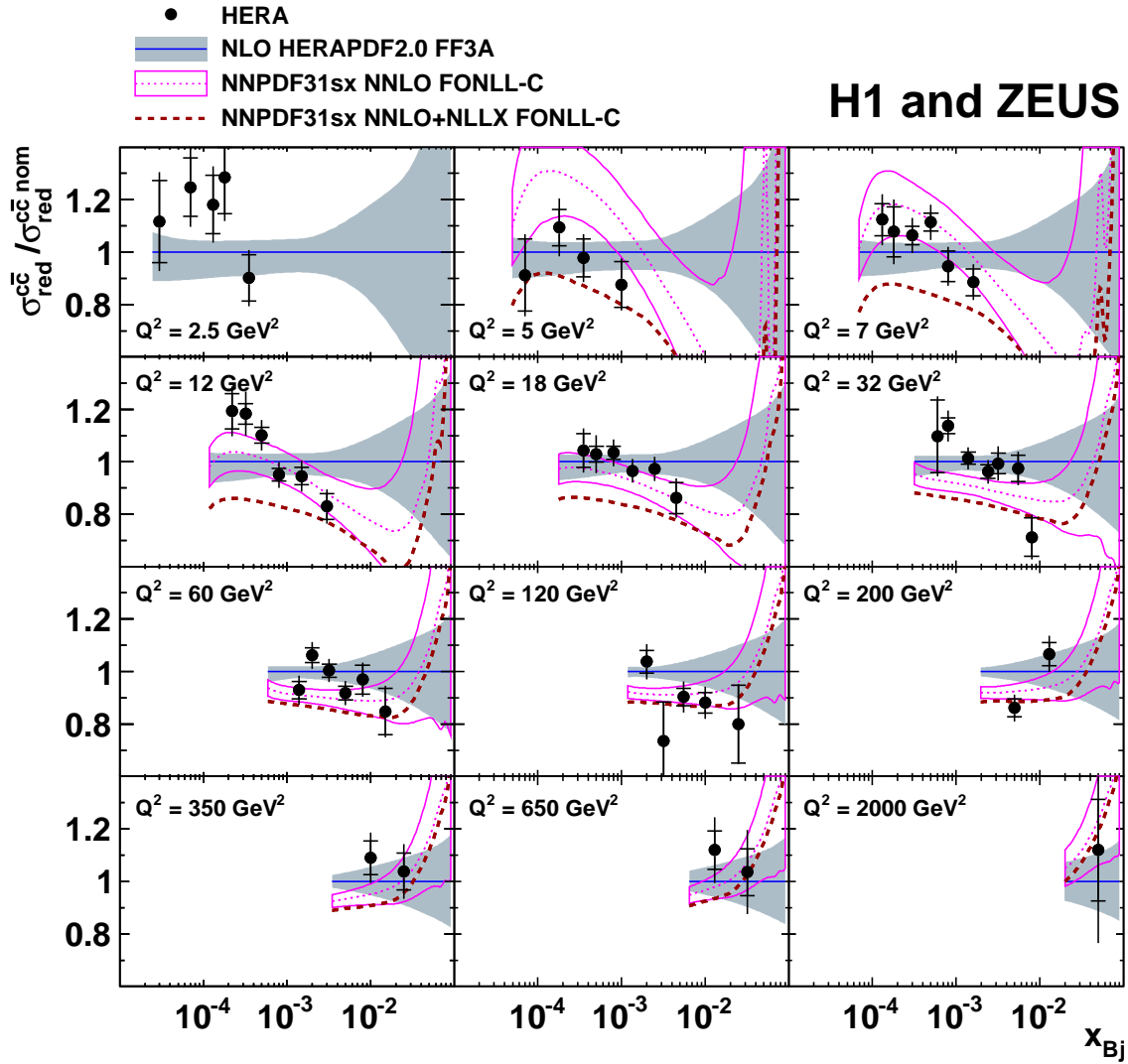 ,width=1.0\textwidth}
\caption{
Ratio of reduced charm cross sections, \redcc, as a function of \xbj for given values of $Q^2$ for the combined data~(full circles) and the NNLO VFNS predictions of the NNPDF group with respect to the FFNS predictions, $\sigma^{ c\overline{c}\ {\rm nom}}_{\rm red}$, obtained using HERAPDF2.0 FF3A (solid lines with shaded uncertainty bands).
Results from two different calculations are shown: without (FONLL-C, dotted lines with uncertainty bands) and with low-$x$ resummation (FONLL-C+NLLsx, dashed lines). For the calculations the NNPDF3.1sx PDF set is used.
For better clarity of the presentation the uncertainties of the FONLL+NLLsx calculations are not shown. These are in size similar to those shown for the FONLL calculations.
 No FONLL predictions based on NNPDF3.1sx are shown at $Q^2=2.5$~GeV$^2$ because this value lies below the starting scale of the QCD evolution in the calculation ($2.6$~GeV$^2$).}  
\label{fig:data-charm-theory-nnpdf-ratio} 
\end{figure}

The calculations considered so far generally show some tension
in describing the $x_{\rm Bj}$ slopes of the measured charm data over a large range in $Q^2$. Therefore the charm data are compared in figure~\ref{fig:data-charm-theory-nnpdf-ratio} to recent calculations\cite{nnlonllx,nnpdfComm} in the FONLL-C scheme 
 with (NNLO+NLLsx) and without (NNLO) low-$x$ resummation in both $O(\alpha_s^2)$ matrix elements and $O(\alpha_s^3)$ PDF evolution, using the NNPDF3.1sx framework, which aim for a better description of the proton structure functions at low $\xbj$ and $Q^2$.
The charm data from the previous combination have already been used for the determination  of the NNPDF3.1sx PDFs.
Both calculations provide a better description of the $x_{\rm Bj}$ shape of the measured charm cross sections for $Q^2< 32$~GeV$^2$. However, the predictions lie significantly below the data in most of the phase space. This is especially the case for the NNLO+NLLsx calculations. Overall, the description is not improved with respect to the FFNS reference calculations.

\subsection{Summary of the comparison to theoretical predictions}
\label{sect:theory-summary}
 
 The comparison to data of the different predictions considered is summarised in table~\ref{tab:qcd-chi2pdf} in which the agreement with data is expressed in terms of $\chi^2$ and the corresponding fit probabilities ($p$-values). 
The table also includes a comparison to the previous combined charm data~\cite{HERAcharmcomb}. 
The agreement of the various predictions with the charm cross-section measurements of the current analysis is poorer than with the results of the previous combination, for which consistency between theory and data within the experimental uncertainties is observed for most of the calculations. As shown in section~\ref{sec:results}, the charm cross sections of the current analysis agree well with the previous measurements but have considerably smaller uncertainties. 
The observed changes in the $\chi^2$ values 
are consistent with the improvement in data precision if the predictions do not fully describe reality.
The tension observed between the central theory predictions and the charm data ranges from $\sim 3\sigma$ to more than  $6\sigma$, depending on the prediction.
Among the calculations considered, the NLO FFNS calculations provide the best description of the charm data. 
For the beauty cross sections, good agreement of theory and data is observed within the larger experimental uncertainties. In all cases, the effect of the PDF uncertainties on the $\chi^2$ values is negligible. 

\section{QCD analysis}
\label{sec:qcd}

The combined charm and beauty data are used together with the combined HERA inclusive DIS data~\cite{HERAPDF20} to perform a QCD analysis in the FFNS using the $\overline{\rm MS}$ mass-renormalisation scheme at NLO. The main focus of this analysis is the simultaneous determination of the running heavy-quark masses $m_c(m_c)$ and $m_b(m_b)$. The theory description of the $x_{\rm Bj}$ dependence of the reduced charm cross section is also investigated.
\subsection{Theoretical formalism and settings}
\label{sec:qcd-settings}

The analysis is performed with the \textsc{xFitter}~\cite{xfitter}  programme, in which  
the scale evolution of partons is calculated through DGLAP 
equations~\cite{dglap} at NLO, 
as implemented in the \textsc{QCDNUM} programme~\cite{qcdnum}. 
The theoretical FFNS predictions for the HERA data are obtained using the OPENQCDRAD programme~\cite{openqcdrad} 
interfaced in the~\textsc{xFitter} framework. 
The number of active flavours is set to $n_f = 3$ at all scales.
For the heavy-flavour contributions the scales are set to $\mu_r = \mu_f=\sqrt{Q^2+4m_{\rm Q}^2}$.
The heavy-quark masses are left free in the fit unless stated otherwise. 
For the light-flavour contributions to the inclusive DIS cross sections,
the pQCD scales are set to $\mu_r = \mu_f = Q$. 
The massless contribution to the longitudinal structure function $F_{\rm L}$ is calculated to $O(\alpha_s)$.
The strong coupling strength is set to $\alpha_s^{n_f=3}(M_Z) = 0.106$, 
 corresponding to $\alpha_s^{n_f=5}(M_Z) = 0.118$. 
In order to perform the analysis in the kinematic region where pQCD is assumed to be applicable,
the $Q^2$ range of the inclusive HERA data is restricted to $Q^2 \ge Q^2_{\rm min} = 3.5$~GeV$^2$. 
No such cut is applied to the charm and beauty data, since the relevant scales $\mu_r^2 = \mu_f^2=Q^2+4m_{\rm Q}^2$ are above $3.5$~GeV$^2$ for all measurements.

This theory setup is slightly different from that used for the original extraction~\cite{HERAPDF20} of HERAPDF2.0 FF3A. In contrast to the analysis presented here, HERAPDF2.0 FF3A was determined using the on-shell mass (pole-mass) scheme for the calculation of heavy-quark production and $F_{\rm L}$ was calculated to $O(\alpha^2_s)$.

Perturbative QCD predictions were fit to the data using the same $\chi^2$ definition as for the fits to the inclusive DIS data (equation (32) in reference~\cite{HERAPDF20}). 
It includes an additional logarithmic term that is relevant when the estimated statistical 
and uncorrelated systematic uncertainties in the data are rescaled during the fit~\cite{Aaron:2012qi}.
The correlated systematic uncertainties are treated through nuisance parameters. 

The procedure for the determination of the PDFs follows the approach of HERAPDF2.0~\cite{HERAPDF20}. 
At the starting scale $\mu_{\rm f,0}$, the density functions of a parton $f$ of the proton are parametrised using the generic form:
\begin{equation}
xf(x)=Ax^B\left(1-x\right)^C\left(1+Dx+Ex^2\right),
\label{eq:genericPDF}
\end{equation}
where $x$ is the 
fraction of the incoming proton momentum carried by the incoming parton
in the proton's infinite-momentum frame.
The parametrised PDFs are the gluon distribution $xg(x)$, the valence quark distributions $xu_v(x)$ and $xd_v(x)$, and 
the $u$- and $d$-type antiquark distributions $x\overline{U}(x)$ and $x\overline{D}(x)$. 

At the initial QCD evolution scale\footnote{In the FFNS this scale is decoupled from the charm-quark mass.} $\mu_{\rm  f, 0}^2 = 1.9$~GeV$^2$, the default parameterisation of the PDFs has the form:
\begin{eqnarray}
xg(x) &=& A_{g} x^{B_{g}}\,(1-x)^{C_{g}} - A'_{g} x^{B'_{g}}\,(1-x)^{C'_{g}},
\label{eq:g} \nonumber\\
xu_v(x) &=& A_{u_v}x^{B_{u_v}}\,(1-x)^{C_{u_v}}\,(1+E_{u_v}x^2) ,
\label{eq:uv} \nonumber\\
xd_v(x) &=& A_{d_v}x^{B_{d_v}}\,(1-x)^{C_{d_v}},
\label{eq:dv}\\
x\overline{U}(x)&=& A_{\overline{U}}x^{B_{\overline{U}}}\, (1-x)^{C_{\overline{U}}}\, (1+D_{\overline{U}}x),
\label{eq:Ubar} \nonumber\\
x\overline{D}(x)&=& A_{\overline{D}}x^{B_{\overline{D}}}\, (1-x)^{C_{\overline{D}}}.
\label{eq:Dbar} \nonumber
\end{eqnarray}
The gluon density function, $xg(x)$, is different from equation~(\ref{eq:genericPDF}), it includes an additional term $- A'_{g} x^{B'_{g}}\,(1-x)^{C'_{g}}$. The antiquark density functions, $x\overline{U}(x)$ and $x\overline{D}(x)$, are defined as $x\overline{U}(x) = x\overline{u}(x)$ and $x\overline{D}(x) = x\overline{d}(x) + x\overline{s}(x)$, where 
$x\overline{u}(x)$, $x\overline{d}(x)$, and $x\overline{s}(x)$ are the up-, down-, and strange-antiquark distributions, respectively. The total quark density functions are $xU(x)=xu_v(x)+x\overline{U}(x)$ and $xD(X)=xd_v(x)+x\overline{D}(x)$.
The sea-antiquark distribution is defined as $x\Sigma(x)=x\overline{u}(x)+x\overline{d}(x)+x\overline{s}(x)$. 
The normalisation parameters $A_{u_{{v}}}$, $A_{d_{v}}$, and $A_{g}$ are determined by the QCD sum rules. 
The $B$ and $B'$ parameters determine the PDFs at small $x$, 
and the $C$ parameters describe the shape of the distributions as $x\,{\to}\,1$. 
The parameter $C'_{g}=25$ is fixed~\cite{Martin:2009ad}. 
Additional constraints $B_{\overline{{U}}} = B_{\overline{{D}}}$ and 
$A_{\overline{{U}}} = A_{\overline{{D}}}(1 - f_{s})$ are imposed to ensure the same normalisation 
for the $x\overline{u}$ and $x\overline{d}$ distributions as $x \to 0$. 
The strangeness fraction $f_{s} = x\overline{s}/( x\overline{d} + x\overline{s})$ is fixed to 
$f_{s}=0.4$ as in the HERAPDF2.0 analysis~\cite{HERAPDF20}. 

The selection of parameters in equation~(\ref{eq:dv}) from the general form, equation~(\ref{eq:genericPDF}), is made by first fitting with all $D$ and $E$ parameters set to zero, 
and then including them one at a time in the fit. 
The improvement in the $\chi^2$ of the fit is monitored. 
If $\chi^2$ improves significantly, the parameter is added and the procedure is repeated until no further significant improvement is observed. 
This leads to the same $14$ free PDF parameters as in the inclusive HERAPDF2.0 analysis~\cite{HERAPDF20}. 

The PDF uncertainties are estimated according to the general approach of HERAPDF2.0~\cite{HERAPDF20}, in which 
the experimental, model, and parameterisation uncertainties are taken into account. 
The experimental uncertainties are determined from the fit using the tolerance criterion of $\Delta\chi^2 =1$. 
Model uncertainties arise from the variations 
of the strong coupling constant $\alpha_s^{n_f=3}(M_Z) = 0.1060 \pm 0.0015$, 
simultaneous variations of the factorisation and renormalisation scales up or down by a factor of two,
the variation of the strangeness fraction $0.3 \leq f_{s} \leq 0.5$, 
and the value of $2.5$~GeV$^2\leq Q^2_{\textrm{min}}\leq 5.0$~GeV$^2$ imposed on the inclusive HERA data. 
The total model uncertainties are obtained by adding the individual contributions in quadrature.
The parameterisation uncertainty is estimated by extending the functional form in equation~(\ref{eq:dv}) 
of all parton density functions with additional parameters $D$ and $E$ added one at a time. 
An additional parameterisation uncertainty is considered by using the functional form in equation~(\ref{eq:dv}) with $E_{u_v} = 0$. 
The $\chi^2$ in this variant of the fit is only $5$ units worse than that with the released $E_{u_v}$ parameter; changing this parameter noticeably affects the mass determination.
In addition, $\mu_{\rm f,0}^2$ is varied within $1.6$~GeV$^2 < \mu_{\rm f, 0}^2 < 2.2$~GeV$^2$.
The parameterisation uncertainty is determined at each $\xbj$ value from the maximal differences between the PDFs resulting from the central fit and all parameterisation variations. 
The total uncertainty is obtained by adding the fit, model and parameterisation uncertainties in quadrature. 
The values of the input parameters for the fit and their variations considered, to evaluate model and parameterisation uncertainties, are given in table~\ref{tab:mcmb}.

\subsection{QCD fit and determination of the running heavy-quark masses}
\label{sec:qcd-results}

In the QCD fit, 
the running heavy-quark masses are fitted simultaneously 
with the PDF parameters in equation~(\ref{eq:dv}).
The fit yields a total $\chi^2=1435$ for $1208$ degrees of freedom. The ratio $\chi^2/{\rm d.o.f.}=1.19$ is similar in size to the values obtained in the analysis of the HERA combined inclusive data~\cite{HERAPDF20}.  The resulting PDF set is termed \pdfhq.
The central values of the fitted parameters are given in the appendix.

In figure~\ref{fig:fittedpdfs}, the PDFs at the scale $\mu^2_{{\rm f ,0}}=1.9$~GeV$^2$ are presented. Also shown are the PDFs, including experimental uncertainties, obtained by a fit to the inclusive data only with the heavy-quark masses fixed to $m_c(m_c)=1.27$~GeV and $m_b(m_b)=4.18$~GeV~\cite{pdg2016}.  No significant differences between the two PDF sets are observed.
Only a slight enhancement in the gluon density of \pdfhq ~compared to that determined from the inclusive data only can be observed around $x=2\cdot10^{-3}$. This corresponds to the region in $x$ where the charm data are most precise. 
When used together with the inclusive HERA data, the heavy-flavour data have only little influence on the shape of the PDFs determined with quark masses fixed to their expected values.
This confirms the findings \cite{HERAPDF20} made with the previously published combined charm data.
However, the smaller uncertainties of the new combination reduce the uncertainty of the charm-quark mass determination with respect to the previous result\footnote{The previous analysis did not consider scale variations and a less flexible PDF parameterisation was used.} \cite{HERAcharmcomb}. The beauty-quark mass determination improves the previous result based on a single dataset \cite{zeusvtx}. 
The running heavy-quark masses are determined as: 
\begin{eqnarray}
m_c(m_c) = 1.290^{+0.046}_{-0.041} {\rm (exp/fit)} {}^{+0.062}_{-0.014}  {\rm (model)} {}^{+0.003}_{-0.031} {\rm (parameterisation)}~{\rm GeV}, \nonumber\\
m_b(m_b) = 4.049^{+0.104}_{-0.109} {\rm (exp/fit)} {}^{+0.090}_{-0.032} {\rm (model)} {}^{+0.001}_{-0.031} {\rm (parameterisation)}~{\rm GeV}. 
\label{eq:finalmcmb}
\end{eqnarray}
The individual contributions to the uncertainties are listed in table~\ref{tab:mcmb}.
The model uncertainties are dominated by those arising from the QCD scale variations. 
In the case of the charm-quark mass, the variation in $\alpha_s$ also yields a sizeable contribution while the 
other sources lead to uncertainties of typically a few MeV, both for $m_c(m_c)$ and $m_b(m_b)$. 
The main contribution to the parameterisation uncertainties comes from the fit variant in which the term $E_{u_v}$ is set to zero,  other contributions are negligible. 
Both mass values are in agreement with the corresponding PDG values~\cite{pdg2016} and
the value of $m_c(m_c)$  determined here agrees well with the result from the previous analysis of HERA combined charm cross sections~\cite{HERAcharmcomb}.

A cross check is performed using the Monte Carlo method~\cite{Giele:1998gw,Giele:2001mr}. It is based on analysing
a large number of pseudo datasets called replicas. For this cross check, $500$ replicas are
created by taking the combined data and fluctuating the values of the reduced cross sections
randomly within their statistical and systematic uncertainties taking into account correlations. 
All uncertainties are assumed to follow a Gaussian distribution. The central values for the fitted parameters 
and their uncertainties are estimated using the mean and RMS values over the replicas.
The obtained heavy-quark masses and their experimental/fit uncertainties are in agreement with those quoted in equation~(\ref{eq:finalmcmb}). 

In order to study the influence of the inclusive data on the mass determination, fits to the combined inclusive data only are also tried. 
In this case, the fit results are very sensitive to the choice of the PDF parameterisation. When using the
default 14 parameters, the masses are determined to be
$m_c(m_c) = 1.80^{+0.14}_{-0.13} {\rm (exp/fit)}~{\rm GeV}$, $m_b(m_b) = 8.45^{+2.28}_{-1.81} {\rm (exp/fit)}~{\rm GeV}$, where only the experimental/fit uncertainties are quoted.
In the variant of the fit using the inclusive data only and the reduced parameterisation with $E_{u_v} = 0$, 
the central fitted values for the heavy-quark masses are: $m_c(m_c) = 1.45~{\rm GeV}$, $m_b(m_b) = 4.00~{\rm GeV}$. The sensitivity to the PDF parameterisation and the large experimental/fit uncertainties for a given parameterisation demonstrate that attempts to extract heavy quark masses from inclusive HERA data alone are not reasonable in this framework. The large effect on the fitted masses observed here, when setting $E_{u_v} = 0$, motivates the  $E_{u_v}$ variation in the \pdfhq~fit.

\begin{figure}[h]
\center
\epsfig{file=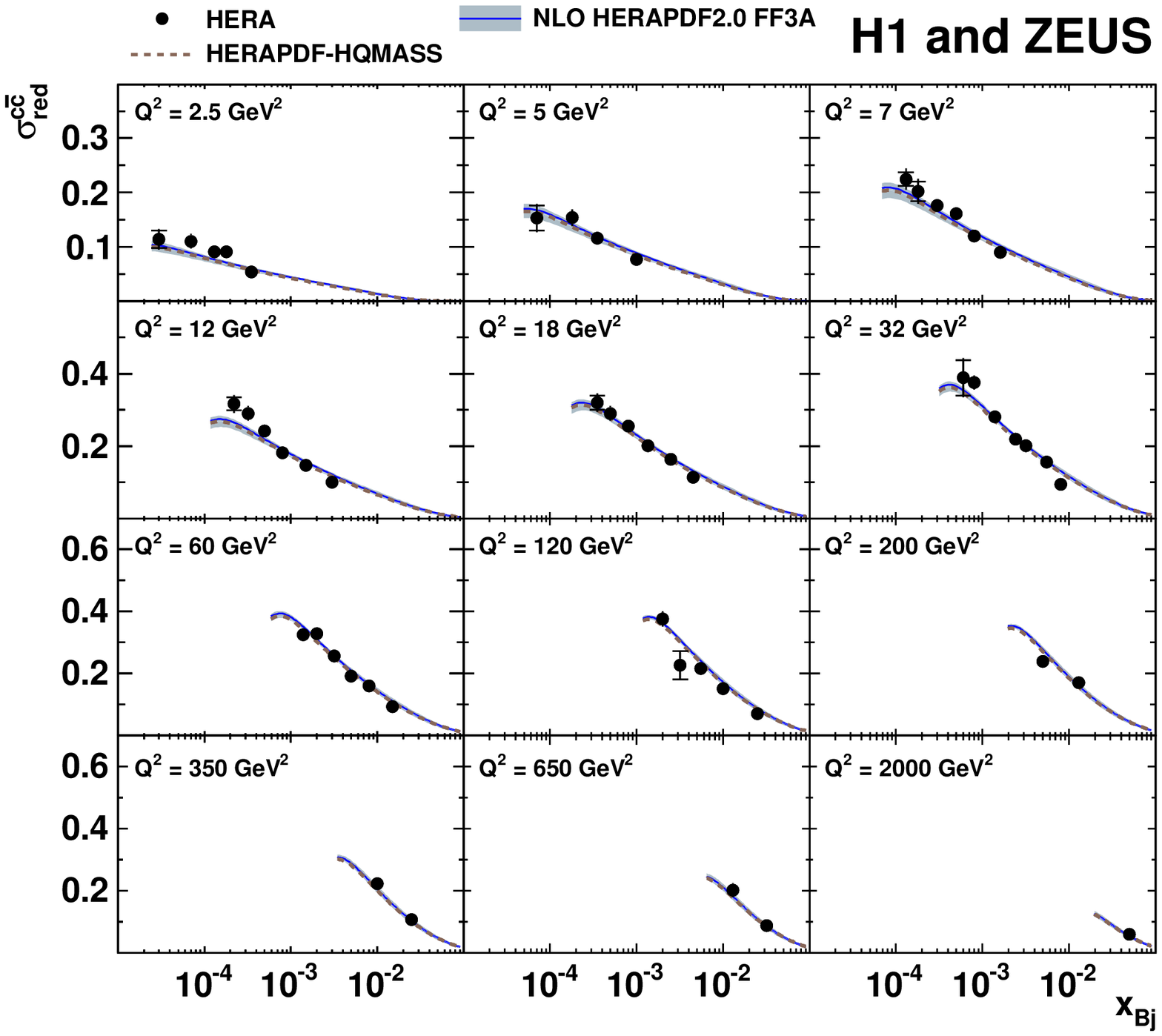 ,width=1.0\textwidth}
\caption{Combined reduced charm cross sections, \redcc,~(full circles) as a function of \xbj for given values of $Q^2$, 
  compared to the NLO QCD FFNS predictions based on \pdfhq~(dashed lines) and on HERAPDF2.0 FF3A (solid lines). The shaded bands on the HERAPDF2.0 FF3A predictions show the theory uncertainties obtained by adding PDF, scale and charm-quark mass uncertainties in quadrature.
 }
\label{fig:charm-fitted-PDF} 
\end{figure}

\begin{figure}[h]
\center
\epsfig{file=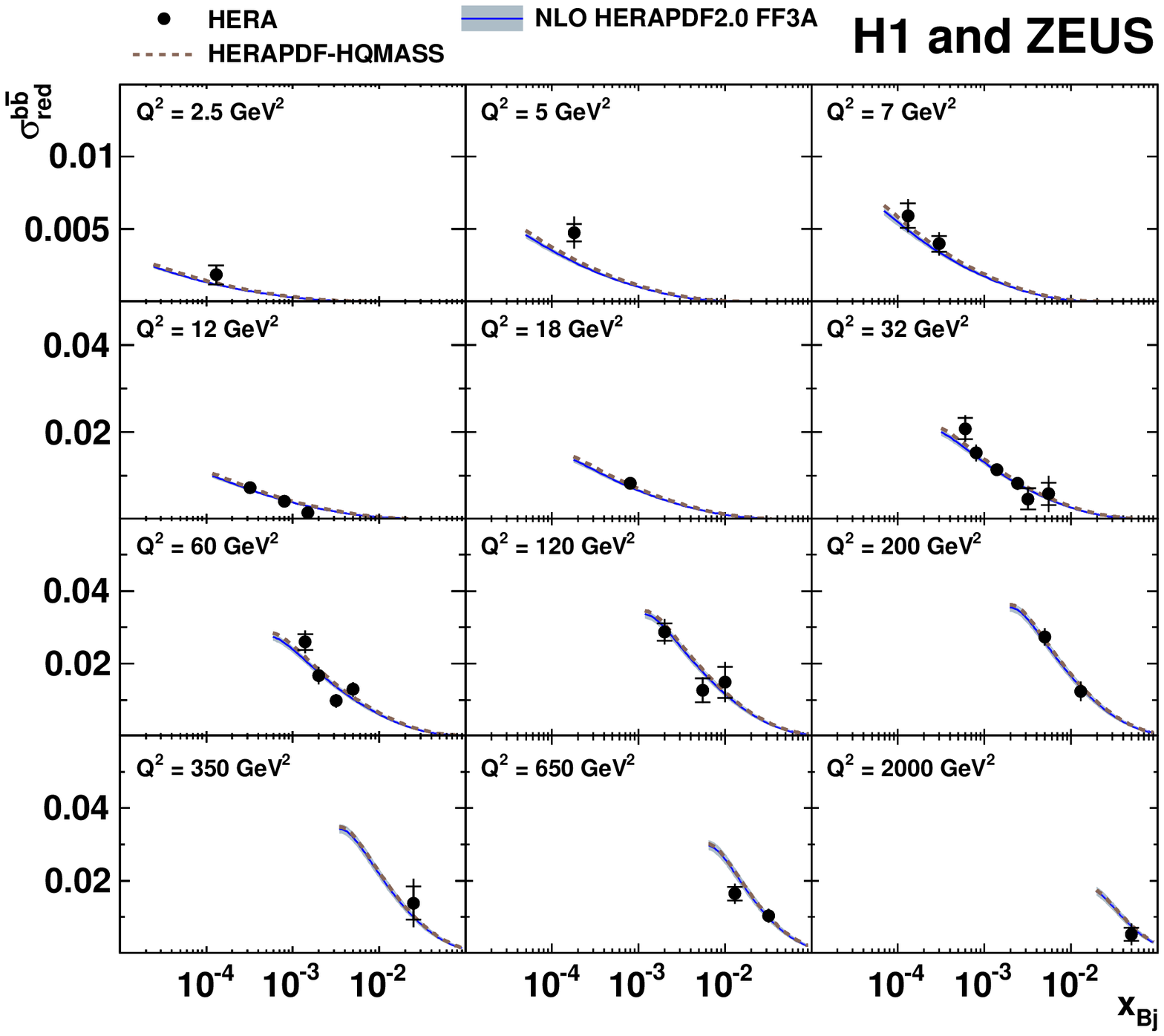 ,width=1.0\textwidth}
\caption{Combined reduced beauty cross sections, \redbb,~(full circles) as a function of \xbj for given values of $Q^2$, 
  compared to the NLO QCD FFNS predictions based on \pdfhq~(dashed lines) and on HERAPDF2.0 FF3A (solid lines). The shaded bands on the predictions using the fitted PDF set show the theory uncertainties obtained by adding PDF, scale and beauty-quark mass uncertainties in quadrature.
 }
\label{fig:beauty-fitted-PDF} 
\end{figure}

The NLO FFNS predictions based on \pdfhq ~are compared to the combined charm and beauty cross sections in figures~\ref{fig:charm-fitted-PDF} and \ref{fig:beauty-fitted-PDF}, respectively. The predictions based on the HERAPDF2.0 set are included in the figures. Only minor differences between the different predictions can be observed.
This is to be expected because of the similarities of the PDFs, in particular that of the gluon and the values of the heavy-quark masses. The description of the data is similar to that observed for the predictions based on the HERAPDF2.0 FF3A set. 

\begin{figure}[h]
\center
\epsfig{file=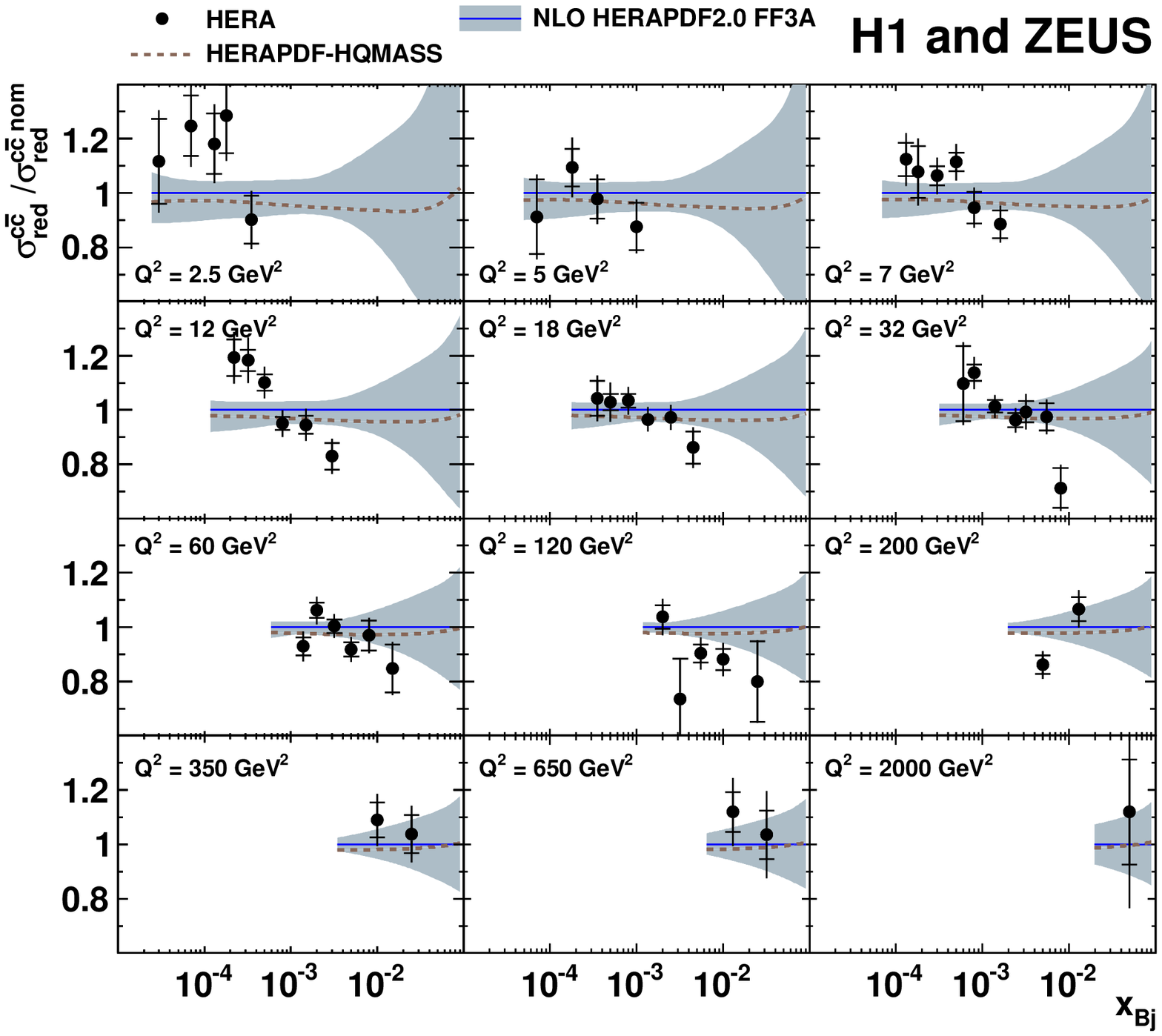 ,width=1.0\textwidth}
\caption{Ratio of reduced charm cross sections, \redcc, as a function of \xbj for given values of $Q^2$ for the combined data~(full circles) and the NLO FFNS predictions using \pdfhq~(dashed lines) with respect to the reference cross sections, $\sigma^{ c\overline{c}\ {\rm nom}}_{\rm red}$, based on HERAPDF2.0 FF3A (solid lines with uncertainty bands).
  }
\label{fig:ratio-charm-fittedPDF} 
\end{figure}

In figure~\ref{fig:ratio-charm-fittedPDF}, the ratios of data and predictions based \pdfhq ~to the predictions based on HERAPDF2.0 FF3A are shown for charm production. 
The description of the data is almost identical for both calculations.
The data show a steeper \xbj dependence than expected in NLO FFNS. 
The partial $\chi^2$ value of $116$ for the heavy-flavour data\footnote{It is not possible to quote the charm and the beauty contribution to this $\chi^2$ value separately because of the correlations between the combined charm and beauty measurements. 
} (d.o.f.~$=79$) in the fit presented is somewhat large. It corresponds to a $p$-value\footnote{The $\chi^2$ and the $p$-value given here do not correspond exactly to the statistical definition of $\chi^2$ or $p$-value because the data have been used in the fit to adjust theoretical uncertainties. Therefore the theory is somewhat shifted towards the measurements. However this bias is expected to be small because the predictions are mainly constrained by the much larger and more precise inclusive data sample.}  of $0.004$, which is equivalent to $2.9\sigma$. 
A similar behaviour can be observed already for the charm cross sections from the previous combination~\cite{HERAcharmcomb}, albeit at lower significance due to the larger uncertainties.

\begin{figure}[h]
\center
\epsfig{file=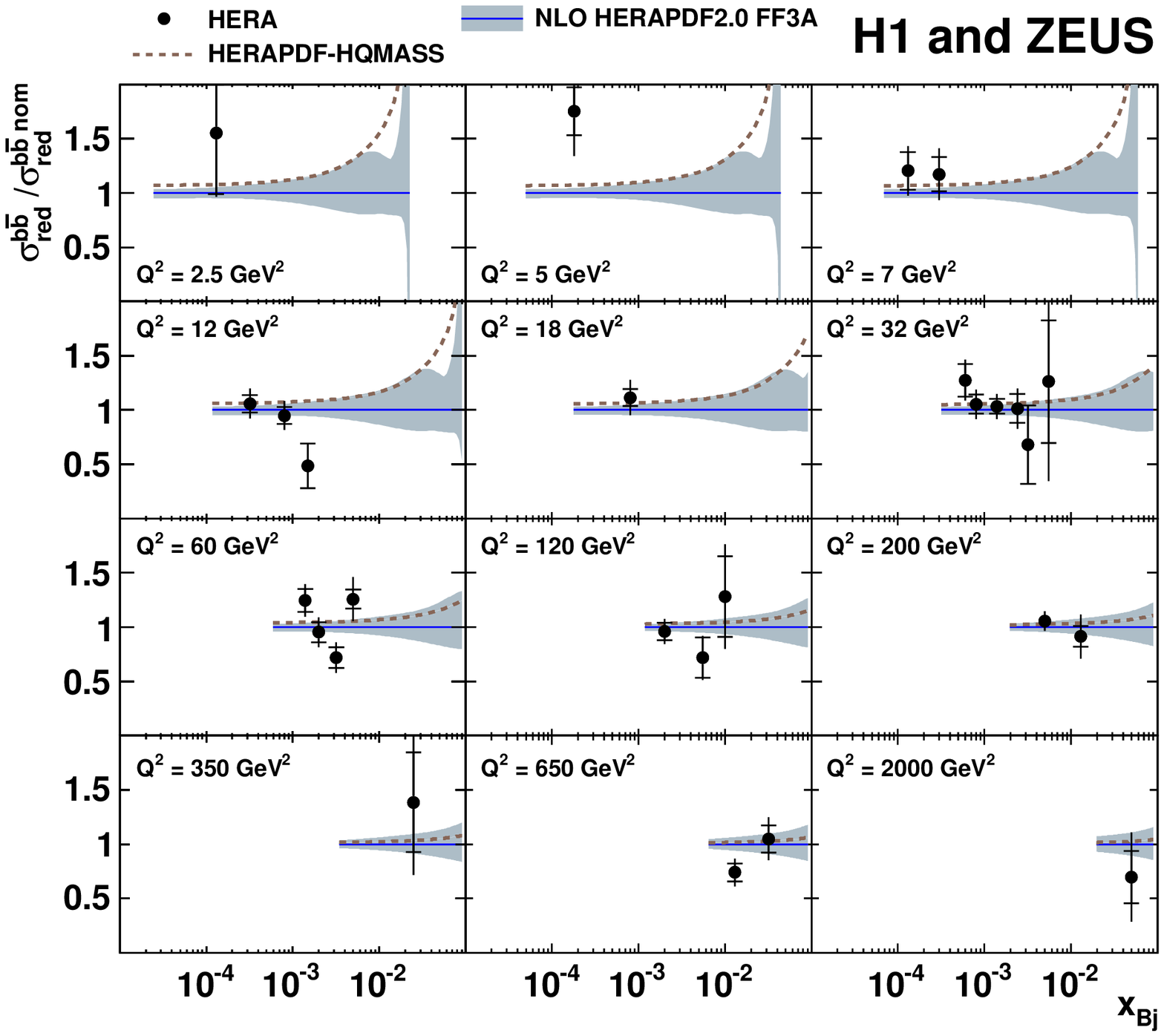 ,width=1.0\textwidth}
\caption{Ratio of reduced beauty cross sections, \redbb, as a function of \xbj for given values of $Q^2$ for the combined data~(full circles) and the NLO FFNS predictions using \pdfhq~(dashed lines) with respect to the reference cross sections, $\sigma^{ b\overline{b}\ {\rm nom}}_{\rm red}$, based on HERAPDF2.0 FF3A (solid lines with uncertainty bands).
 }
\label{fig:ratio-beauty-fittedPDF} 
\end{figure}

In figure~\ref{fig:ratio-beauty-fittedPDF}, the same ratios as in figure~\ref{fig:ratio-charm-fittedPDF} are shown for beauty production. Agreement is observed between theory and data within the large uncertainties of the measurements.

\subsection{Reduced heavy-flavour cross sections as a function of the partonic $\boldsymbol x$ }
Since in LO QCD heavy-flavour production proceeds via boson-gluon-fusion, at least two partons, the heavy-quark pair, are present in the final state. Therefore, already in LO, the $x$ of the incoming parton is different from \xbj measured at the photon vertex. 
At LO, the gluon $x$ is given by
\begin{equation}x=\xbj\cdot\left(1+\frac{\hat{s}}{Q^2}\right).\end{equation}
It depends on the kinematic DIS variables \xbj and $Q^2$ and on the invariant mass $\hat{s}$ of the heavy-quark pair.
At higher orders, the final state contains additional partons, such that $x$ cannot be expressed in a simple way. 
Independent of the order of the calculations,
only an average $\langle x\rangle$ can be determined at a given $(x_{\rm Bj},Q^2)$ point by the integration over all contributions to the cross section in the vicinity of this phase space point. In figure~\ref{fig:xgluon}, the ratio of the measured reduced cross sections to the NLO FFNS predictions based on \pdfhq ~is shown as a function of $\langle x\rangle$ instead of $x_{\rm Bj}$, where $\langle x\rangle$ is the geometric mean calculated at NLO with HVQDIS.
While the charm measurements cover the range $0.0005\lesssim\langle x\rangle\lesssim0.1$ the beauty data are limited to a higher $x$ range, $0.004\lesssim \langle x\rangle\lesssim0.1$, because of the large beauty-quark mass.  For the charm data, a deviation from the reference calculation is evident, showing a steeper slope in $\langle x\rangle$ in the range $0.0005\lesssim\langle x\rangle\lesssim 0.01$, consistent with being independent of $Q^2$. Due to the larger experimental uncertainties, no conclusion can be drawn for the beauty data.  

\subsection{Increasing the impact of the charm data on the gluon density}
\label{sect:xbjdep}

While inclusive DIS cross sections constrain the gluon density indirectly via scaling violations, and directly only through higher order corrections, heavy-flavour production probes the gluon directly already at leading order. 
Contributions to heavy-flavour production from light-flavour PDFs are small. 
For charm production they amount to five to eight per cent, varying only slightly with \xbj or $Q^2$~\cite{daum_fl}. Because of the high precision of \redcc~reached in this analysis, a study is performed to enhance the impact of the charm measurement on the gluon determination in the QCD fit.

\begin{figure}[h]
\center
\epsfig{file=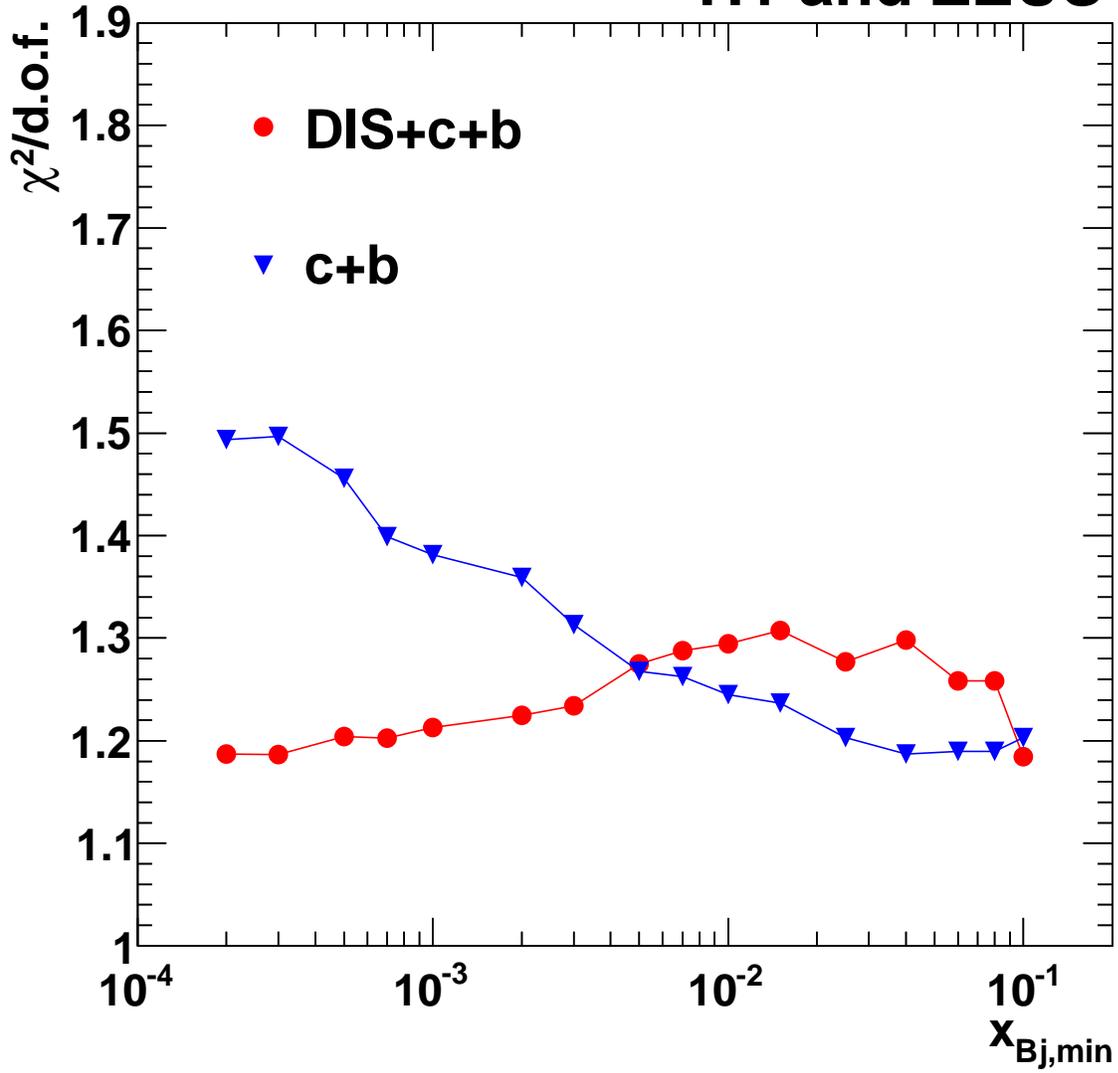 ,width=1.0\textwidth}
\caption{
The  values of $\chi^2$ per degree of freedom of the QCD fit to the inclusive and heavy-flavour data: (triangles) for the heavy-flavour data only and (dots) for the inclusive plus heavy-flavour data
when including in the fit only inclusive data with $x_{\rm Bj}\ge x_{\rm Bj,min}$.}
\label{fig:x-scan} 
\end{figure}

To reduce the impact of the inclusive data in the determination of the gluon density function, a series of fits is performed by requiring a minimum $\xbj\ge x_{\rm Bj,min}$ for the inclusive data included in the fit, with $x_{\rm Bj,min}$
varying from $2\cdot10^{-4}$ to $0.1$. No such cut is applied to the heavy-flavour data. The $\chi^2/{\rm d.o.f.}$ values for the inclusive plus heavy-flavour data and the partial $\chi^2/{\rm d.o.f.}$ for the heavy-flavour data only are presented in figure~\ref{fig:x-scan} as a function of $x_{{\rm Bj,min}}$. The partial $\chi^2/{\rm d.o.f.}$ for the heavy-flavour data improves significantly with rising $x_{{\rm Bj,min}}$ cut reaching a minimum at $x_{{\rm Bj,min}}\approx0.04$, while the $\chi^2/{\rm d.o.f.}$ for the inclusive plus heavy-flavour data sample is slightly larger than that obtained without a cut in $x_{\rm Bj}$.
For further studies $x_{{\rm Bj,min}}=0.01$ is chosen.
The total $\chi^2$ is $822$ for $651$ degrees of freedom. The partial $\chi^2$ of the heavy-flavour data improves to $98$ for $79$ degrees of freedom (corresponding to a $p$-value of $0.07$ or $1.8\sigma$). The resulting gluon density function, shown in figure~\ref{fig:fittedpdfs-xmin-scan} at the scale $\mu_{\rm f}^2=1.9$~GeV$^2$, is significantly steeper than the gluon density function determined when including all inclusive measurements in the fit. The other parton density functions are consistent with the result of the default fit. 

\begin{figure}[h]
\center
\epsfig{file=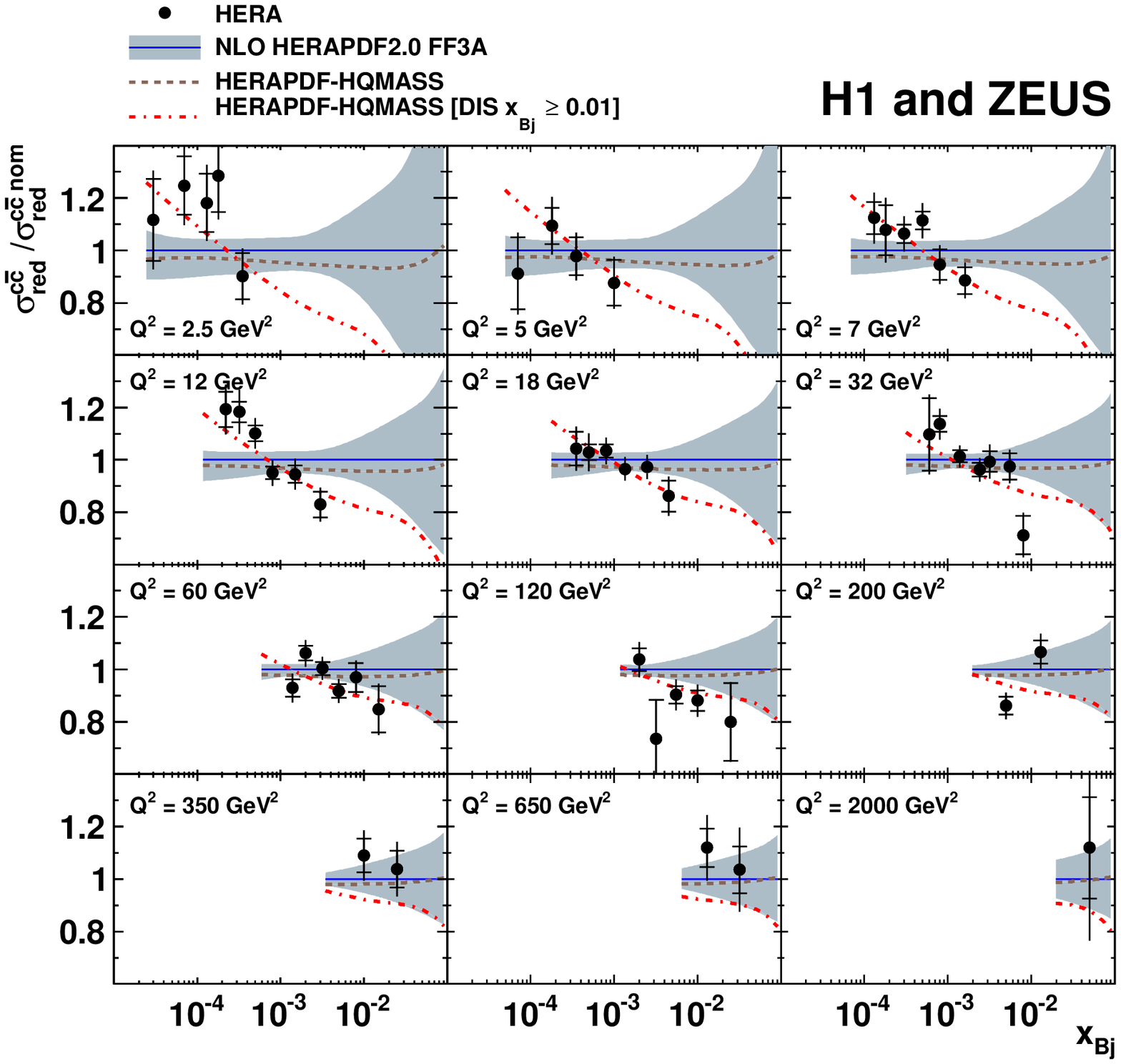 ,width=1.0\textwidth}
\caption{Ratio of combined reduced charm cross sections, \redcc,~(full circles) as a function of \xbj for given values of $Q^2$, 
  compared to the NLO FFNS predictions based on \pdfhq~(dashed lines) and those resulting from the alternative fit when requiring $x_{{\rm Bj}}\ge0.01$  for the inclusive data (dashed dotted lines), 
  with respect to the reference cross sections, $\sigma^{ c\overline{c}\ {\rm nom}}_{\rm red}$, based on HERAPDF2.0 FF3A (full line with uncertainty bands).}
\label{fig:ratio-charm-fittedPDF-xcut} 
\end{figure}

In figure~\ref{fig:ratio-charm-fittedPDF-xcut}, a comparison is presented of the ratios of the combined reduced charm cross section and the  cross section 
as calculated from the alternative fit, in which the inclusive data are subject to the cut $x_{\rm Bj}\ge0.01$,
to the reference cross sections based on HERAPDF2.0 FF3A. 
The predictions from \pdfhq ~are also shown.
As expected, the charm cross sections fitted
with the $x_{\rm Bj}$ cut imposed on the inclusive data rise more strongly towards small \xbj and describe the data  better than the other predictions. In general, the predictions from the fit with $x_{\rm Bj}$ cut follow nicely the charm data.  A similar study for beauty is also made but no significant \reading{improvement in the description of the beauty data is observed. The heavy-quark masses extracted from the fit with $x_{{\rm Bj}} \ge 0.01$ are consistent with those quoted in equation~(\ref{eq:finalmcmb}).}

\begin{figure}[h]
\center
\epsfig{file=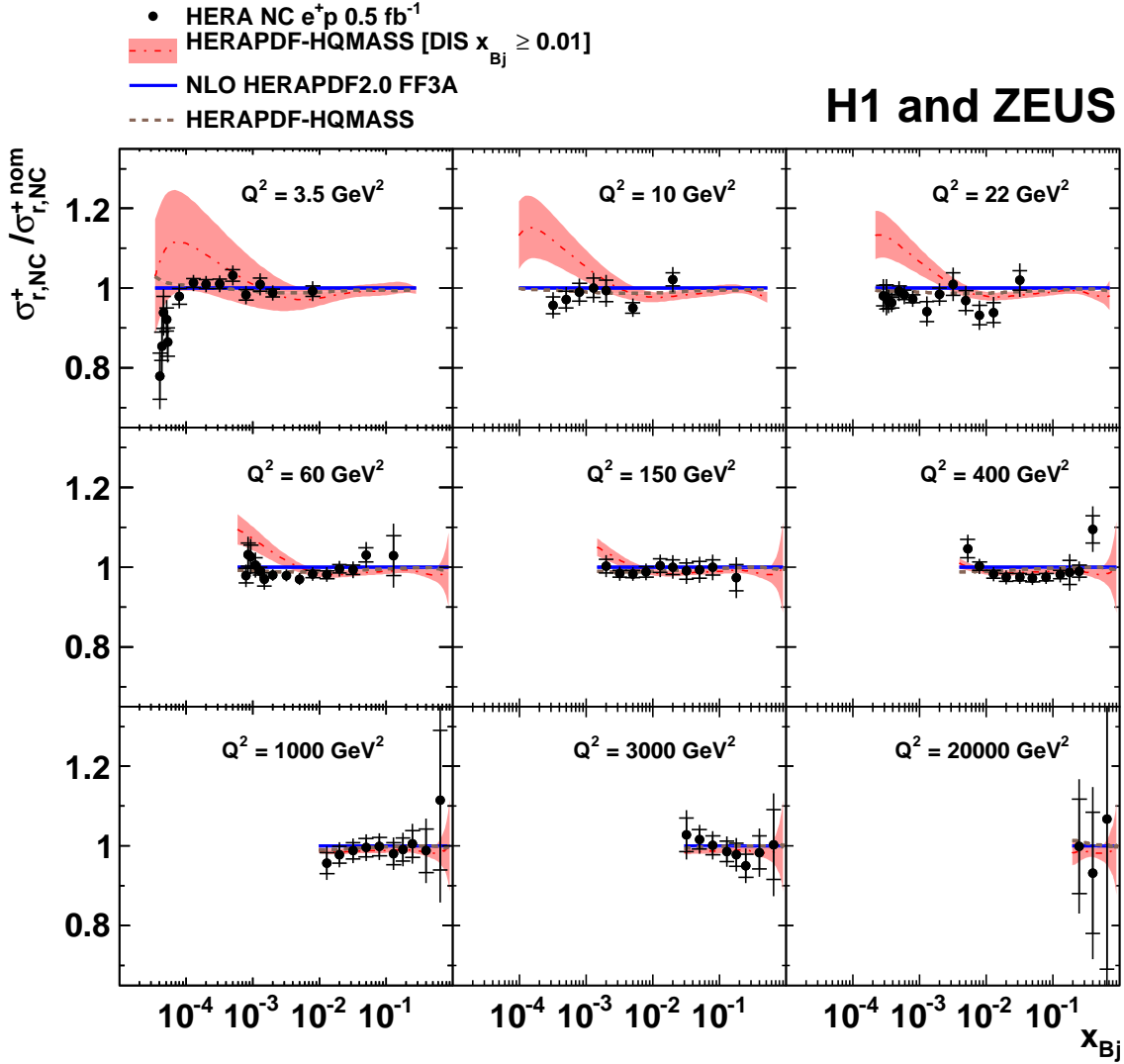 ,width=1.0\textwidth}
\caption{Ratio of combined reduced NC cross sections, $\sigma^{+}_{\rm r,NC}$,~(full circles) as a function of \xbj for selected values of $Q^2$, 
  compared to the NLO FFNS predictions based on  \pdfhq ~(dashed lines) and those resulting from the alternative fit with 
  $x_{{\rm Bj}}\ge0.01$ required for the inclusive data (dashed-dotted lines), 
   with respect to the reference cross sections, $\sigma^{+~{\rm nom}}_{\rm r,NC}$, based on HERAPDF2.0 FF3A (solid lines with uncertainty bands). 
}
\label{fig:ratio-inclusive-fittedPDF-xcut} 
\end{figure}

Cross-section predictions based on the three PDF sets, discussed in the previous paragraph, are calculated for inclusive DIS. 
In figure \ref{fig:ratio-inclusive-fittedPDF-xcut}, these predictions are compared to the inclusive reduced cross sections~\cite{HERAPDF20} for NC $e^+p$ DIS. 
The predictions based on HERAPDF2.0 FF3A and  on \pdfhq ~agree with the inclusive measurement. The calculations based on the PDF set determined by requiring $x_{\rm Bj}\ge 0.01$ for the inclusive data predict significantly larger inclusive reduced cross sections at
small $x_{\rm Bj}$. 

This study shows that a better description of the charm data can be achieved 
by excluding the low-$x_{\rm Bj}$ inclusive data in the fit. However, the calculations then fail to describe the inclusive data at low $x_{\rm Bj}$. 
In the theoretical framework used in this analysis, it seems impossible 
to resolve the $2.9\sigma$ difference in describing simultaneously the inclusive and charm measurements from HERA, using this simple approach of changing the gluon density. The comparison of various theory predictions to the charm data in section~\ref{sect:comp-thoery} suggests that the situation is unlikely to improve at NNLO because the NNLO predictions presented provide a poorer description of the charm data than that observed at NLO. 
The combined inclusive analysis~\cite{HERAPDF20} already revealed some 
tensions in the theory description of the inclusive DIS data.
The current analysis reveals some additional tensions in 
describing simultaneously the combined charm data and the combined inclusive data.

\section{Summary}
\label{sec:summary}

Measurements of charm and beauty production cross sections in deep inelastic $ep$ scattering by the H1 and ZEUS experiments 
are combined at the level of reduced cross sections, accounting for their statistical and systematic correlations. 
The beauty cross sections are combined for the first time.
The datasets are found to be consistent and the combined data have significantly reduced uncertainties.
The combined charm cross sections presented in this paper are significantly more precise than those previously published.

Next-to-leading and approximate next-to-next-to-leading-order QCD predictions of different schemes are compared to the data. 
The calculations are found to be 
in fair agreement with the charm data. 
The next-to-leading-order calculations in the fixed-flavour-number scheme provide the best description of the heavy-flavour data. 
The beauty data, which have larger experimental uncertainties, are well described by all QCD predictions. 

The new combined heavy-flavour data together with the previously published combined inclusive data from HERA are subjected to a next-to-leading-order QCD analysis in the fixed-flavour-number scheme using the ${\rm \overline{MS}}$ running-mass definition. The running heavy-quark masses are determined as:
\begin{eqnarray}
m_c(m_c) = 1.290^{+0.046}_{-0.041} {\rm (exp/fit)} {}^{+0.062}_{-0.014}  {\rm (model)} {}^{+0.003}_{-0.031} {\rm (parameterisation)}~{\rm GeV}, \nonumber\\
m_b(m_b) = 4.049^{+0.104}_{-0.109} {\rm (exp/fit)} {}^{+0.090}_{-0.032} {\rm (model)} {}^{+0.001}_{-0.031} {\rm (parameterisation)}~{\rm GeV}. 
\nonumber
\end{eqnarray}
The simultaneously determined parton density functions are found to agree well with HERAPDF2.0 FF3A.

The QCD analysis reveals some tensions, at the level of $3 \sigma$, in describing simultaneously the inclusive and the heavy-flavour HERA DIS data. 
The measured reduced charm cross sections show a stronger \xbj dependence than 
obtained in the combined QCD fit of charm and inclusive data, in which the PDFs are dominated by the fit of the inclusive data.
A study in which inclusive data with $x_{\rm Bj}<0.01$ are excluded from the fit is carried out. A better description of the charm data can be achieved this way. However, the resulting PDFs fail to describe
the inclusive data in the excluded \xbj region. 
Alternative next-to-leading-order and next-to-next-leading-order QCD calculations considered, including those with low-$x$ resummation, do not provide a better description of the combined heavy-flavour data.

\section*{Acknowledgements}

We are grateful to the HERA machine group whose outstanding
efforts have made these experiments possible.
We appreciate the contributions to the construction, maintenance and operation of the H1 and ZEUS detectors of many people who are not listed as authors.
We thank our funding agencies for financial 
support, the DESY technical staff for continuous assistance and the 
DESY directorate for their support and for the hospitality they 
extended to the non-DESY members of the collaborations. 
We would like to give credit to all partners contributing to the EGI computing infrastructure for their support.

\noindent
\begin{flushleft}

\end{flushleft}

\newpage
\begin{table}[t]
\begin{center}
\tabcolsep1.5mm
\renewcommand*{\arraystretch}{1.2}
	\begin{tabular}{|rl|l|rcr|r|r|r|r|}\hline 
		   \multicolumn{2}{|c|}{Dataset}               & Tagging 
		 & \multicolumn{3}{c|}{$Q^2$ range}       & \multicolumn{1}{c|}{${\cal L}$}& \multicolumn{1}{c|}{${\sqrt{s}}$} & $N_c$ & $N_b$ \\ 
		 &                                      &                 &
											   \multicolumn{3}{c|}{[GeV$^2$]}
											   & [pb$^{-1}$] & [GeV] &  & \\ \hline
	1&H1 VTX \cite{h1ltt_hera2}   &VTX&  $5$ &--& $2000$                & $245$ & $318$ & $29$ & $12$ \\
	2&H1 $D^{*\pm}$ HERA-I \cite{h1dstar_hera1}      &$D^{*+}$    &  $2$ &--& $100$                  & $47$& $318$ & $17$ & \\
	3&H1 $D^{*\pm}$ HERA-II (medium $Q^2$) \cite{h1dstar_hera2}     &$D^{*+} $     &  $5$ &--& $100$                  & $348$& $318$ &  $25$ & \\
	4&H1 $D^{*\pm}$ HERA-II (high $Q^2$) \cite{h1dstarhighQ2}     &$D^{*+}$     &  $100$ &--& $1000$            & $351$& $318$  &  $6$ &  \\
	5&ZEUS $D^{*+}$ 96-97 \cite{zd97}           &$D^{*+}$      & $1$ &--& $200$                  & $37$& $300$  & $21$ &  \\
	6&ZEUS $D^{*+}$ 98-00 \cite{zd00}           &$D^{*+}$      &  $1.5$ &--& $100$0            & $82$& $318$  & $31$ &  \\
	7&ZEUS $D^0$ 2005 \cite{zd0dp}                  &$D^{0}$ &$ 5$ &--& $1000$    & $134$& $318$ & $9$  &  \\
	8&ZEUS $\mu$ 2005 \cite{zmu} &$\mu$         & $20$ &--& $10000$          &  $126$& $318$ & $8$  & $8$  \\
	9&ZEUS $D^{+}$ HERA-II \cite{zeusdplus_hera2} &$D^+$         & $5$ &--& $1000$               & $354$& $318$ & $14$   & \\
	10&ZEUS $D^{*+}$ HERA-II \cite{zeusdstar_hera2}                  &$D^{*+}$         & $5$ &--& $1000$               & $363$& $318$ & $31$   & \\
	11&ZEUS VTX HERA-II \cite{zeusvtx}                  &VTX         & $5$ &--& $1000$               & $354$& $318$ & $18$   & $17$ \\
	12&ZEUS $e$ HERA-II \cite{zel} & $e$        & $10$ &--& $1000$               & $363$& $318$ &   &  $9$ \\
	13&ZEUS $\mu + {\rm jet}$ HERA-I \cite{zmu_hera1} & $\mu$      & $2$ &--& $3000$               & $114$& $318$ &   &  $11$ \\
	\hline
	\end{tabular}
\end{center}
\caption{Datasets used in the combination. 
For each dataset, the tagging method, the $Q^2$ range, integrated luminosity (${\cal L}$), centre-of-mass energy ($\sqrt{s}$) 
and the numbers of charm ($N_c$) and beauty ($N_b$) measurements are given. The tagging method VTX denotes inclusive measurements based on lifetime information using a silicon vertex detector. Charge conjugates are always implied for the particles  given in the column 'Tagging'.
} 
\label{tab:data}
\end{table}

\newpage
\begin{table}[t]
\renewcommand{\arraystretch}{1.3}
\begin{center}
\begin{tabular}{|c|c|c|c|r|r|r|r|}
\hline
         \#&$Q^2$ [GeV$^2$]&         \xbj&\redcc&        $\delta_{\rm stat} [\%]$&       $\delta_{\rm uncor} [\%]$&         $\delta_{\rm cor} [\%]$&         $\delta_{\rm tot} [\%]$\\
\hline
           1&         2.5&     0.00003&      0.1142&         8.9&        10.7&         9.4&        16.9\\
           2&         2.5&     0.00007&      0.1105&         5.8&         6.7&         8.2&        12.1\\
           3&         2.5&     0.00013&      0.0911&         7.1&         6.2&         7.9&        12.3\\
           4&         2.5&     0.00018&      0.0917&         4.8&         9.6&         7.2&        12.9\\
           5&         2.5&     0.00035&      0.0544&         5.3&         8.2&         6.9&        12.0\\
           6&         5.0&     0.00007&      0.1532&        11.6&         9.6&         8.2&        17.1\\
           7&         5.0&     0.00018&      0.1539&         5.3&         3.4&         7.8&        10.0\\
           8&         5.0&     0.00035&      0.1164&         5.2&         5.3&         5.7&         9.3\\
           9&         5.0&     0.00100&      0.0776&         4.8&         8.7&         5.6&        11.4\\
          10&         7.0&     0.00013&      0.2249&         4.3&         3.3&         6.7&         8.6\\
          11&         7.0&     0.00018&      0.2023&         6.8&         5.7&         7.2&        11.4\\
          12&         7.0&     0.00030&      0.1767&         2.3&         2.4&         5.4&         6.4\\
          13&         7.0&     0.00050&      0.1616&         2.5&         1.8&         5.2&         6.0\\
          14&         7.0&     0.00080&      0.1199&         4.6&         4.0&         4.9&         7.8\\
          15&         7.0&     0.00160&      0.0902&         4.1&         3.9&         5.2&         7.7\\
          16&        12.0&     0.00022&      0.3161&         4.9&         2.9&         5.7&         8.0\\
          17&        12.0&     0.00032&      0.2904&         2.9&         1.5&         6.3&         7.1\\
          18&        12.0&     0.00050&      0.2410&         2.4&         1.3&         4.6&         5.3\\
          19&        12.0&     0.00080&      0.1813&         2.1&         1.4&         4.5&         5.1\\
          20&        12.0&     0.00150&      0.1476&         3.2&         1.5&         5.1&         6.2\\
          21&        12.0&     0.00300&      0.1010&         4.4&         4.0&         5.1&         7.8\\
          22&        18.0&     0.00035&      0.3198&         5.2&         3.3&         5.2&         8.1\\
          23&        18.0&     0.00050&      0.2905&         2.6&         1.4&         6.4&         7.0\\
          24&        18.0&     0.00080&      0.2554&         2.2&         1.2&         4.2&         4.9\\
          25&        18.0&     0.00135&      0.2016&         2.0&         1.1&         4.1&         4.7\\
          26&        18.0&     0.00250&      0.1630&         1.9&         1.3&         4.2&         4.7\\
          27&        18.0&     0.00450&      0.1137&         5.5&         4.1&         5.4&         8.7\\
          \hline
          \end{tabular}
          \end{center}
          \caption{Reduced cross section for charm production, \redcc, obtained by the combination of H1 and ZEUS measurements. The cross-section values are given together with the statistical $(\delta_{\rm stat})$ and the uncorrelated $(\delta_{\rm uncor})$ and correlated $(\delta_{\rm cor})$ systematic  uncertainties. The total uncertainties $(\delta_{\rm tot})$ are obtained through adding the statistical, uncorrelated and correlated systematic uncertainties in quadrature.  
  }
          \label{tab:sigcc}

\end{table}   
\newpage
\begin{table}[t]
\renewcommand{\arraystretch}{1.3}
\begin{center}
\begin{tabular}{|c|c|c|c|r|r|r|r|}
\hline
         \#&$Q^2$ [GeV$^2$]&         \xbj&\redcc&        $\delta_{\rm stat} [\%]$&       $\delta_{\rm uncor} [\%]$&         $\delta_{\rm cor} [\%]$&         $\delta_{\rm tot} [\%]$\\
\hline
       
          28&        32.0&     0.00060&      0.3885&         8.5&         9.3&         5.8&        13.9\\
          29&        32.0&     0.00080&      0.3756&         2.3&         1.4&         4.4&         5.2\\
          30&        32.0&     0.00140&      0.2807&         2.0&         1.1&         3.4&         4.1\\
          31&        32.0&     0.00240&      0.2190&         2.3&         1.4&         3.9&         4.7\\
          32&        32.0&     0.00320&      0.2015&         3.6&         1.6&         5.4&         6.6\\
          33&        32.0&     0.00550&      0.1553&         4.2&         3.0&         4.1&         6.6\\
          34&        32.0&     0.00800&      0.0940&         8.7&         5.4&         6.0&        11.9\\
          35&        60.0&     0.00140&      0.3254&         3.2&         1.4&         4.8&         5.9\\
          36&        60.0&     0.00200&      0.3289&         2.3&         1.2&         4.1&         4.9\\
          37&        60.0&     0.00320&      0.2576&         2.2&         1.2&         3.6&         4.4\\
          38&        60.0&     0.00500&      0.1925&         2.3&         1.6&         4.1&         5.0\\
          39&        60.0&     0.00800&      0.1596&         4.8&         3.1&         3.4&         6.7\\
          40&        60.0&     0.01500&      0.0946&         8.1&         6.5&         4.9&        11.5\\
          41&       120.0&     0.00200&      0.3766&         3.3&         2.6&         5.0&         6.5\\
          42&       120.0&     0.00320&      0.2274&        14.6&        13.7&         2.7&        20.2\\
          43&       120.0&     0.00550&      0.2173&         3.3&         1.6&         5.4&         6.5\\
          44&       120.0&     0.01000&      0.1519&         3.9&         2.3&         5.2&         6.9\\
          45&       120.0&     0.02500&      0.0702&        13.6&        12.6&         4.4&        19.1\\
          46&       200.0&     0.00500&      0.2389&         3.1&         2.4&         4.5&         6.0\\
          47&       200.0&     0.01300&      0.1704&         3.4&         2.3&         5.0&         6.5\\
          48&       350.0&     0.01000&      0.2230&         5.1&         3.0&         6.4&         8.7\\
          49&       350.0&     0.02500&      0.1065&         6.1&         2.9&         7.4&        10.0\\
          50&       650.0&     0.01300&      0.2026&         5.4&         3.7&         9.1&        11.2\\
          51&       650.0&     0.03200&      0.0885&         7.8&         3.8&        12.8&        15.4\\
          52&      2000.0&     0.05000&      0.0603&        16.0&         6.7&        26.4&        31.6\\          \hline
\end{tabular}      
 \end{center}
 \captcont{continued}
 \end{table}

\newpage
\begin{table}[t]
\renewcommand{\arraystretch}{1.3}
\begin{center}
\begin{tabular}{|c|c|c|c|r|r|r|r|}
\hline
          \#&$Q^2$ [GeV$^2$]&         \xbj&\redbb&        $\delta_{\rm stat} [\%]$&       $\delta_{\rm uncor}$ [\%]&         $\delta_{\rm cor}$ [\%]&         $\delta_{\rm tot} [\%]$\\
\hline
           1&         2.5&     0.00013&      0.0018&        28.4&        22.4&        11.4&        37.9\\
           2&         5.0&     0.00018&      0.0048&        10.5&         7.1&        19.8&        23.5\\
           3&         7.0&     0.00013&      0.0059&         8.8&        11.2&        12.7&        19.1\\
           4&         7.0&     0.00030&      0.0040&         8.5&        10.3&        15.2&        20.2\\
           5&        12.0&     0.00032&      0.0072&         4.9&         5.8&        10.5&        13.0\\
           6&        12.0&     0.00080&      0.0041&         4.6&         6.9&        11.1&        13.9\\
           7&        12.0&     0.00150&      0.0014&        32.2&        26.9&         3.6&        42.1\\
           8&        18.0&     0.00080&      0.0082&         4.8&         5.0&        12.8&        14.5\\
           9&        32.0&     0.00060&      0.0207&         8.9&         7.8&         8.9&        14.8\\
          10&        32.0&     0.00080&      0.0152&         5.8&         6.1&        10.0&        13.1\\
          11&        32.0&     0.00140&      0.0113&         3.9&         5.3&         9.0&        11.2\\
          12&        32.0&     0.00240&      0.0082&         9.0&         9.5&        12.9&        18.4\\
          13&        32.0&     0.00320&      0.0046&        32.2&        41.9&         3.0&        52.9\\
          14&        32.0&     0.00550&      0.0058&        39.8&        20.4&        57.4&        72.8\\
          15&        60.0&     0.00140&      0.0260&         4.8&         6.9&         8.8&        12.2\\
          16&        60.0&     0.00200&      0.0167&         7.5&         6.5&        10.5&        14.4\\
          17&        60.0&     0.00320&      0.0097&        10.7&         7.7&        14.4&        19.5\\
          18&        60.0&     0.00500&      0.0129&         5.4&         4.2&        14.7&        16.2\\
          19&       120.0&     0.00200&      0.0288&         6.3&         5.4&         9.0&        12.2\\
          20&       120.0&     0.00550&      0.0127&        21.2&        14.9&        10.9&        28.1\\
          21&       120.0&     0.01000&      0.0149&        20.5&        20.6&        23.6&        37.5\\
          22&       200.0&     0.00500&      0.0274&         3.8&         3.7&         6.9&         8.7\\
          23&       200.0&     0.01300&      0.0123&         9.5&         4.8&        19.5&        22.2\\
          24&       350.0&     0.02500&      0.0138&        20.4&        26.2&        35.0&        48.2\\
          25&       650.0&     0.01300&      0.0164&         8.1&         7.5&        13.1&        17.1\\
          26&       650.0&     0.03200&      0.0103&         8.1&         8.7&        14.6&        18.8\\
          27&      2000.0&     0.05000&      0.0052&        30.6&        15.2&        47.6&        58.6\\
          \hline
          \end{tabular}
          \end{center}
          \caption{Reduced cross section for beauty production, \redbb, obtained by the combination of H1 and ZEUS measurements. The cross-section values are given together with the statistical $(\delta_{\rm stat})$ and the uncorrelated $(\delta_{\rm uncor})$ and correlated $(\delta_{\rm cor})$ systematic  uncertainties. The total uncertainties $(\delta_{\rm tot})$ are obtained through adding the statistical, uncorrelated and correlated systematic uncertainties in quadrature. }
          \label{tab:sigbb}

          \end{table}

\newpage
\begin{table}
\renewcommand*{\arraystretch}{1.2}
\begin{center}
 \begin{tabular}{|l|l|c|}
\hline
Dataset & PDF (scheme) & $\chi^2$ $[p$-value$]$ 
\\\hline
\hline
\multirow{4}{*}{charm~\cite{HERAcharmcomb}} & HERAPDF20\_NLO\_FF3A (FFNS) & $59$ $[0.23]$ 
\\
& {\color{black}ABKM09} (FFNS) & {\color{black}$59$ $[0.23]$} \\
& ABMP16\_3\_nlo (FFNS) & $61$ $[0.18]$ 
\\
& ABMP16\_3\_nnlo (FFNS) & $70$ $[0.05]$ 
\\
& HERAPDF20\_NLO\_EIG (RTOPT) & $71$ $[0.04]$ 
\\
(${\rm N}_{\rm data}$ = 52)  \multirow{1}{*}{} & HERAPDF20\_NNLO\_EIG (RTOPT) & $66$ $[0.09]$ 
\\
\hdashline
  \multirow{1}{*}{} & NNPDF31sx NNLO (FONLL-C) & $106$ $[1.5\cdot 10^{-6}]$ \\
(${\rm N}_{\rm data}$ = 47)  \multirow{1}{*}{} & NNPDF31sx NNLO+NLLX (FONLL-C) & $71$ $[0.013]$ \\
\hline
\multirow{5}{*}{charm, } & HERAPDF20\_NLO\_FF3A (FFNS) & $86$ $[0.002]$ 
\\
& {\color{black}ABKM09} (FFNS) & {\color{black}$82$ $[0.005]$} \\
& ABMP16\_3\_nlo (FFNS) & $90$ $[0.0008]$ 
\\
this analysis& ABMP16\_3\_nnlo (FFNS) & ${109}$ ${[6\cdot 10^{-6}]}$ 
\\
& HERAPDF20\_NLO\_EIG (RTOPT) & ${99}$ ${[9\cdot 10^{-5}]}$ 
\\
(${\rm N}_{\rm data}$ = 52) \multirow{1}{*}{} & HERAPDF20\_NNLO\_EIG (RTOPT) & ${102}$ ${[4\cdot 10^{-5}]}$ 
\\
\hdashline
  \multirow{1}{*}{} & NNPDF31sx NNLO (FONLL-C) & $140$ $[1.5\cdot 10^{-11}]$ \\
(${\rm N}_{\rm data}$ = 47)  \multirow{1}{*}{} & NNPDF31sx NNLO+NLLX (FONLL-C) & $114$ $[5\cdot 10^{-7}]$ \\
\hline
\hline
  & HERAPDF20\_NLO\_FF3A (FFNS) & $33 [0.20]$ 
\\
\multirow{1}{*}{beauty,}& ABMP16\_3\_nlo (FFNS) & $37$ $[0.10]$ 
\\
 this analysis & ABMP16\_3\_nnlo (FFNS) & {$41$ $[0.04]$} 
\\
& HERAPDF20\_NLO\_EIG (RTOPT) & $33$ $[0.20]$ 
\\
(${\rm N}_{\rm data}$ = 27)& HERAPDF20\_NNLO\_EIG (RTOPT) & $45$ $[0.016]$ 
\\
\hline
\end{tabular}
\end{center}
\caption{The $\chi^2$, $p$-values and number of data points of the charm and beauty data with respect to the NLO and
approximate NNLO calculations using various PDFs as described in the text. The measurements at $Q^2=2.5$~GeV$^2$ are excluded in the calculations of the $\chi^2$ values for the NNPDF3.1sx predictions, by which the number of data points is reduced to $47$, as detailed in the caption of figure~\ref{fig:data-charm-theory-nnpdf-ratio}.
}
\label{tab:qcd-chi2pdf}
\end{table}

\clearpage
\begin{table}[!hp]
  \renewcommand{\arraystretch}{1.8}
  \centering
  \begin{tabular}{|l|c|c|c|}
  	\hline
  	Parameter               &         Variation          &                     $m_c(m_c)$ uncertainty & $m_b(m_b)$ uncertainty \\
  	                        &                            & [GeV]  & [GeV] \\ \hline
  	\multicolumn{4}{|c|}{Experimental$~$/$~$Fit uncertainty}                      \\ \hline
  	Total                   &      $\Delta\chi^2=1$      &              $^{+0.046}_{-0.041}$ & $^{+0.104}_{-0.109}$ \\ \hline
  	\multicolumn{4}{|c|}{Model uncertainty}                \\ \hline
  	$f_s$                   & $0.4^{+0.1}_{-0.1}$     &              $^{-0.003}_{+0.004}$ & $^{-0.001}_{+0.001}$ \\
  	$Q^2_{\text{min}}$      & $3.5^{+1.5}_{-1.0}$ GeV$^2$ &        $^{-0.001}_{+0.007}$ & $^{-0.005}_{+0.007}$ \\
  	$\mu_{r,f}$      & ${\mu_{r,f}}^{\times 2.0}_{\times 0.5}$ &        $^{+0.030}_{+0.060}$ & $^{-0.032}_{+0.090}$ \\
  	$\alpha_s^{n_f=3}(M_Z)$      & $0.1060^{+0.0015}_{-0.0015}$ &        $^{-0.014}_{+0.011}$ & $^{+0.002}_{-0.005}$ \\
  	\hline
    Total                   &                            &              $^{+0.062}_{-0.014}$ & $^{+0.090}_{-0.032}$ \\ \hline
    \multicolumn{4}{|c|}{PDF parameterisation uncertainty}  \\ \hline
  	$\mu_{\rm f,0}^2$                 & $1.9 \pm 0.3$ GeV$^2$ &        $^{+0.003}_{-0.001}$ & $^{-0.001}_{+0.001}$ \\
  	$E_{u_v}$               &        set to $0$         &           {\footnotesize $-0.031$} & {\footnotesize $-0.031$} \\
  	\hline
    Total                   &                            &              $^{+0.003}_{-0.031}$ & $^{+0.001}_{-0.031}$ \\ \hline
  \end{tabular}
  \caption{List of uncertainties for the charm- and beauty-quark mass determination. The PDF parameterisation uncertainties not shown have no effect on $m_c(m_c)$ and $m_b(m_b)$.}
  \label{tab:mcmb}
\end{table}

\newpage

\newpage

\newpage

\newpage

\newpage

\newpage

\newpage

\newpage

\newpage

\newpage

\newpage

\newpage

\newpage
\setlength{\unitlength}{1cm}
\center
\begin{figure}[h]
\begin{picture}(15.,17.)
\put(0,8){\epsfig{file=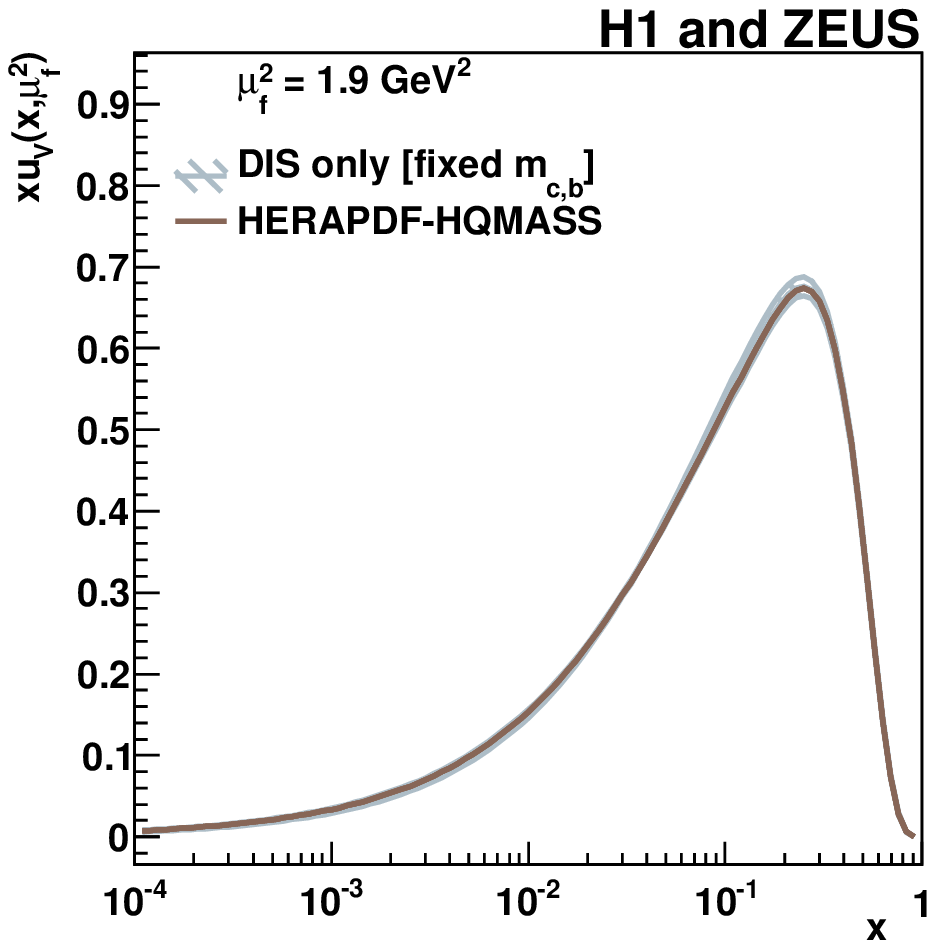,width=0.5\textwidth}}
\put(6.5,14.5){\large(a)}
\put(8,8){\epsfig{file=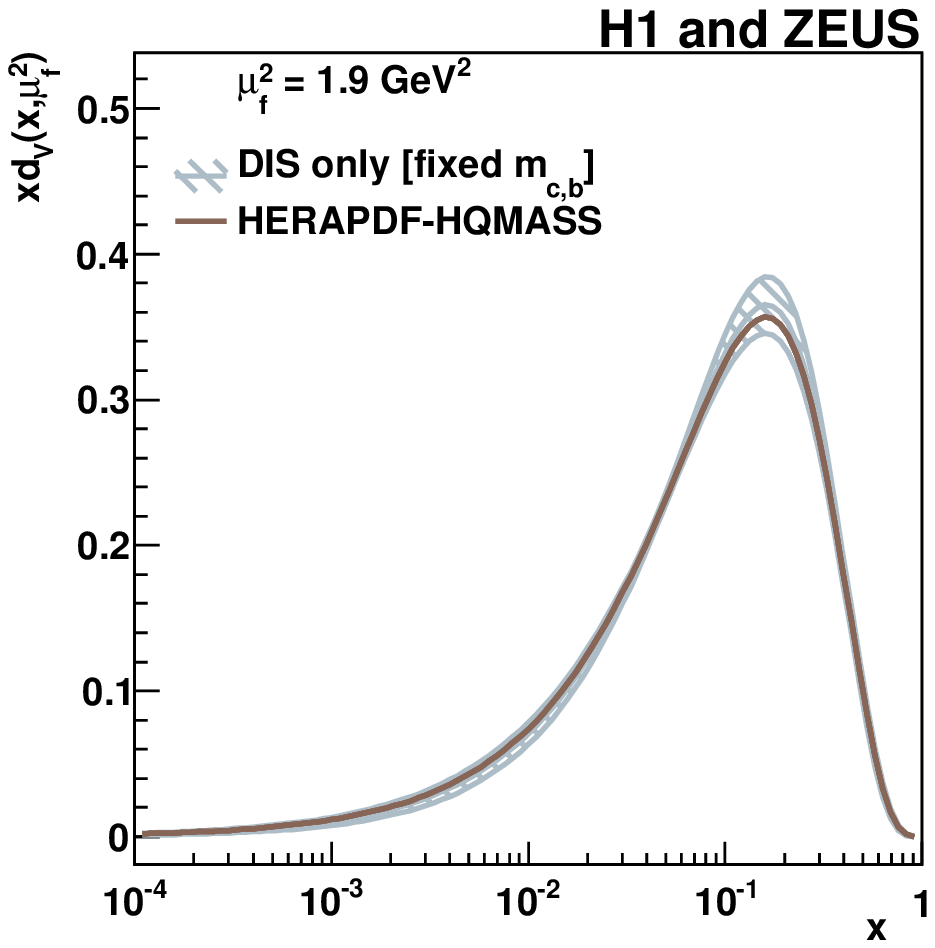,width=0.5\textwidth}}
\put(14.5,14.5){\large(b)}
\put(0,0){\epsfig{file=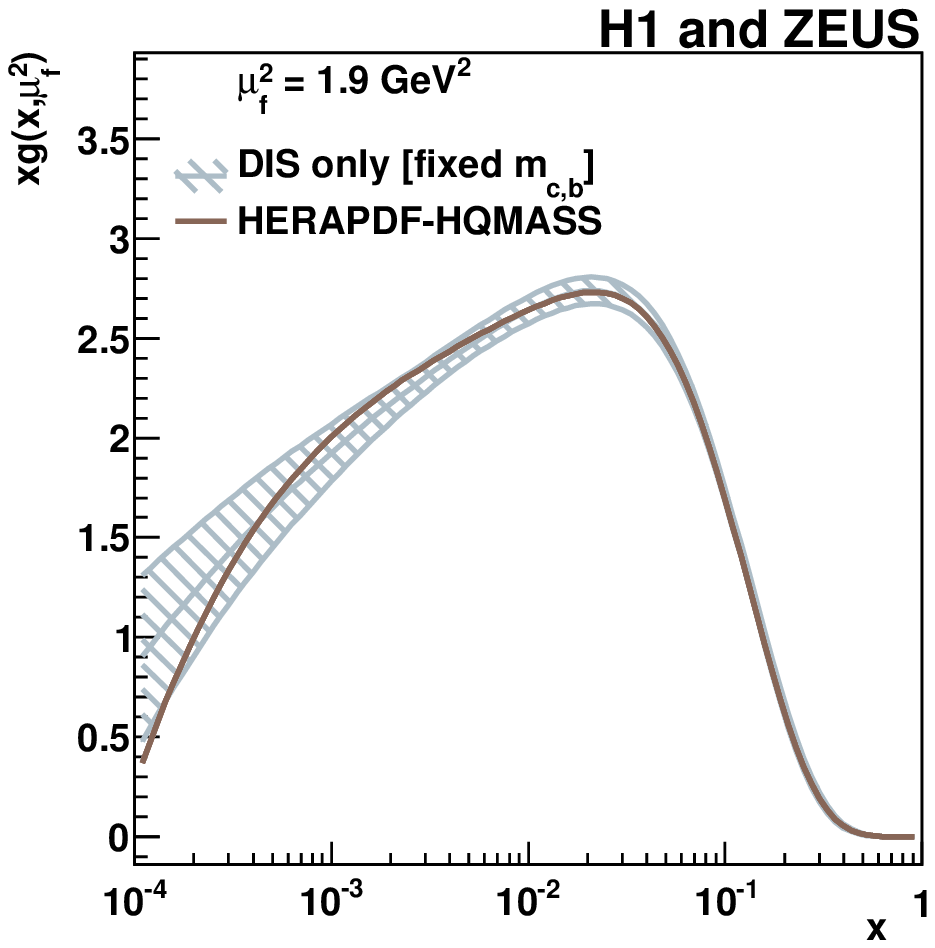,width=0.5\textwidth}}
\put(6.5,6.5){\large(c)}
\put(8,0){\epsfig{file=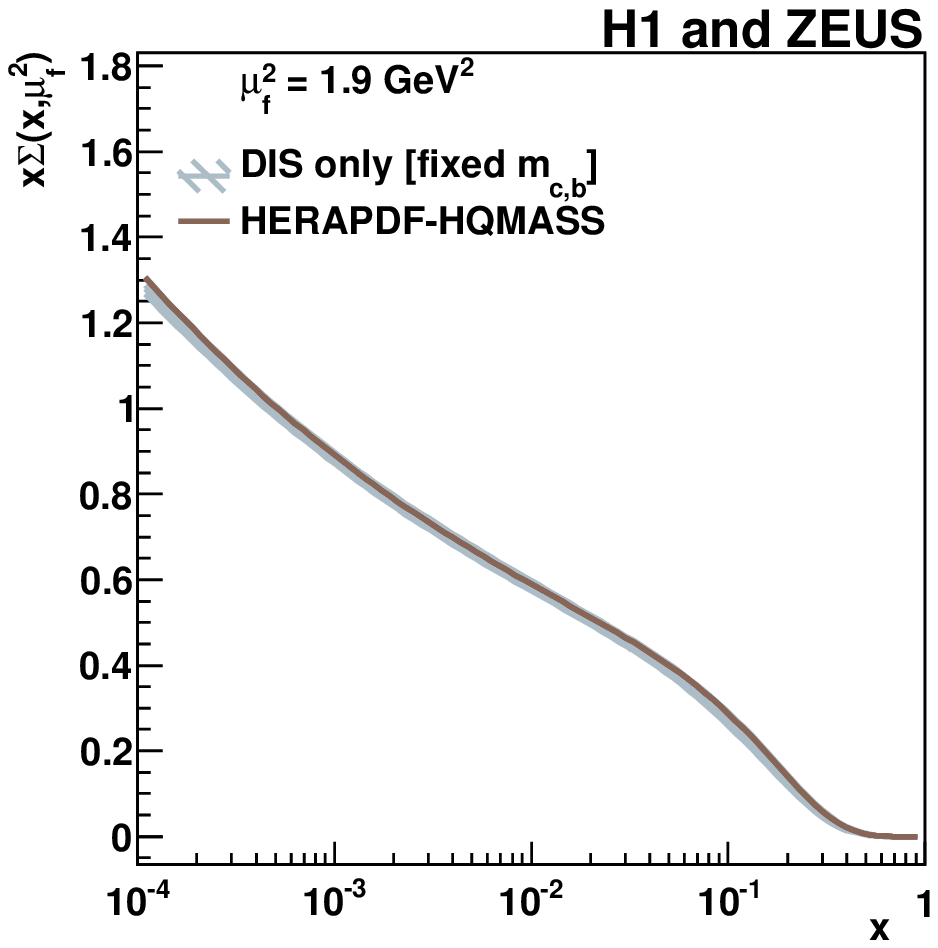,width=0.5\textwidth}}
\put(14.5,6.5){\large(d)}
\end{picture}
\caption{
Parton density functions $xf(x,Q^2)$ at the starting scale $\mu^2_{\rm f} = \mu^2_{\rm f,0}=1.9$~GeV$^2$ with
$f=u_v,d_v,g,\Sigma$
 for the valence up quark (a), the valence down quark (b), the gluon (c) and the sea quarks (d)
 of \pdfhq~(solid dark lines) and obtained from  fit to the combined inclusive data only (light grey lines).
 The experimental/fit uncertainties obtained from the fit to inclusive data only are indicated by the hatched bands. For better visibility the uncertainties for HERAPDF-HQMASS, which are of similar size, are not shown.
}
\label{fig:fittedpdfs} 
\end{figure}
\newpage

\newpage

\newpage

\newpage

\newpage
\setlength{\unitlength}{1cm}
\begin{figure}[h]
\center
\begin{picture}(15.,20.)
\put(01,9){\epsfig{file=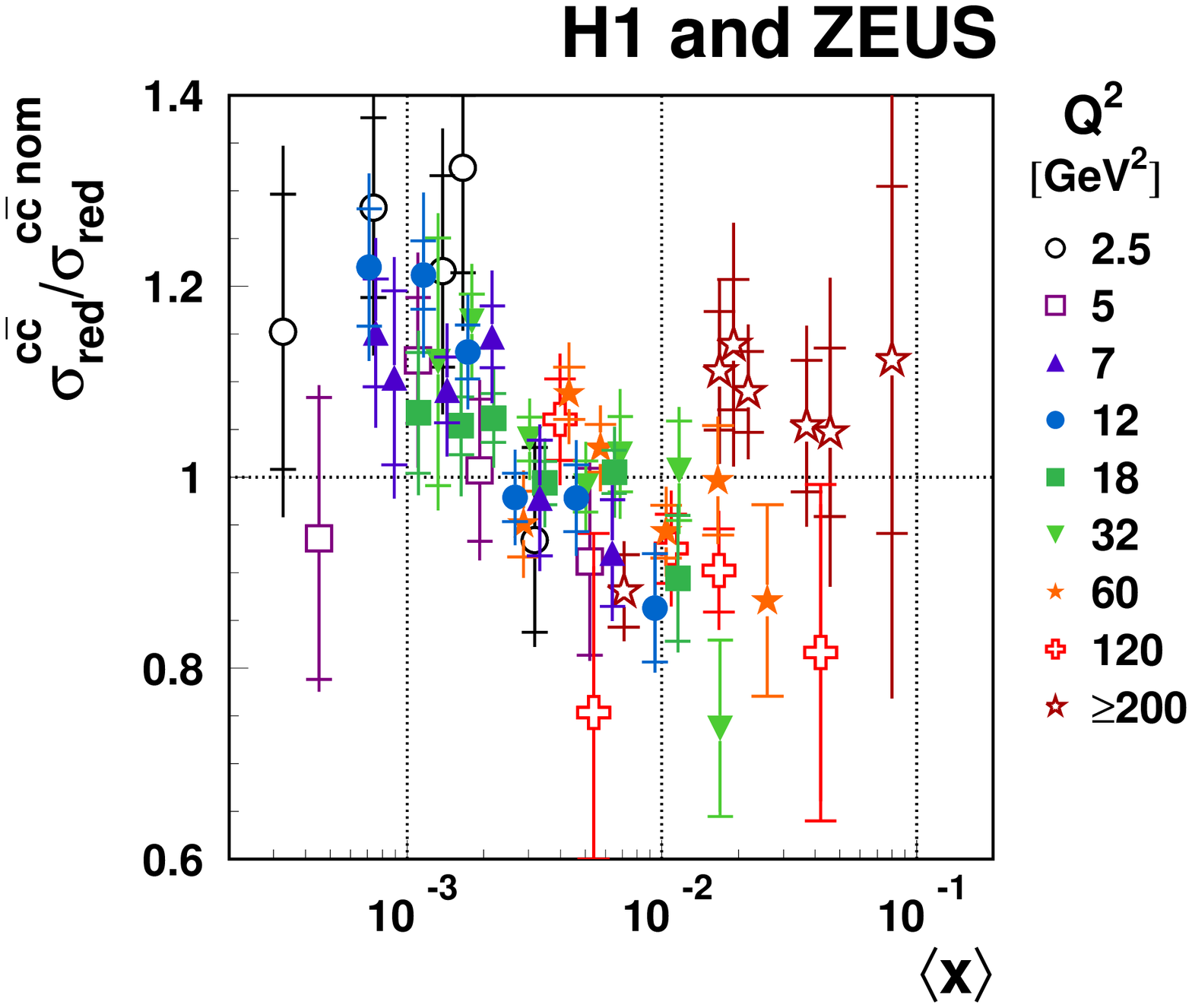,width=0.75\textwidth}}
\put(01,-1){\epsfig{file=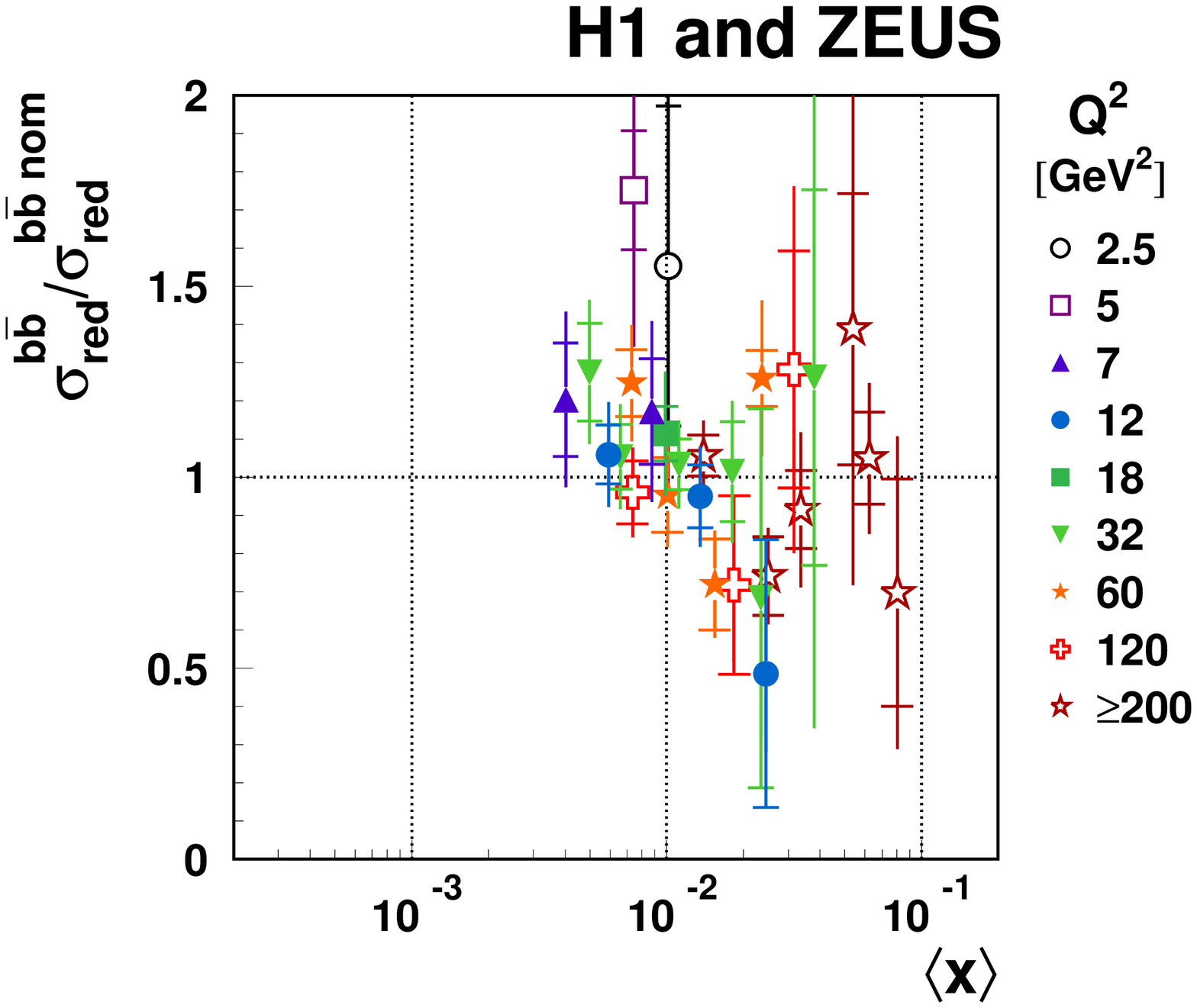,width=0.75\textwidth}}
\put(4,2.2){\Large (b)}
\put(4,12.2){\Large (a)}
\end{picture}
\caption{Ratio of the combined reduced cross sections, (a)~\redcc and (b)~\redbb, to the respective NLO FFNS cross-section predictions,
$\sigma_{\rm red}^{\rm nom}$, based on \pdfhq, as a function of the partonic $\langle x\rangle$ for different values of $Q^2$.
 }
\label{fig:xgluon} 
\end{figure}

\newpage
\newpage
\setlength{\unitlength}{1cm}
\center
\begin{figure}[h]
\begin{picture}(15.,17.)
\put(0,8){\epsfig{file=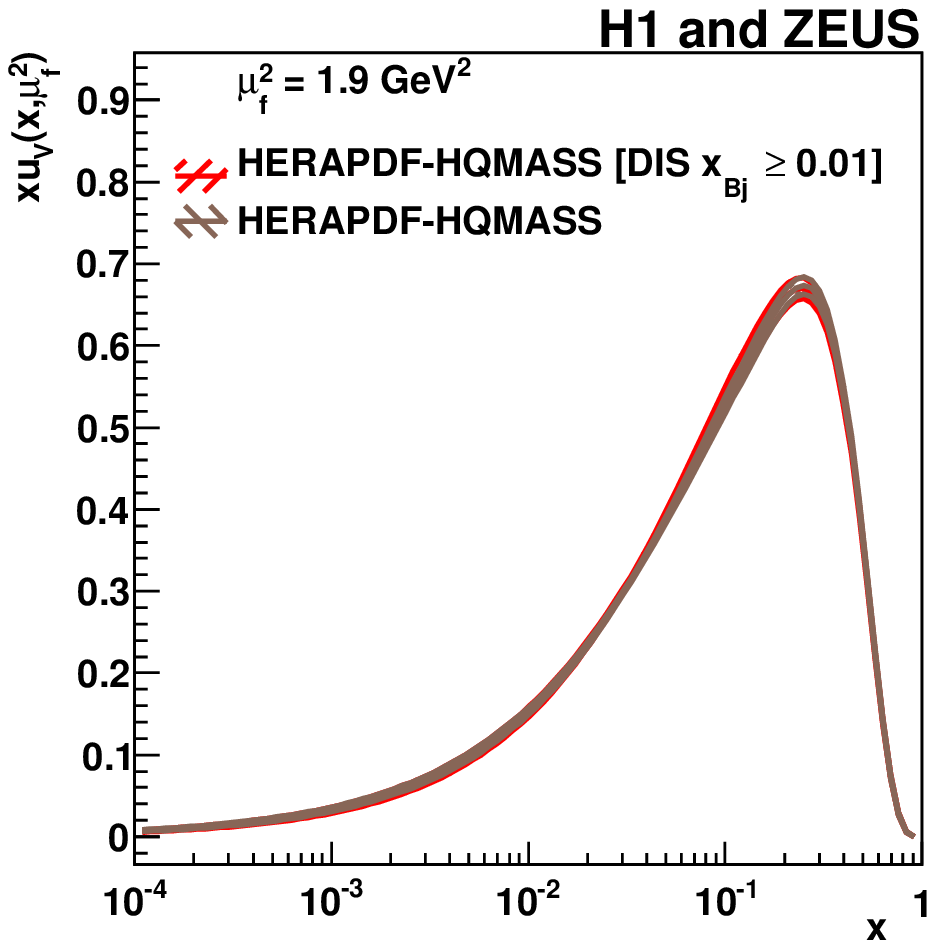,width=0.5\textwidth}}
\put(6.5,14.92){\large(a)}
\put(8,8){\epsfig{file=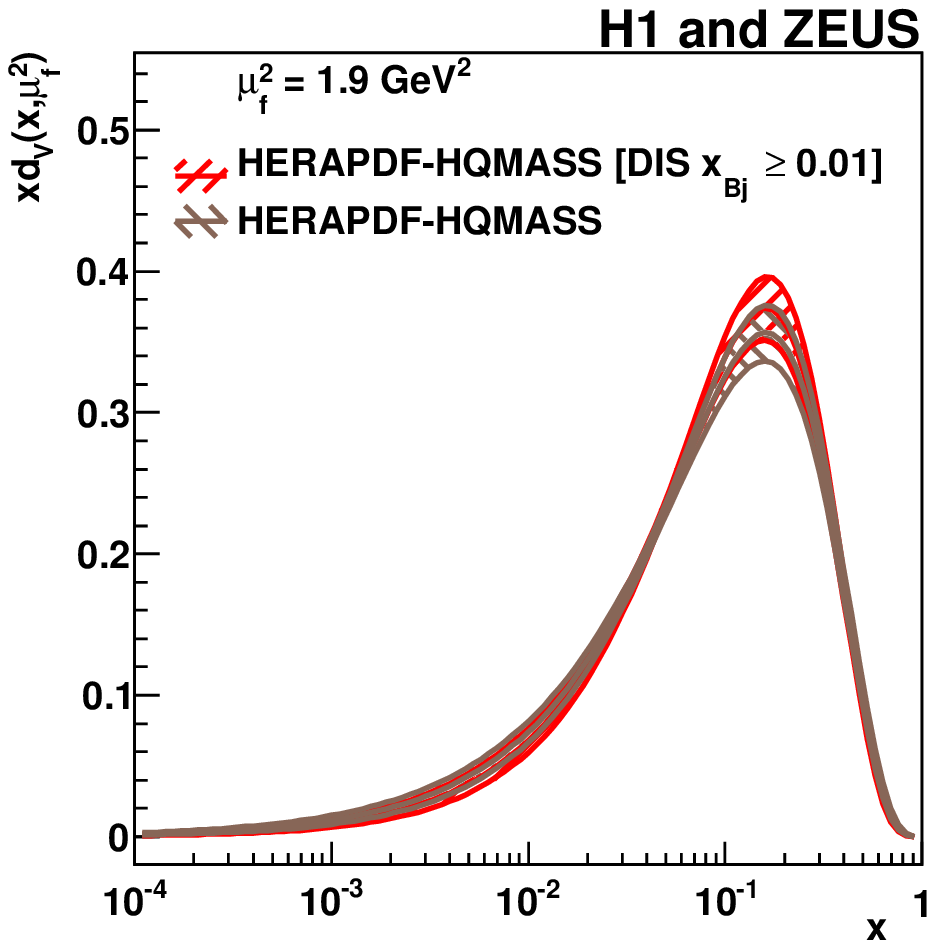,width=0.5\textwidth}}
\put(14.5,14.92){\large(b)}
\put(0,0){\epsfig{file=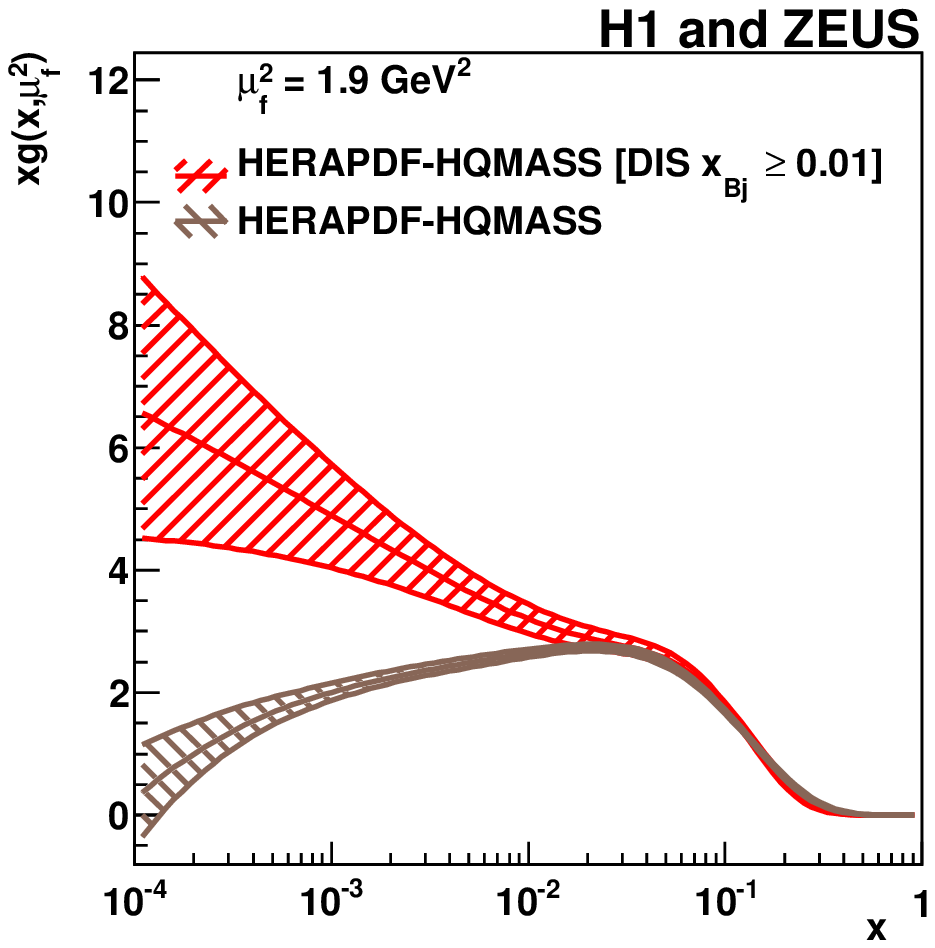,width=0.5\textwidth}}
\put(6.5,6.92){\large(c)}
\put(8,0){\epsfig{file=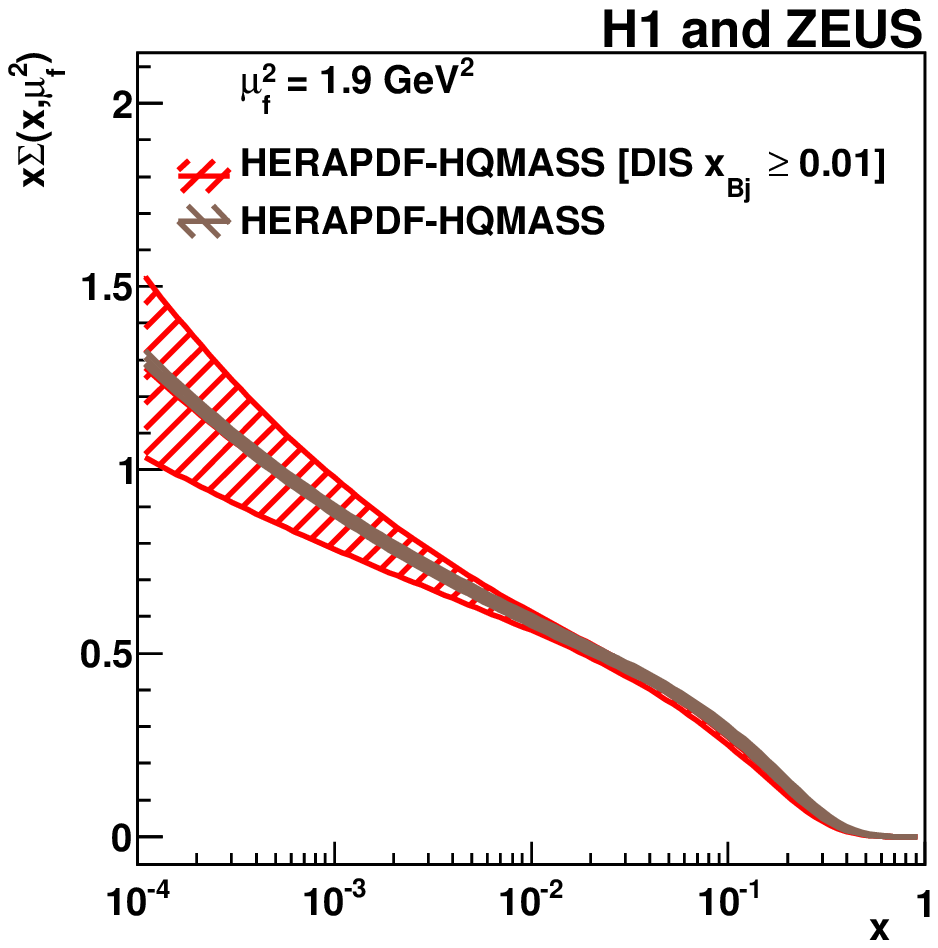,width=0.5\textwidth}}
\put(14.5,6.92){\large(d)}
\end{picture}
\caption{
Parton density functions $xf(x,Q^2)$ at the starting scale $\mu^2_{\rm f} = \mu^2_{\rm f,0}=1.9$~GeV$^2$ with
$f=u_v,d_v,g,\Sigma$
 for the valence up quark (a), the valence down quark (b), the gluon (c) and the sea quarks (d) of \pdfhq~(full lines) and
 obtained from the QCD fit to the combined inclusive and heavy-flavour data with imposing a minimum cut of $x_{\rm Bj}\ge0.01$ to the inclusive data included in the fit.
The experimental/fit uncertainties are shown by the hatched bands. 
}
\label{fig:fittedpdfs-xmin-scan} 
\end{figure}
\newpage

\newpage


\appendix
\refstepcounter{section}
\section*{Appendix}
\label{appendix}
\setcounter{table}{0}
\renewcommand\thetable{\thesection.\arabic{table}}
\begin{flushleft}
Table~\ref{tab:sys} lists the sources of correlated uncertainties together with the shifts and reductions obtained as a result of the combination. 
Table~\ref{tab:fitpars} provides the central values of the fitted parameters.
\end{flushleft}

\begin{table}[t]
\begin{center}
\tabcolsep2.1mm
\renewcommand*{\arraystretch}{1.3}
	\begin{tabular}{|l|l|r|c|} 
\hline Dataset& Name & shift $[\sigma]$ & reduction factor \\
\hline
2--7,8c,9,10,11c, & theory, $m_c$ & $0.29$ & $0.65$ \\\hline
2--13 & theory $\mu_r,\mu_f$ variation & $-0.82$ & $0.45$ \\\hline
2--13 & theory, $\alpha_s(M_Z)$ & $0.17$ & $0.95$ \\\hline
1--7,8c,9,10 & theory, $c$ fragmentation $\alpha_K$ & $-0.82$ & $0.80$ \\\hline
2--7,8c,9,10 & theory, $c$ fragmentation $\hat{s}$ & $-1.44$ & $0.83$ \\\hline
2--7,8c,9,10 & theory, $c$ transverse fragmentation & $-0.10$ & $0.90$ \\\hline
2--7,10 & $f(c \to D^{*+})$ & $0.43$ & $0.92$ \\\hline
2--6,10 & ${\rm BR}(D^{*+} \to D^0 \pi^{+})$ & $0.14$ & $0.99$ \\\hline
2--7,10 & ${\rm BR}(D^{0} \to K^{-} \pi^{+})$ & $0.47$ & $0.98$ \\\hline
 1--4 & H1 CJC efficiency & $0.29$ & $0.78$ \\\hline
 2 & H1 integrated luminosity (1998-2000) & $-0.05$ & $0.97$ \\\hline
 2 & H1 trigger efficiency (HERA-I) & $-0.07$ & $0.94$ \\\hline
 2--4 & H1 electron energy & $0.29$ & $0.67$ \\\hline
 2--4 & H1 electron polar angle & $0.23$ & $0.74$ \\\hline
 2 & H1 MC alternative fragmentation & $-0.09$ & $0.68$ \\\hline
 3,4 & H1 primary vertex fit & $0.31$ & $0.98$ \\\hline
 1,3,4 & H1 hadronic energy scale & $-0.06$ & $0.81$ \\\hline
 3,4 & H1 integrated luminosity (HERA-II) & $-0.19$ & $0.77$ \\\hline
 3,4 & H1 trigger efficiency (HERA-II) & $-0.06$ & $0.98$ \\\hline
 3,4 & H1 fragmentation model in MC & $-0.17$ & $0.87$ \\\hline
 1,3,4 & H1 photoproduction background & $0.31$ & $0.91$ \\\hline
 3,4 & H1 efficiency using alternative MC model & $0.30$ & $0.71$ \\\hline
 1 & H1 vertex resolution & $-0.53$ & $0.88$ \\\hline
 1 & H1 CST efficiency & $-0.34$ & $0.89$ \\\hline
 1 & H1 B multiplicity & $0.26$ & $0.79$ \\\hline
 1 & H1 $D^{+}$ multiplicity & $-0.30$ & $0.94$ \\\hline
 1 & H1 $D^{*+}$ multiplicity & $-0.02$ & $0.98$ \\\hline
 1 & H1 $D_s^{+}$ multiplicity & $0.09$ & $0.97$ \\\hline
\end{tabular}
\end{center}
\caption{Sources of bin-to-bin correlated systematic uncertainties considered in the combination. 
For each source, the affected datasets are given, together with the cross-section shift induced by this source and the reduction factor of the correlated uncertainty in units of $\sigma$ after the first iteration. For those measurements which have simultaneously extracted charm and beauty cross sections, a suffix $b$ or $c$ indicates that the given systematic source applies only to the charm or beauty measurements, respectively. } 
\label{tab:sys}
\end{table}
\newpage
\begin{table}[t]
\begin{center}
\tabcolsep2.1mm
\renewcommand*{\arraystretch}{1.3}
	\begin{tabular}{|l|l|r|c|} 
\hline  Dataset& Name & shift $[\sigma]$ & reduction factor \\
\hline
 1 & H1 $b$ fragmentation & $-0.05$ & $0.96$ \\\hline
 1 & H1 VTX model: $x$ reweighting & $-0.20$ & $0.92$ \\\hline
 1 & H1 VTX model: $p_T$ reweighting & $-0.31$ & $0.68$ \\\hline
 1 & H1 VTX model: $\eta(c)$ reweighting & $-0.36$ & $0.80$ \\\hline
 1 & H1 VTX $uds$ background & $-0.14$ & $0.43$ \\\hline
 1 & H1 VTX $\phi$ of $c$ quark & $0.05$ & $0.84$ \\\hline
 1 & H1 VTX $F_2$ normalisation & $-0.05$ & $0.93$ \\\hline
 9,10,11 & ZEUS integrated luminosity (HERA-II) & $-1.24$ & $0.88$ \\\hline
 9,10,11 & ZEUS tracking efficiency & $0.03$ & $0.88$ \\\hline
 11 & ZEUS VTX decay length smearing (tail) & $-0.23$ & $0.96$ \\\hline
 9,10,11 & ZEUS hadronic energy scale & $0.08$ & $0.54$ \\\hline
 9,10,11 & ZEUS electron energy scale & $0.24$ & $0.55$ \\\hline
 11 & ZEUS VTX $Q^2$ reweighting in charm MC & $-0.10$ & $1.00$ \\\hline
 11 & ZEUS VTX $Q^2$ reweighting in beauty MC & $0.04$ & $1.00$ \\\hline
 11 & ZEUS VTX $\eta({\rm jet})$ reweighting in charm MC & $-0.57$ & $0.97$ \\\hline
 11 & ZEUS VTX $\eta({\rm jet})$ reweighting in beauty MC & $0.10$ & $0.99$ \\\hline
 11 & ZEUS VTX $E_T({\rm jet})$ reweighting in charm MC & $0.48$ & $0.96$ \\\hline
 11 & ZEUS VTX $E_T({\rm jet})$ reweighting in beauty MC & $-0.43$ & $0.92$ \\\hline
 11 & ZEUS VTX light-flavour background & $0.48$ & $0.85$ \\\hline
 11 & ZEUS VTX charm fragmentation fucntion & $-0.91$ & $0.87$ \\\hline
 11 & ZEUS VTX beauty fragmentation fucntion & $-0.17$ & $0.95$ \\\hline
 9 & $f(c \to D^{+})$ & $-0.11$ & $0.94$ \\\hline
 9 & $BR(D^{+} \to K^{-}\pi^{+}\pi^{+})$ & $-0.10$ & $0.95$ \\\hline
 9 & ZEUS $D^{+}$ decay length smearing & $0.05$ & $0.99$ \\\hline
 9,10 & ZEUS beauty MC normalisation & $0.67$ & $0.85$ \\\hline
 9 & ZEUS $D^{+}$ $\eta$ MC reweighting & $0.23$ & $0.85$ \\\hline
 9 & ZEUS $D^{+}$ $p_T$, $Q^2$ MC reweighting & $0.92$ & $0.66$ \\\hline
 9 & ZEUS $D^{+}$ MVD hit efficiency & $-0.04$ & $0.99$ \\\hline
 9 & ZEUS $D^{+}$ secondary vertex description & $-0.08$ & $0.97$ \\\hline
 5,13 & ZEUS integrated luminosity (1996-1997) & $0.57$ & $0.95$ \\\hline
\end{tabular}
\end{center}
\captcont{continued} 
\end{table}
\newpage
\begin{table}[t]
\begin{center}
\tabcolsep2.1mm
\renewcommand*{\arraystretch}{1.3}
	\begin{tabular}{|l|l|r|c|} 
\hline  Dataset& Name & shift $[\sigma]$ & reduction factor \\
\hline
 6,13 & ZEUS integrated luminosity (1998-2000) & $0.42$ & $0.87$ \\\hline
 10 & ZEUS $D^{*+}$ $p_T(\pi_s)$ description & $0.84$ & $0.92$ \\\hline
 10 & ZEUS $D^{*+}$ beauty MC efficiency & $-0.17$ & $0.97$ \\\hline
 10 & ZEUS $D^{*+}$ photoproduction background & $0.39$ & $0.96$ \\\hline
 10 & ZEUS $D^{*+}$ diffractive background & $-0.35$ & $0.92$ \\\hline
 10 & ZEUS $D^{*+}$ $p_T$, $Q^2$ MC reweighting & $-0.45$ & $0.91$ \\\hline
 10 & ZEUS $D^{*+}$ $\eta$ MC reweighting & $0.34$ & $0.77$ \\\hline
 10 & ZEUS $D^{*+}$ $\Delta(M)$ window efficiency & $-0.77$ & $0.92$ \\\hline
 7 & $f(c \to D^{0})$ & $0.32$ & $0.99$ \\\hline
 7,8,12 & ZEUS integrated luminosity (2005) & $0.66$ & $0.91$ \\\hline
 8c & $BR(c \to l)$ & $-0.10$ & $0.97$ \\\hline
 8 & ZEUS $\mu$: B/RMUON efficiency & $0.54$ & $0.90$ \\\hline
 8 & ZEUS $\mu$: FMUON efficiency & $0.15$ & $0.95$ \\\hline
 8 & ZEUS $\mu$: energy scale & $-0.01$ & $0.67$ \\\hline
 8 & ZEUS $\mu$: $p_T^{\rm miss}$ calibration & $0.13$ & $0.66$ \\\hline
 8 & ZEUS $\mu$: hadronic resolution & $0.62$ & $0.58$ \\\hline
 8 & ZEUS $\mu$: IP resolution & $-0.70$ & $0.83$ \\\hline
 8 & ZEUS $\mu$: MC model & $-0.08$ & $0.75$ \\\hline
 1b & H1 VTX beauty: $Q^2$ charm reweighting & $-0.02$ & $1.00$ \\\hline
 1b & H1 VTX beauty: $Q^2$ beauty reweighting & $-0.02$ & $0.99$ \\\hline
 1b & H1 VTX beauty: $x$ reweighting  & $0.09$ & $0.89$ \\\hline
 1b & H1 VTX beauty: $p_T$ reweighting & $-1.06$ & $0.82$ \\\hline
 1b & H1 VTX beauty: $\eta$ reweighting & $0.01$ & $0.91$ \\\hline
 1b & H1 VTX beauty: BR($D^{+})$ & $-0.21$ & $0.99$ \\\hline
 1b & H1 VTX beauty: BR($D^0$) & $0.16$ & $1.00$ \\\hline
 8b,11b,12,13 & theory, $m_b$ & $0.60$ & $0.93$ \\\hline
 8b,12,13& theory, $b$ fragmentation & $-0.71$ & $0.97$ \\\hline
 8b,12,13, & $BR(b \to l)$ & $-0.60$ & $0.97$ \\\hline
 13 & ZEUS muon efficiency (HERA-I) & $-1.02$ & $0.91$ \\\hline
\end{tabular}
\end{center}
\captcont{continued} 
\end{table}

\newpage
\begin{table}[!hp]
  \renewcommand{\arraystretch}{1.8}
  \centering
  \begin{tabular}{|c|c|c|c|c|c|c|c|}
  \hline
   & $A$ & $B$ & $C$ & $D$ & $E$ & $A'$ & $B'$ \\
  \hline
  $xg$ & $2.81$ & $-0.198$ & $8.14$ & $$ & $$ & $1.39$ & $-0.273$ \\
  $xu_v$ & $3.66$ & $0.678$ & $4.87$ & $$ & $14.7$ & $$ & $$ \\
  $xd_v$ & $3.38$ & $0.820$ & $4.27$ & $$ & $$ & $$ & $$ \\
  $x\overline{U}$ & $0.102$ & $-0.172$ & $8.27$ & $13.9$ & $$ & $$ & $$ \\
  $x\overline{D}$ & $0.170$ & $-0.172$ & $5.83$ & $$ & $$ & $$ & $$ \\
  \hline
  $m_c(m_c)$ [GeV] & \multicolumn{7}{c|}{$1.29$} \\
  $m_b(m_b)$ [GeV] & \multicolumn{7}{c|}{$4.05$} \\
  \hline
  \end{tabular}
  \caption{Central values of the fitted parameters of \pdfhq~(see equation~\ref{eq:dv}).}
  \label{tab:fitpars}
\end{table}

\unitlength1cm
\end{document}